\documentclass[a4paper, 10pt]{article}
\usepackage{amssymb}
\usepackage{amsmath}
\usepackage{graphics}
\usepackage{epsfig}
\usepackage{subfigure}
\usepackage{cite}
\usepackage{titling}
\usepackage{color}
\usepackage[font=footnotesize,labelfont=bf]{caption}
\usepackage[left=2.5cm, right=2.5cm]{geometry}
\usepackage{bm}
\usepackage[utf8]{inputenc}
\usepackage[T1]{fontenc}
\usepackage{mathptmx}

\graphicspath{{figure/}}   
\begin{document}

\title{The critical behavior of Hegselmann-Krause opinion model with smart agents}
\author{Yueying Zhu$^{1,*}$, Jian Jiang$^{1}$, and Wei Li$^{2,3}$ \vspace{0.2cm}\\
\small $^1$ Research Center of Nonlinear Science, College of Mathematics and Computer Science, \\
\small Wuhan Textile University, 430200 Wuhan, China\\
\small $^2$ Complexity Science Center \& Institute of Particle Physics,\\
\small Central China Normal University, 430079 Wuhan, China\\
\small $^3$ Max-Planck Institute for Mathematics in the Sciences,\\
\small Inselst. 22, 04103 Leipzig, Germany\\
\small $^*$ Correspondence author: yyzhu@wtu.edu.cn
}

\date{}
\maketitle

\abstract{
The Hegselmann-Krause (HK) model allows one to characterize the continuous change of agents' opinions with the bounded confidence threshold $\varepsilon$. To consider the heterogeneity of agents in characteristics, we study the HK model on homogeneous and heterogeneous networks by introducing a kind of smart agent. Different from the averaging rule in opinion update of HK model, smart agents will consider, in updating their opinions, the environmental influence following the fact that an agent's behavior is often coupled with environmental changes. The environment is characterized by a parameter that represents the biased resource allocation between different cliques. We focus on the critical behavior of the underlying system. A phase transition point separating a complete consensus from the coexistence of different opinions is identified, which occurs at a critical value $\varepsilon_c$ for the bounded confidence threshold. We state analytically that $\varepsilon_c$ can take only one of two possible values, depending on the behavior of the average degree $k_a$ of a social graph, when agents are homogeneous in characteristics. Results also suggest that the phase transition point weakly depends on the network structure but is strongly correlated with the fraction of smart agents and the environmental parameter. We finally establish the finite size scaling law that stresses the role that the system size has in the underlying opinion dynamics. Meanwhile, introducing smart agents does not change the functional dependence between the time to reach a complete consensus and the system size. However, it can drive a complete consensus to be reached faster, for homogeneous networks that are far from the mean field limit.
}

\section{Introduction}\label{sec1}

The dynamics regarding human social behavior has been widely studied in many disciplines, as well as in their intersections \cite{xia2012evaluating,huang2015experimental,an2018spontaneous}. In this field, one hot topic is the opinion dynamics that has been largely studied in the literature, both analytically and by means of large-scale numerical simulations \cite{chen2017first,Shang2019Opinion,nguyen2020dynamics}. In the field of opinion research, people have been trying to find out how a common opinion or decision is formed on top of a given population \cite{dong2018consensus,sood2005voter}. Historically, the classical modeling of opinion evolution assumes homogeneously mixed populations, which implies that any two individuals in the population are connected at the same probability \cite{liggett2013stochastic,Benatti2020}. However, such an approach leads to fluctuations in fitting to real systems, due to the complexity in individual's characteristics and the stochastic nature in communications among people \cite{castellano2009statistical}. With the advent of modern network science, we have witnessed significant influence of heterogeneously mixed patterns in real systems on the critical behavior of opinion dynamics \cite{suchecki2005voter,holzer2017pattern}.

A set of agents are considered with each one assigned a special opinion from a real space. The opinion dynamics is then formed by the fact that an arbitrary agent may update its opinion when it is aware of the opinions of others. Discrete opinion space is first focused by researchers especially those in physics because of its simplicity to analytic work \cite{Zhou10052,guo2013heterogeneity,chen2015nucleation,min2019multilayer}. It represents a reasonable description in several instances but fails to explain how an individual varies smoothly its attitude from one extreme of the range of possible choices to the other. The political orientation of an individual, as an example, is not limited to the binary choices of left or right but may take any one possible position in between. Hence, several mathematical models of continuous opinion dynamics are then designed to reveal the underlying mechanism that determines the smooth change in human psychology \cite{lorenz2007continuous,Martins2008Bayesian}.

As a typical example of continuous opinion dynamics, HK model was first proposed by Hegselmann and Krause in 2002 \cite{hegselmann2002opinion}. It introduces a realistic aspect of human communications that a discussion between two people usually happens only if their opinions differ from each other no larger than a certain parameter $\varepsilon$. This parameter is named the bounded confidence threshold. In HK model, an agent $i$ takes a real value between 0 and 1 as its initial opinion. The update rule for its opinion at time $t$ is defined mathematically by
\begin{equation}
o_i(t+1)=\frac{1}{|N_i(t)|}\sum_{j\in N_i(t)}o_j(t),
\label {eq1}
\end{equation}
where $|\cdot|$ for a finite set denotes the number of elements. $N_i(t)$ collects compatible neighbors of agent $i$, saying
\begin{equation}
N_i(t)=\{j, j\in N||o_i(t)-o_j(t)|\leq \varepsilon \text{ and } a(i,j)=1\},
\label{eq2}
\end{equation}
where $a(i,j)$ is the element of the adjacency matrix $\bm{A}$ that mathematically formulates the network of agents and their interpersonal relations. $a(i,j)=1$ if there is an edge from agent $i$ to agent $j$ and $a(i,j)=0$ otherwise.

The study of opinion dynamics has broad applications to maintaining social stability and to keeping people safe, mainly including the public decision prediction in political election, the rumor spread tracking on online social communicating network, and the malignant group event prevention in real social systems \cite{salathe2008effect,morarescu2010opinion,burghardt2016competing}. Heterogeneity of agents in real communications, usually represented by topologically heterogeneous networks, has been well studied in opinion dynamics \cite{Shao2009Dynamic,Yang2015Opinion,anderson2019recent}. In recent years, the heterogeneity of real agents in characteristics also comes to be widely considered, such as the introduction of leader \cite{dong2017managing,wen2017bipartite} and stubborn agent \cite{Tian2018Opinion,han2019opinion}, the consideration of heterogeneous bounded confidence threshold \cite{liang2013opinion}. Most available literature concerning heterogeneous characteristics of agents in HK model mainly focuses on the Monte Carlo simulation or empirical analysis of opinion formation by neglecting the coupling effect between agents' behaviors and environmental changes \cite{meng2018opinion,yin2019agent,zhu2020neural}.

In recent social opinion research, some numerical results demonstrate the great effect of environmental changes on the evolution of agents' opinions which also in turn reacts on the social environment  \cite{liu2019coevolution,Freitas2020Imperfect,li2020effect}. With this consideration, we introduce in this work a virtual gambling game to design the coupling mechanism between environmental changes and the evolution of agents' opinions. The opinion of an agent represents the cost or the degree of attention that it pays for the issue of gambling. We then devote ourselves to the study of the critical behavior of opinion dynamics in HK model, after introducing both the heterogeneity of agent's characteristics and that of social communications. The social environment considered here is represented by a resource allocation among agents that is directly determined by agents' opinions. The resource that an agent shares from the system decides whether it will win or not. A rewards and punishment system is then proposed, with which agents can accumulate scores and the scores further affect their opinions. Actually, similar ideas to the concept of score have been proposed in earlier modeling of social behaviors. In the modeling of discrete opinion dynamics, for instance, a parameter is introduced to indicate the convincing power of each agent \cite{Nuno2014The}. It may increase or decrease with time, similar to the concept of score, and decides whether the majority opinion succeeds or not in a randomly selected group. Another example is the naming game where the concept of score is introduced to represent the agent's reputation which is variable in time and partially controls the behavior of the underlying system \cite{Brigatti2008Consequence}.

In HK model an agent updates its opinion by adopting the average opinion computed over its compatible neighbors. In real-world systems, however, agents have different abilities to persuade others as well as different tendencies to change or keep their own opinions. In general, human beings are recognised as "higher animal" because of self-awareness and the freedom from nature's determinism that allows us to choose, whether for good or ill. Some people, however, may behave irrationally in facing some activities. This maybe because the underlying activity does not interest them or they do not care about their personal gains. Based on this phenomenon, we consider in our system two kinds of agents: smart ones and general ones. In the opinion evolution, smart agents will instinctively learn from their friends who have earned the most scores from the gambling activity, because of their primary intend of wining money and/or material goods. General agents will follow the averaging rule of HK model in updating their opinions, as illustrated in Eq. (\ref{eq1}). With the model, we attempt to describe the formation of a common opinion in a real-world system of agent's heterogeneity and the coupling between agents' behaviors and environmental changes, from both numerical and theoretical viewpoints. Furthermore, finite size effect analysis is also conducted to help understand the critical behavior of the underlying system.

The remainder of the paper is organized as follows. In the second section we introduce smart agents to HK model and define the opinion update rule for heterogeneous agents. In section 3, we state the consensus threshold $\varepsilon_c$ of the underlying opinion dynamics, from both analytic and numerical points of view. Section 4 focuses on the critical behavior of the underlying system. We establish the scaling law of order parameter, and explain analytically the influence of environmental and heterogeneous parameters. Some conclusions are drawn in section 5.

\section{Model definition}\label{sec2}
Opinions of agents are assigned initially by real values between -1 and 1. Agents in our system are then divided into two cliques based on the sign of their opinions: $G_+$ (for positive opinions) and $G_-$ (for negative opinions). This setting, similar to the El Farol game of agents having two possible choices, i.e.  whether to go or not to go to a bar \cite{Marsili2004Shedding}, has certain practical significance.

The coupling mechanism between the evolution of agents' opinions and environmental changes is designed by a virtual gambling game. The environment of the underlying system is quantified by the resource allocation among agents. Considering the biased resource allocation in a real-world system, we introduce a parameter $\gamma$ to quantify the resource discrepancy between two cliques: $\gamma=R_+/R_-$, where $R_+$ labels the resource allocated to $G_+$ and $R_-$ to $G_-$. The value of $\gamma$ is restricted to the closed region [0, 1] as the system's behavior is symmetric with respect to $\gamma=1$. In our system, we can regard the absolute value of an agent's opinion as the cost or the degree of attention that it pays for the activity of gambling. Moreover, an agent who pays more cost or attention will share more resource (or information) according to the general rule of survival: the more you contribute, the more you get \cite{Roberta2013The,Hadani2017The,Proserpio2018You}. On this basis, the resource $r_i(t)$ that agent $i$ can share, at time $t$, from its clique is proportional to the absolute value of its current opinion $o_i(t)$:
\begin{equation}
r_i(t)=\left\{
\begin{array}{ll}
\frac{o_i(t)}{\Omega^+(t)}\cdot R_+ & \text{if $o_i(t)>0$}\\
\\
\frac{o_i(t)}{\Omega^-(t)}\cdot R_- & \text{if $o_i(t)\leq 0$}
\end{array}
\right.
\label{resource}
,
\end{equation}
where
\begin{equation}
\Omega^+(t)=\sum_{j\in G_+}o_j(t), \qquad \Omega^-(t)=\sum_{j\in G_-}o_j(t).
\end{equation}
The total resource $R (=R_++R_-)$ in the system is supposed to be conserved (unchanged with time evolution and system size). For simplicity, $R$ is set to be 1000 in numerical simulations, but its value does not influence the presented results.

Generally speaking, the primary intend of people joining in a gambling game is to gain money and/or material goods \cite{weibull1997evolutionary}. In our model, smart agents have the ability to gain more scores by learning from their friends. Parameter $p$ is used to control the fraction of general agents who follow the averaging rule of HK model in updating their opinions. The fraction of smart agents in the system then is $(1-p)$. $p$ can take any one real value between 0 and 1. In the system, opinions determine the resource distribution among agents and further control the scores that agents gain from the gambling activity. Following the primary intend of smart agents to gain money and/or material goods, a smart agent will update its opinion by the average one of compatible friends who have gained the most scores. If its scores are the highest compared to its compatible friends, it will keep the current opinion. A rewards and punishment mechanism is then designed based on the resource that an agent shares from the system. An agent is regarded as a winner and to be rewarded with fixed scores $f_0$ if its resource is no less than the global average $A_r$ ($=R/N$). Otherwise, it is labeled as a loser and to be punished with losing $f_0$ scores. The update rule for the number of cumulative scores of agent $i$ at time $t$ reads
\begin{equation}
f_i(t+1)=\left \{
\begin{array}{ll}
f_i(t)+f_0 & \text{if $r_i(t)\geq A_r$}\\
\\
f_i(t)-f_0 & \text{if $r_i(t)< A_r$}
\end{array}
\right.
.
\end{equation}
$f_0$ is constant for each agent and set to be 5 in our numerical simulations, but its value does not influence the results that we presented in this paper.

The update rule for the opinion of smart agent $i$ is given mathematically by
\begin{equation}
o_i(t+1)=\delta_{f_i(t)m_i(t)}o_i(t)+(1-\delta_{f_i(t)m_i(t)})\frac{1}{|M_i(t)|}\sum_{j\in M_i(t)}o_j(t),
\end{equation}
where $\delta_{xy}$ is the Kronecker delta function with $\delta_{xy}=1$ if $x=y$ and $\delta_{xy}=0$ otherwise.
\begin{equation}
m_i(t)=\max(f_i(t), \{f_j(t), j\in N_i(t)\}),
\end{equation}
\begin{equation}
M_i(t)=\{j, j\in N_i(t)|f_j(t)=m_i(t)\},
\end{equation}
where $\max()$ is a function by returning the maximum value in a set of real numbers. $N_i(t)$ is defined in Eq.~(\ref{eq2}). For a general agent $i$, specifically, its opinion will be updated by the averaging rule of HK model, as illustrated in Eq.~(\ref{eq1}).

It is of interest to identify the critical behavior of the extended HK model, and to explore how smart agents, biased resource allocation, as well as graph topology influence the behavior of the underlying system. Basically, we carry out simulations of the model on different network topologies and determine in each case the value of the consensus threshold and the critical behavior. We analyze three different types of networks: (1) a complete network where any two agents can talk to each other (the mean field limit); (2) a regular network of the nearest-connections (an example of homogeneous networks); (3) a Barab\'asi-Albert (BA) scale-free network (an example of heterogeneous networks). In Monte Carlo simulations, we chose to update the opinions of agents in ordered sweeps over the population. The system is considered to reach a stable state if any opinion changes by less than $10^{-8}$. Specifically, a complete consensus is defined whenever the opinion distance between any two agents is less than 0.01. Simulated results illustrated hereafter are all obtained by averaging over 1000 independent realizations.

\section{The consensus threshold}\label{sec3}

A bifurcation diagram has been presented in previous work for HK model with uniform initial density in the opinion space \cite{lorenz2007continuous}. It states that the dynamics of HK model leads to the pattern of stationary states with the number of final opinion clusters decreasing with the increase of $\varepsilon$. Thus, it is of interest to identify the consensus threshold $\varepsilon_c$ of the bounded confidence parameter $\varepsilon$, that specifies a phase transition from disordered state (coexistence of different opinions) to an ordered one (a complete consensus). In particular, for $\varepsilon$ above consensus threshold $\varepsilon_c$, a group opinion is certainly formed. Most investigations regarding the opinion formation of HK model often set the opinion space to be $[0, 1]$. Fortunato, on this basis, has claimed based on numerical results that the threshold $\varepsilon_c$ for a complete consensus can only take one of two possible values, depending on the behavior of the average degree $k_a$ of the underlying social graph, when the number of agents $N$ approaches infinity \cite{fortunato2005}. If $k_a$ stays finite in the limit of large $N$, $\varepsilon_c=0.5$. Instead, if $k_a \to \infty$ when $N \to \infty$, $\varepsilon_c=0.2$. It is natural to extend the statement to HK model of initial opinion space being [-1, 1] with $\varepsilon_c=1$ for constant $k_a$ and $\varepsilon_c=0.4$ if $k_a$ diverges. This is because the consensus threshold $\varepsilon_c$ linearly depends on the maximal opinion distance $g_o$ at initial time, see Fig. \ref{opinion:gap}$(a)$. An analytic explanation for the underlying results is presented in Appendix \ref{appendix:1}.

\begin{figure}
\centering
	\includegraphics[width=0.4\textwidth]{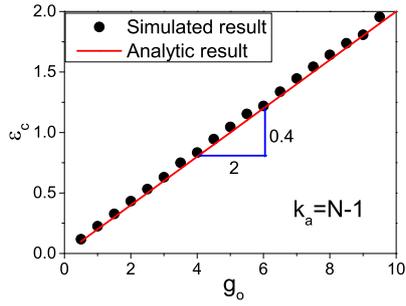}
    \includegraphics[width=0.4\textwidth]{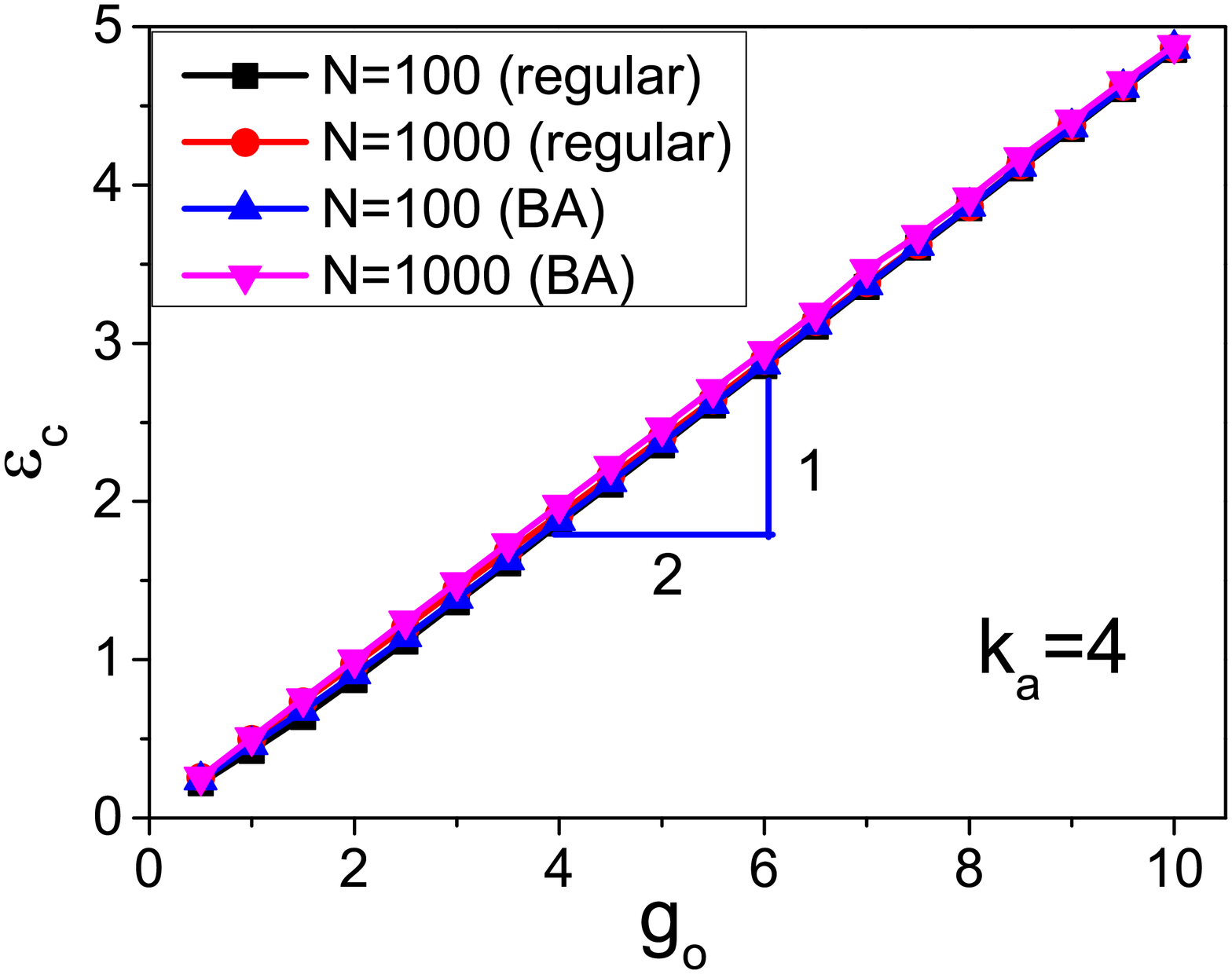}\\
    {\footnotesize $(a)$ $\varepsilon_c$ changes with $g_o$ for $p=1$ (HK model).}\\
    \includegraphics[width=0.4\textwidth]{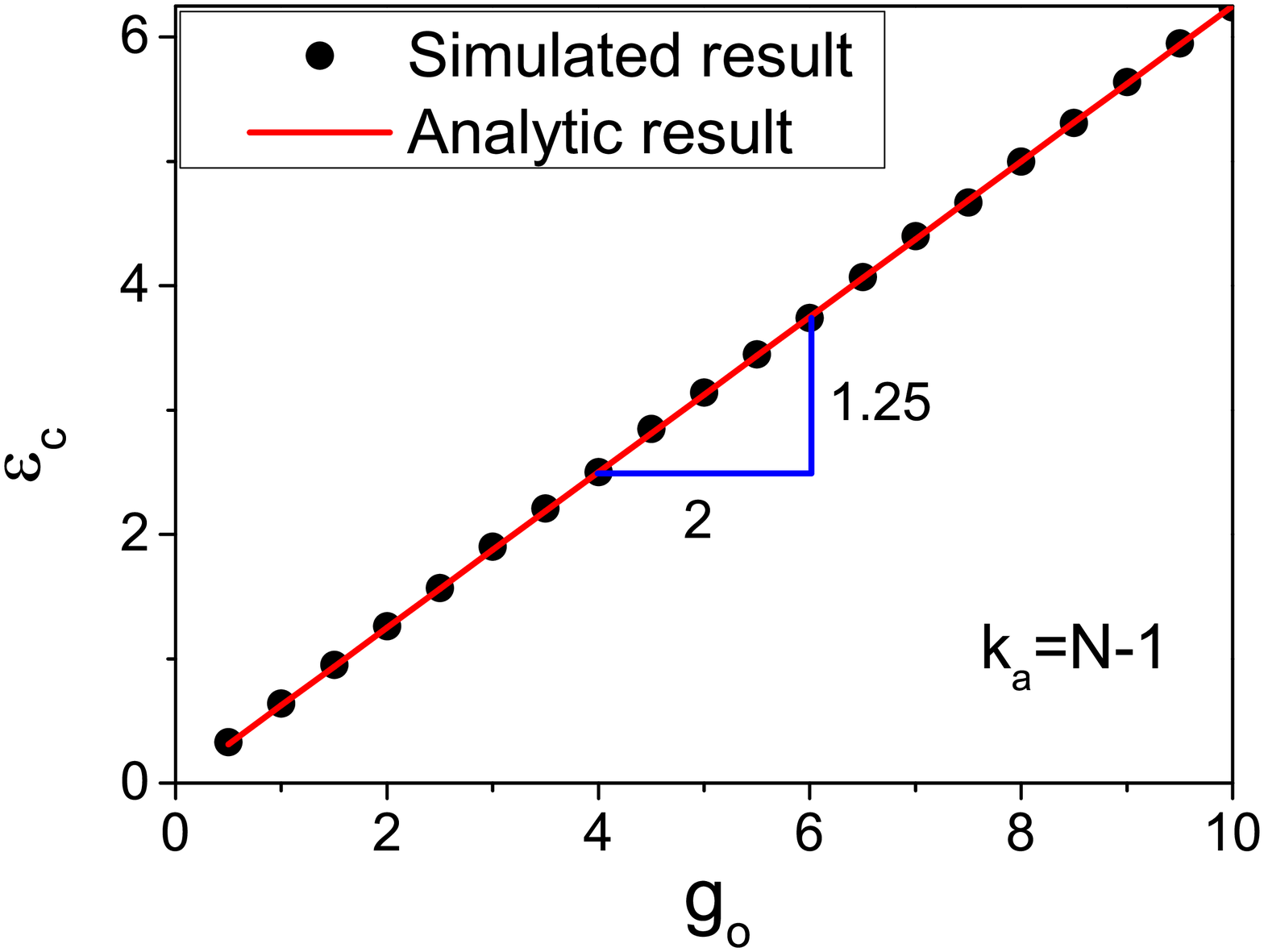}
    \includegraphics[width=0.4\textwidth]{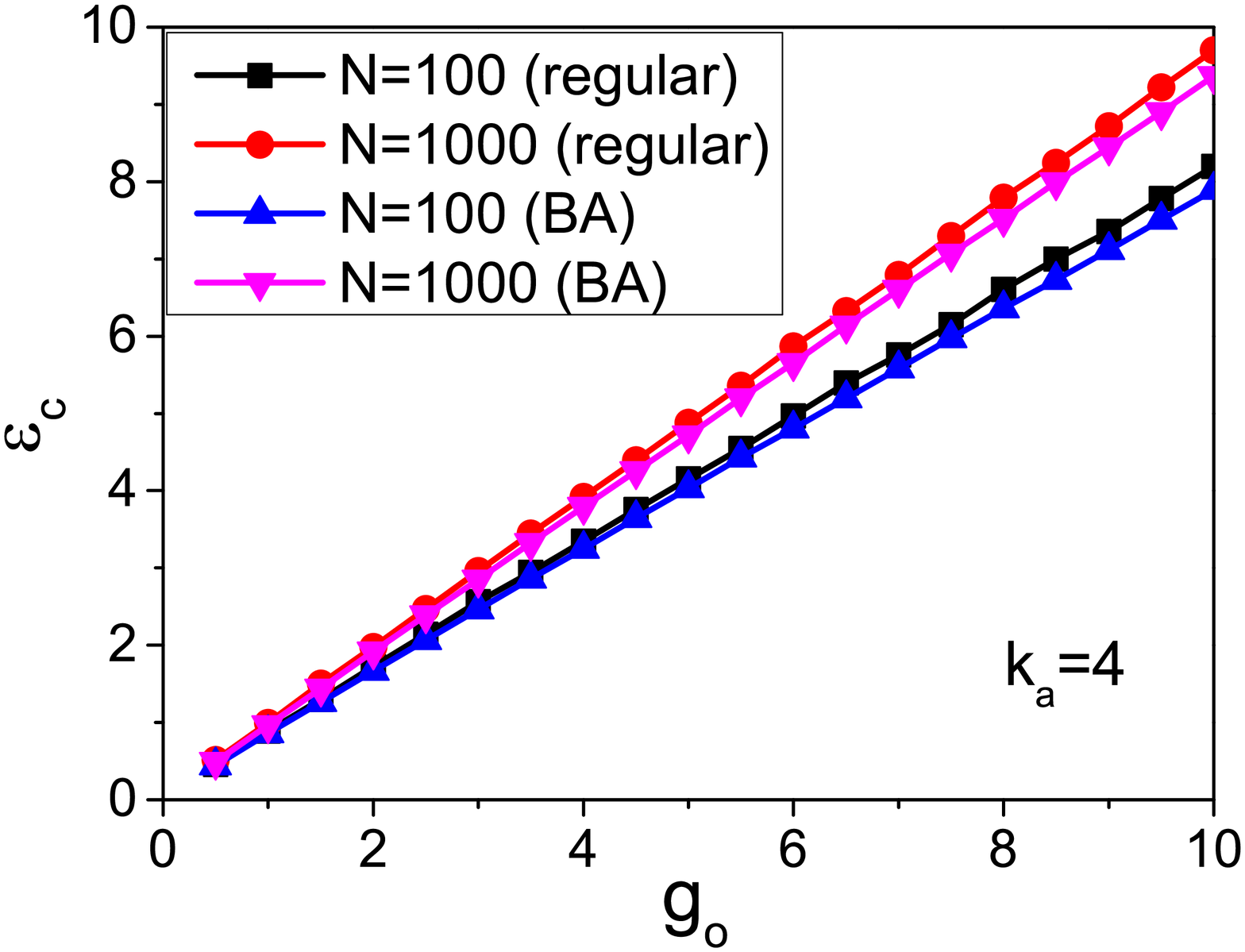}\\
    {\footnotesize $(b)$ $\varepsilon_c$ changes with $g_o$ for $p=0$ (the case of all agents being smart).}\\
	\caption{(color online) Analytic values are compared to numerical ones for the linear correlation between consensus threshold $\varepsilon_c$ and the initial opinion change distance $g_o$ for HK model ($a$) and the case of all agents being smart ($b$) with $\gamma=0$. In the left panels, $N=1000$. Numerical results are averaged over 1000 realizations.}
	\label{opinion:gap}
\end{figure}

We first discuss how smart agents impact the consensus threshold $\varepsilon_c$. Take the case of $p=0$ as a typical example, saying all agents being smart. In this case, agents who gained scores at $t=0$ have the ability to affect the opinions of others. Following this statement, the opinion at initial time is natural to be obtained, with which an agent can gain scores from the beginning of gambling,
\begin{equation}
|o_i|\geq \left \{
\begin{array}{ll}
\frac{(1+\gamma)O_2}{4\gamma} & o_i>0,\\
\\
\frac{(1+\gamma)|O_1|}{4} & o_i \leq 0,
\end{array}
\right.
\end{equation}
where agents' opinions are supposed to be uniformly distributed, at $t=0$, in the real range $[O_1, O_2]$ with $O_2=-O_1$. Considering a particular case where the resource is concentrated to the clique of agents holding negative opinions ($\gamma=0$). The opinion space of agents then can be divided into two adjacent intervals, including $R1$: $[O_1, \frac{O_1}{4}]$ and $R2$: $(\frac{O_1}{4}, O_2]$. Agents with opinions being interval $R1$ will persuade agents holing opinions from $R2$. In the mean filed limit, the gap between $\frac{O_1}{4}$ and $O_2$ should be no larger than the previously specified confidence threshold $\varepsilon$ to form a group opinion, that is
\begin{equation}
O_2-\frac{O_1}{4}\leq \varepsilon.
\end{equation}
Recalling the condition $O_1=-O_2$, we then get the functional dependence between the initial opinion gap $g_o (=O_2-O_1)$and the consensus threshold $\varepsilon_c$:
\begin{equation}
\varepsilon_c=\frac{5}{8}g_o,
\end{equation}
which agrees completely with numerical simulations of the agent-based model, see Fig. \ref{opinion:gap}$(b)$. For sparse networks, this linear correlation slightly depends upon the underlying system size, but is independent of the network structure. This differs from the behavior of HK model where the correlation between $\varepsilon_c$ and $g_o$ depends neither on the system size nor on the network structure, even when the network is very sparse. Furthermore, compared to HK model, the introduction of smart agents can strengthen, in the mean field limit, the linear correlation between $\varepsilon_c$ and $g_o$.

Interestingly, the opinion dynamics of smart agents leads to the pattern of stationary states with the number of final opinion clusters decreasing if $\varepsilon$ increases, similar to that of HK model. The consensus threshold $\varepsilon_c$ can also only take two possible values when the system size $N$ tends to be infinity. $\varepsilon_c=2$ if the average degree $k_a$ is finite in the limit of large $N$. If instead $k_a \to \infty$ when $N \to \infty$, $\varepsilon_c=1.25$. This is verified by the functional relationship between $\varepsilon_c$ and $k_a$ (Ref.\cite{zhu2017formation}):
\begin{equation}
\varepsilon_c=\left\{
\begin{array}{ll}
\frac{5}{4}+\frac{2}{k_a}, & \text{if } k_a\geq\frac{8}{3};\\
\\
2, & \text{if } k_a< \frac{8}{3}.
\end{array}
\right.
\label{equationk}
\end{equation}

\begin{figure}
\centering
	\includegraphics[width=0.4\textwidth]{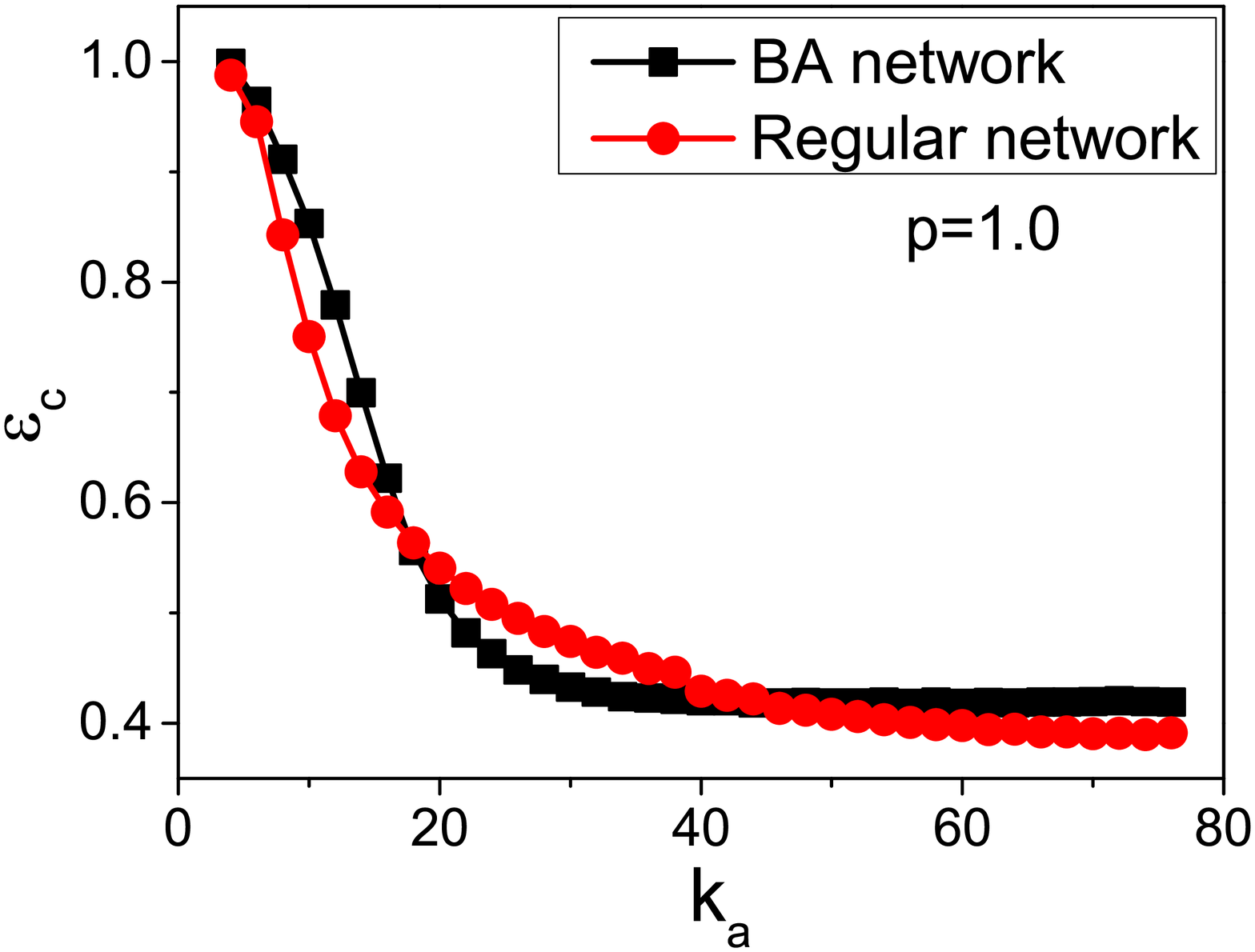}
\includegraphics[width=0.4\textwidth]{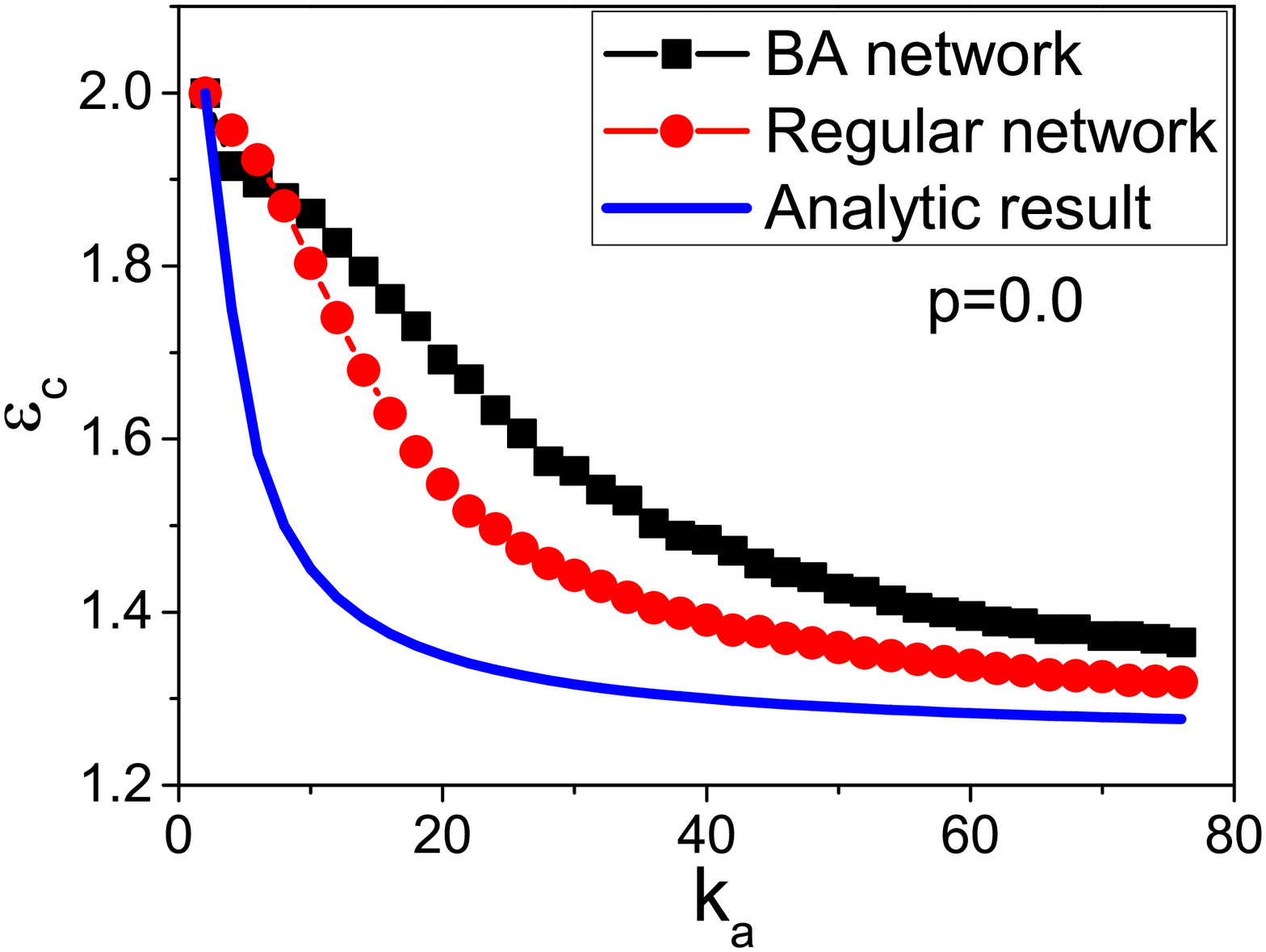}
	\caption{(color online) The consensus threshold $\varepsilon_c$ as a function of the average degree $k_a$ for BA scale-free and regular networks with $\gamma=0$ and $N=1000$. When $k_a$ is quite small, $\varepsilon_c=1$ for $p=1$ and $\varepsilon_c=2$ for $p=0$. If instead $k_a$ is large enough, $\varepsilon_c=0.4$ for $p=1$ and $\varepsilon_c=1.25$ for $p=0$. Each data point is averaged over 1000 realizations and the curve without markers in the right panel is obtained by the mean-field approximation.}
\label{opinion:degree}
\end{figure}

However, for a social system of finite number of agents, the consensus threshold $\varepsilon_c$ will decrease with the increase of $k_a$ in both cases of $p=1$ and $p=0$, as presented in Fig. \ref{opinion:degree}. Homogeneous and heterogeneous network structures are both considered here. In HK model, the reliance of $\varepsilon_c$ on $k_a$ is almost independent of the network structure, as stated previously. When all agents are smart, however, $\varepsilon_c$ of BA scale-free (heterogeneous) networks is larger than that of regular (homogeneous) ones, for each specified $k_a$. This suggests that a lower confidence threshold could guarantee the formation of a group opinion in the homogeneous network, compared to the heterogeneous one. Simulation results will gradually agree with analytic ones (Eq.~(\ref{equationk})) that are obtained by the mean field approximation.

\section{The critical behavior}

\subsection{Opinion clustering}

Compared to the bifurcation diagram, the structure of clusters formed by agents holding the same opinion provides more detailed information in uncovering the behavior of opinion fusion \cite{li2013consensus}. In our model, a cluster is formed by agents whose opinions differ from each other less than 0.01. Denote the number of clusters at stationary state by $M$. At $t=0$, opinions of agents are randomly assigned by real values between -1 and 1, making the maximum of $M$ to be $M_0$ (=2/0.01=200).

\begin{figure}
\centering
    \includegraphics[width=0.3\textwidth]{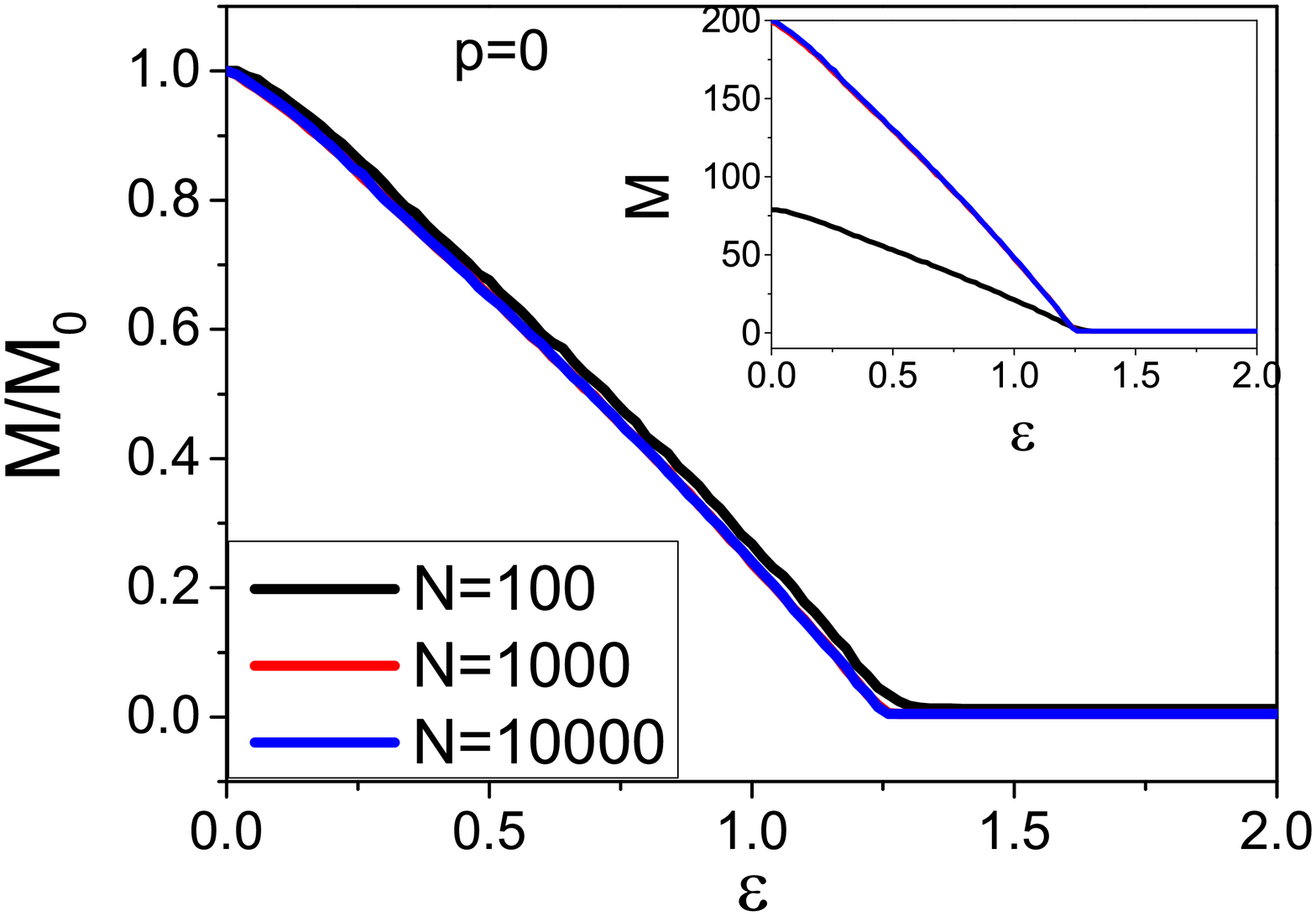}
    \includegraphics[width=0.3\textwidth]{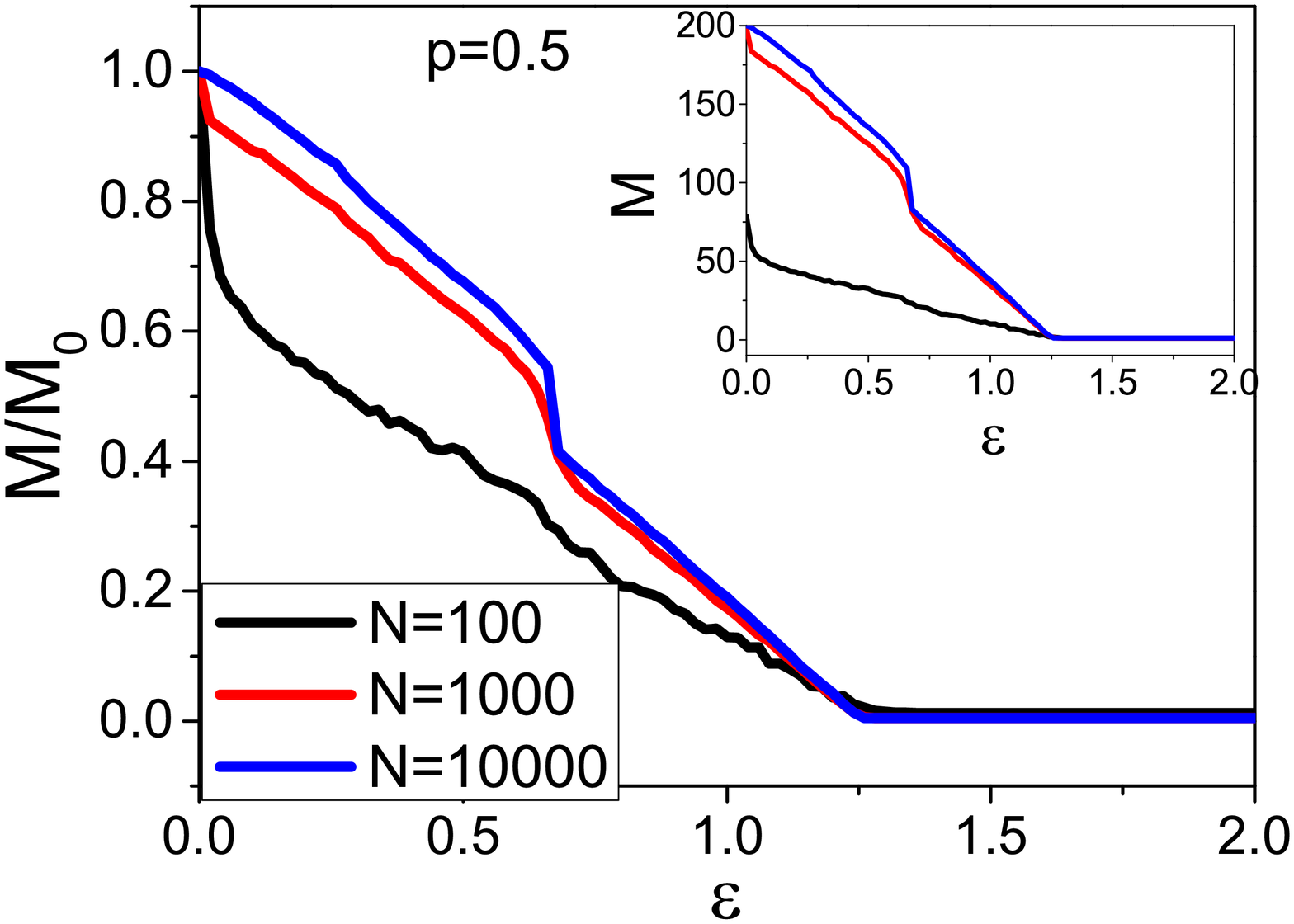}
    \includegraphics[width=0.3\textwidth]{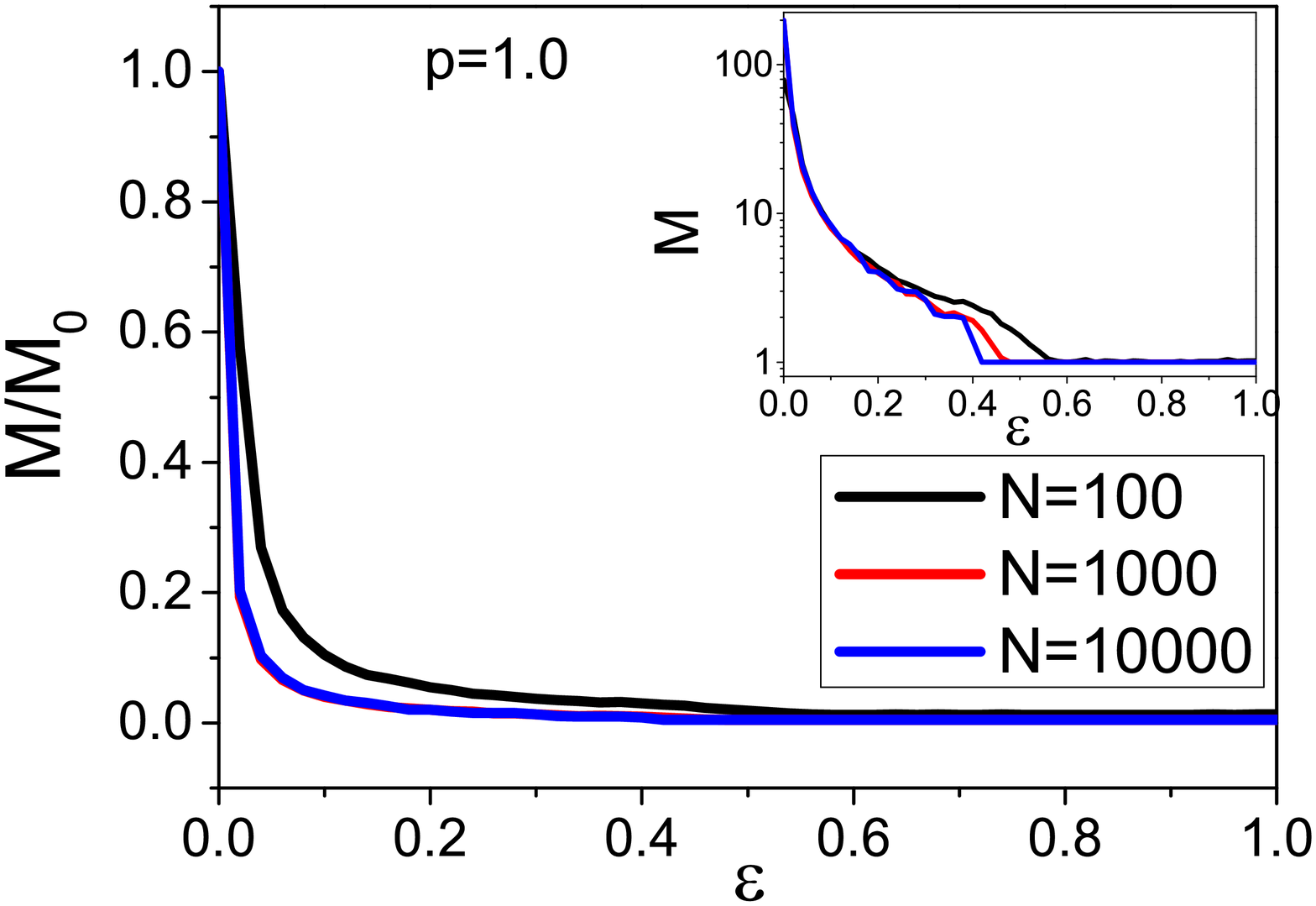}\\
    \includegraphics[width=0.3\textwidth]{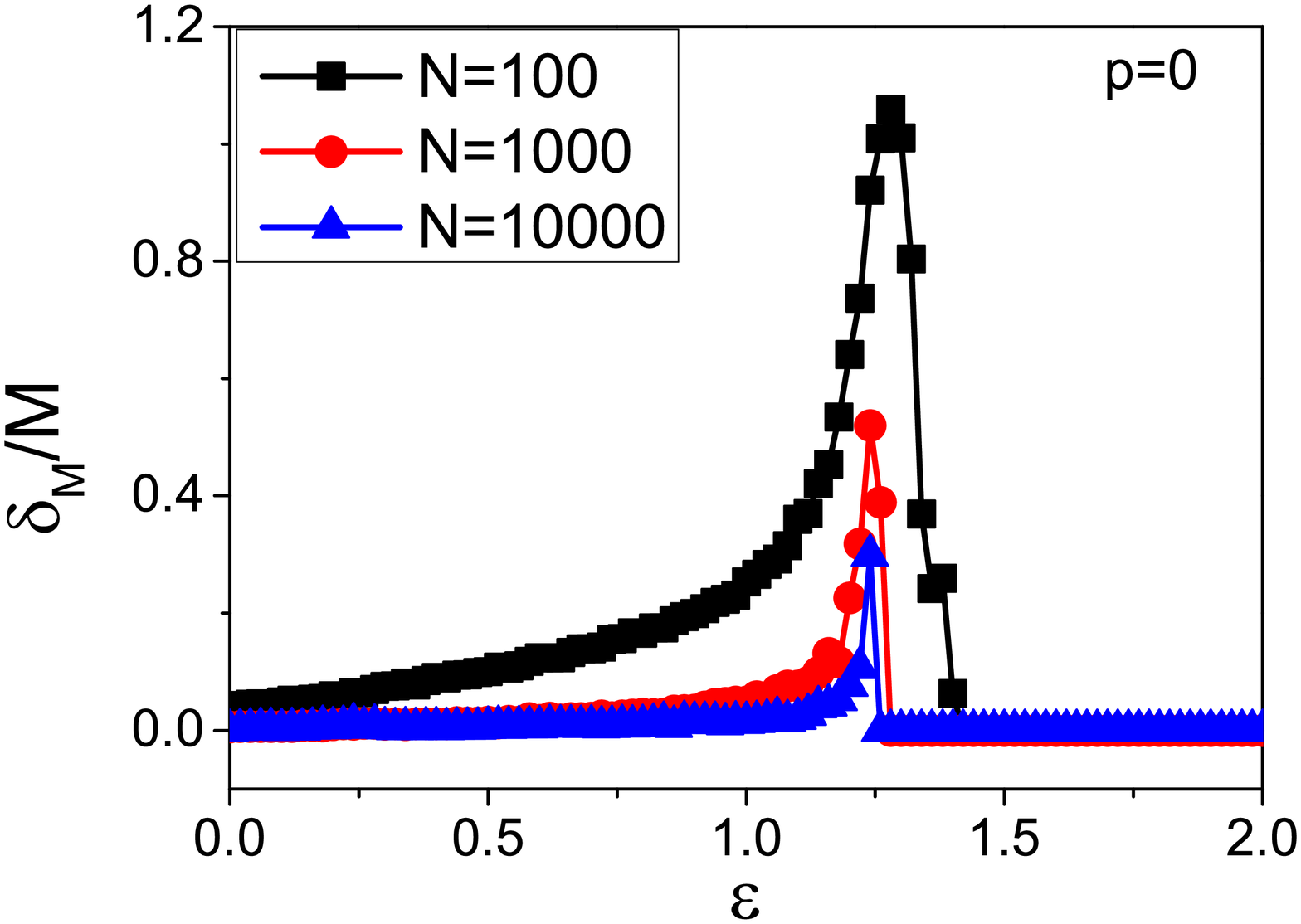}
    \includegraphics[width=0.3\textwidth]{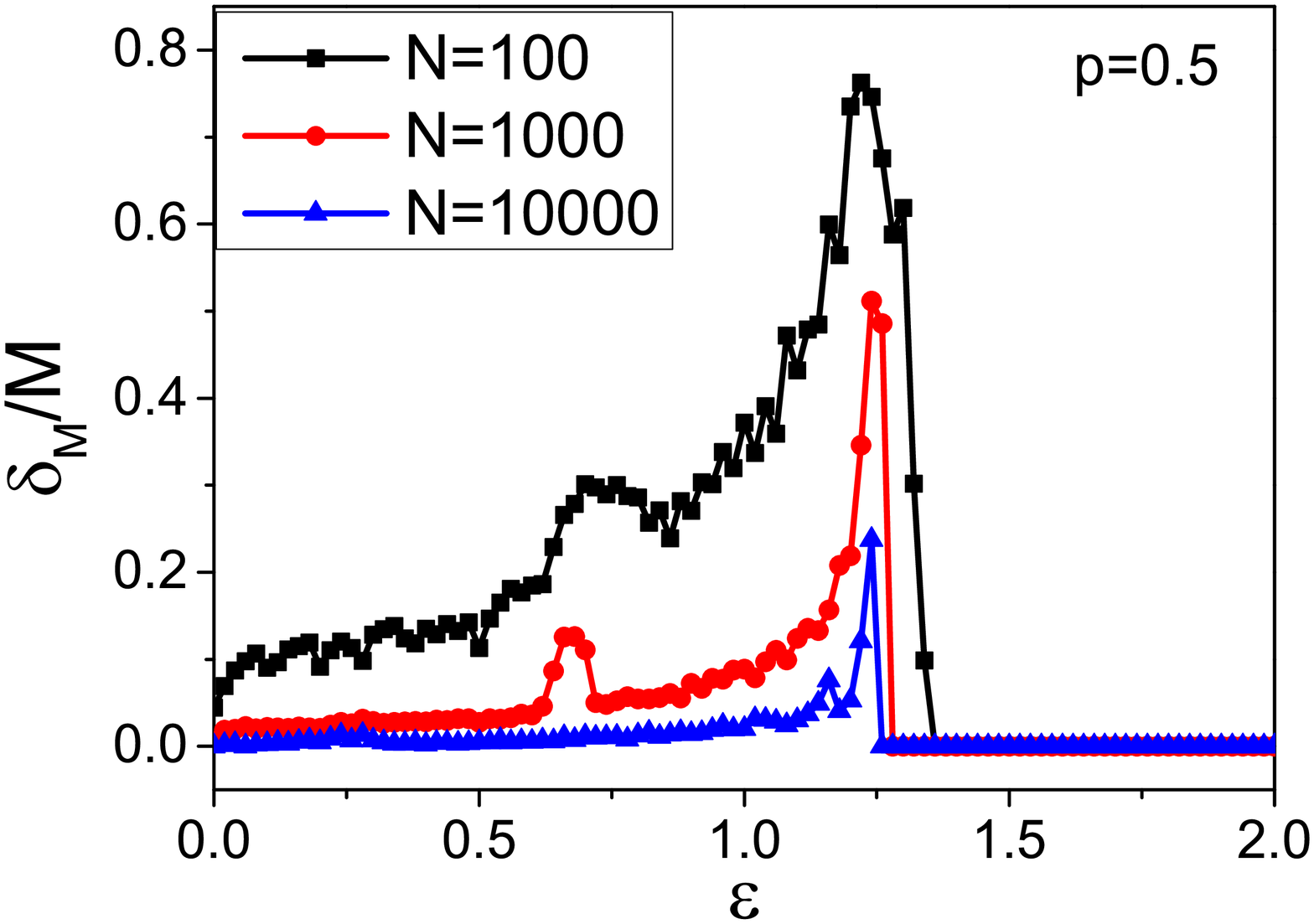}
    \includegraphics[width=0.3\textwidth]{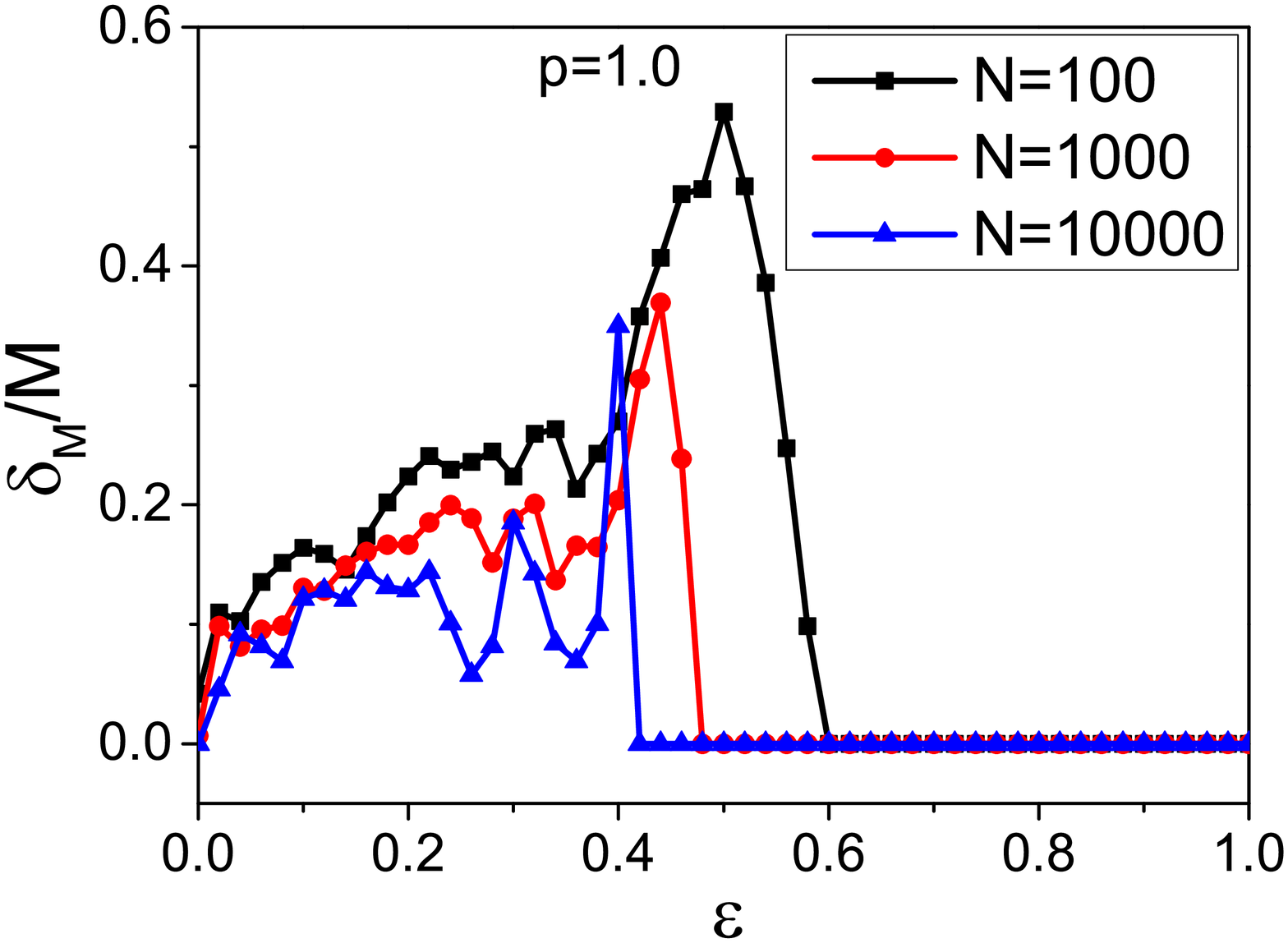}
	\caption{(color online) Distribution of normalized number of opinion clusters $M/M_0$ and its standard variance $\delta_M/M$ along the axis of confidence threshold $\varepsilon$ for the pattern of stationary states. Results are obtained by averaging over 1000 realizations with $\gamma=0$ and $k_a=N-1$.}
	\label{cluster:varepsilon-size}
\end{figure}

Analytically, the underlying system is more disordered when $M/M_0$ is closer to 1 and more ordered if $M/M_0$ is closer to 0. Numerical results for the special case of mean field limit ($k_a=N-1$) are exhibited in Fig. \ref{cluster:varepsilon-size}($a$). In the presence of smart agents, a continuous phase transition point is identified that separates an ordered state from the disordered one. The value of confidence threshold at the transition point is independent of the system size, but weakly relies on the fraction of smart agents. In HK model ($p=1$), however, the phase transition point is connected with system size, shifting along the direction of $\varepsilon$ decrease with the increase of system size. When the system size $N$ approaches infinity, the phase transition point tends to the analytic value 0.4. This is verified by the fluctuation $\delta_M$ of $M$, see Fig. \ref{cluster:varepsilon-size}($b$), where a peak happens at the transition point. More results are displayed in Appendix \ref{opinion:clustering} where the influence of resource allocation parameter $\gamma$ and network structure is considered.

\subsection{The order parameter}\label{subsec1}

\begin{figure}
\centering
    \includegraphics[width=0.4\textwidth]{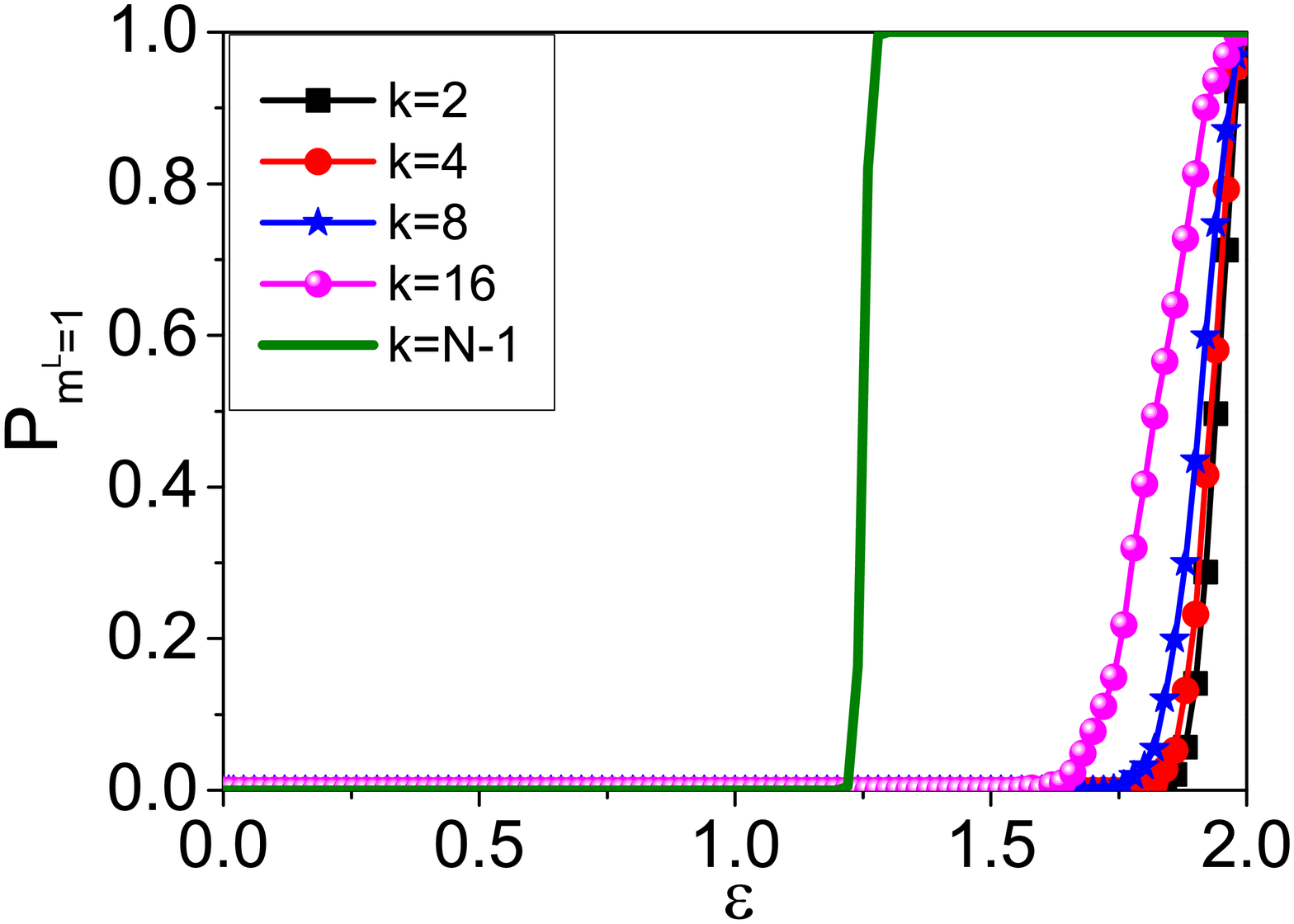}
    \includegraphics[width=0.4\textwidth]{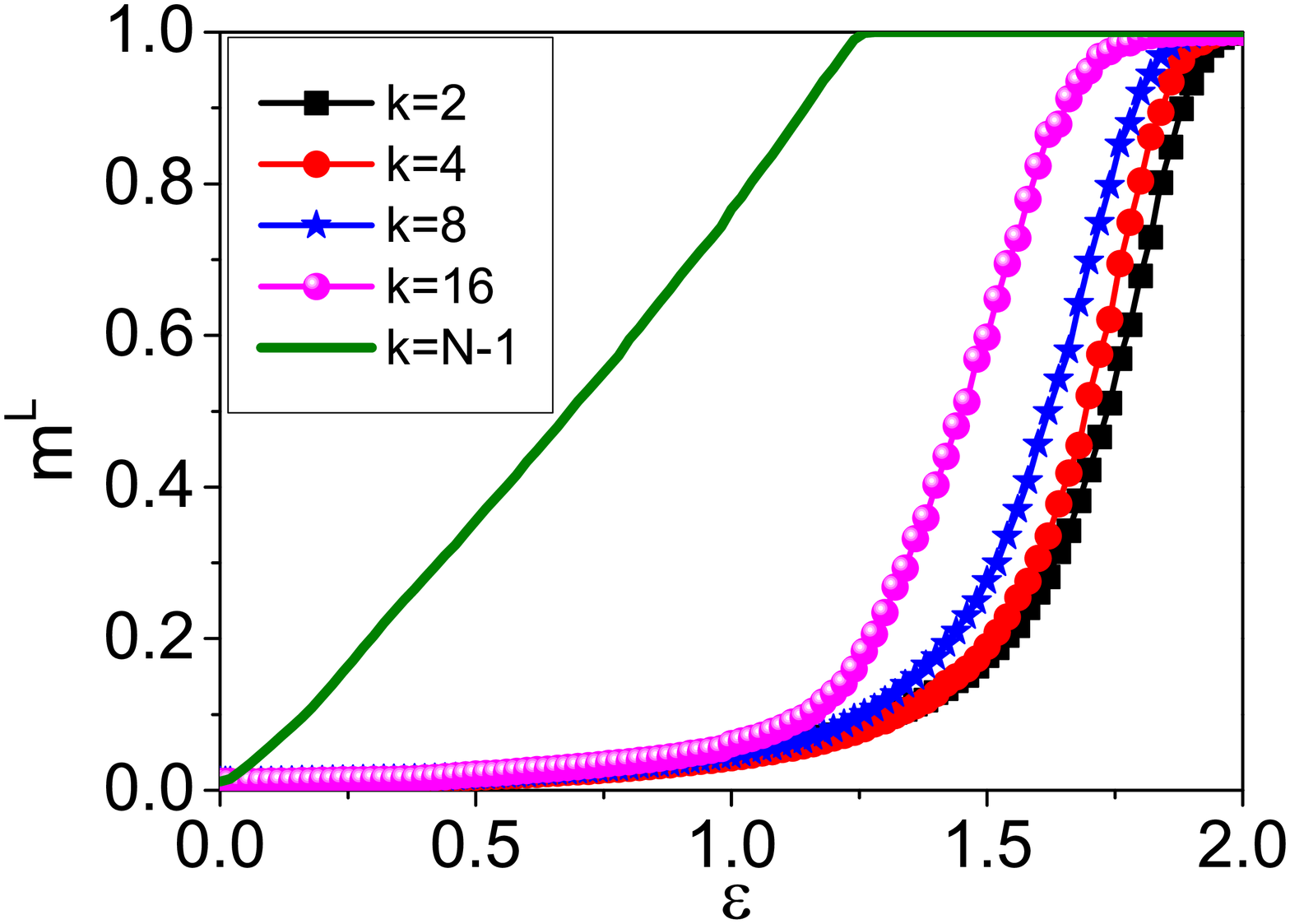}\\
    {\footnotesize $(a)$ $P_{m^L=1}$ and $m^L$ vary with $\varepsilon$ for BA scare-free networks.}\\
    \includegraphics[width=0.4\textwidth]{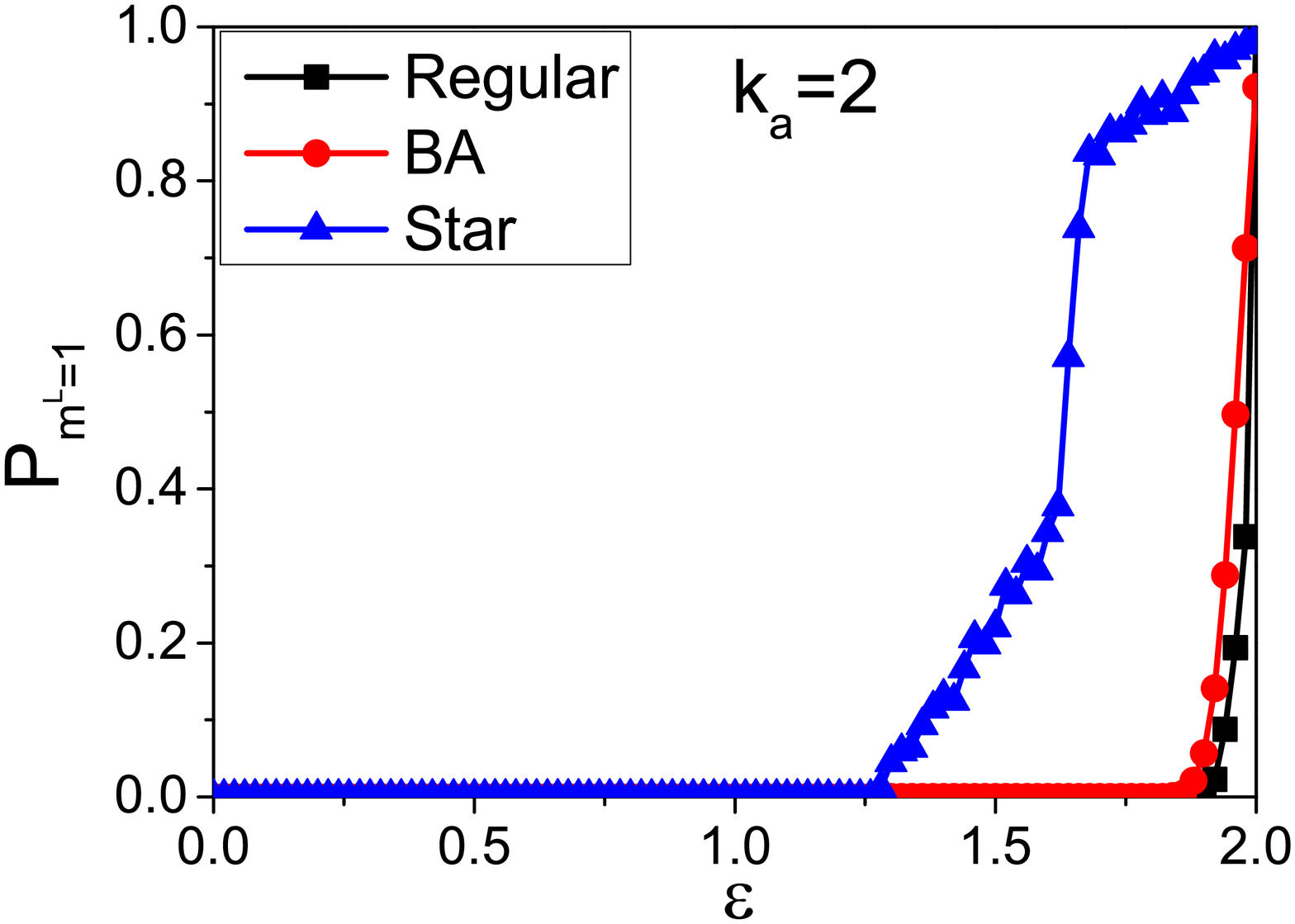}
    \includegraphics[width=0.4\textwidth]{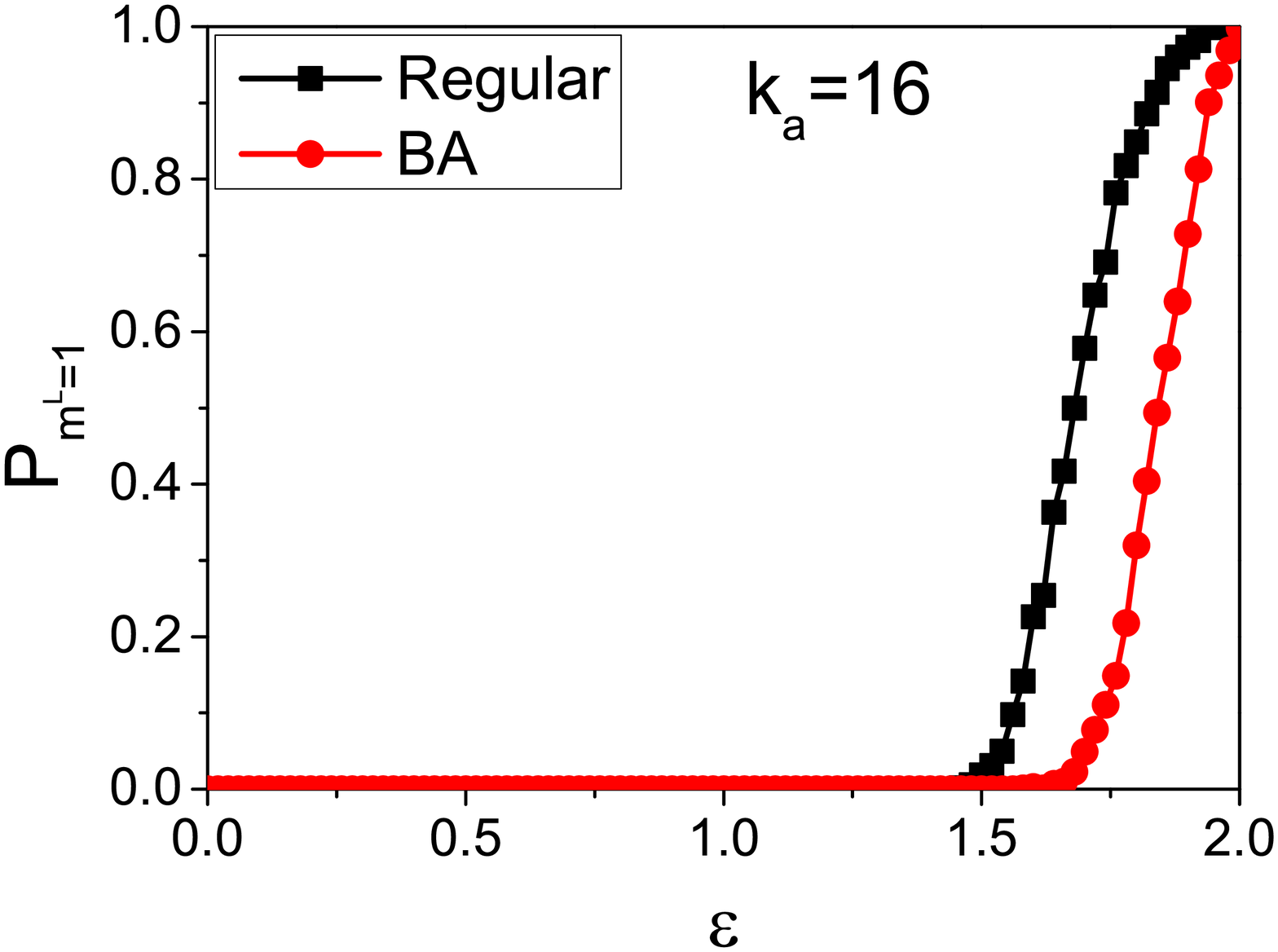}\\
    {\footnotesize $(b)$ The dependence of $P_{m^L=1}$ with $\varepsilon$ for homogeneous and heterogeneous networks.}\\
	\caption{(color online) The order parameter $P_{m^L=1}$ as a function of the confidence threshold $\varepsilon$ for homogeneous and heterogeneous networks of different values of $k_a$ with $\gamma=0$, $p=0$ and $N=1000$. Each data point is averaged over 1000 realizations.}
	\label{consensus:networkdegree}
\end{figure}

Opinion clustering analysis allows one to explain the transition from a coexistent state of many different opinions to that of a few of ones. Actually, it still has some difficulties in determining with precision the transition point separating a complete consensus from the coexistence of different opinions. We label the normalized size of the largest opinion cluster as $m^L$. An order parameter, denoted by $P_{m^L=1}$, is then natural to be introduced for understanding further the critical behavior of the underlying system \cite{toral2006finite}. It is defined as the probability to have complete consensus in different configurations. We first study $P_{m^L=1}$ as a function of $\varepsilon$ in the special case of all agents being smart ($p=0$). In numerical simulations, $P_{m^L=1}$ is calculated as the fraction of samples with all agents sharing the same opinion over 1000 independent configurations. The underlying results are illustrated in Fig. \ref{consensus:networkdegree}. For spare networks, the complete consensus comes to be generated only when the confidence threshold $\varepsilon$ is close to 2, as already discussed in section \ref{sec3}. The curve corresponding to larger average degree $k_a$ is definitely above the curve relative to smaller $k_a$, which suggests that when $k_a$ diverges, $P_{m^L=1}$ will probably attain the value 1 at a fixed position. A discontinuous change happening to the curve for complete network ($k_a=N-1$) indicates this position is $\varepsilon=1.25$. This states again the consensus threshold $\varepsilon_c=2$ for finite average degree $k_a$, and $\varepsilon_c=1.25$ for diverged $k_a$, when all agents in the system are smart.

Furthermore, homogeneous and heterogeneous networks of agents and their interpersonal relations drive some kinds of difference in the distribution of $P_{m^L=1}$. For quite spare networks: $k_a=2$, $P_{m^L=1}$ is almost the same for both regular and BA scale-free networks, which is less than that on star network where complete consensus begins to be generated at $\varepsilon=1.25$. The deviation of results on regular and BA scale-free networks gradually appears with the increase of average degree. This implies a homogeneous network is more benefit than a heterogeneous one for a complete consensus when $k_a$ is large. Interestingly, the normalized size $m^L$ of the largest cluster linearly depends upon the confidence threshold $\varepsilon$ until the consensus state is reached, in the mean field limit. For networks away from the mean field limit, $m^L$ follows a nonlinear function of $\varepsilon$.

\begin{figure}
\centering
    \includegraphics[width=0.4\textwidth]{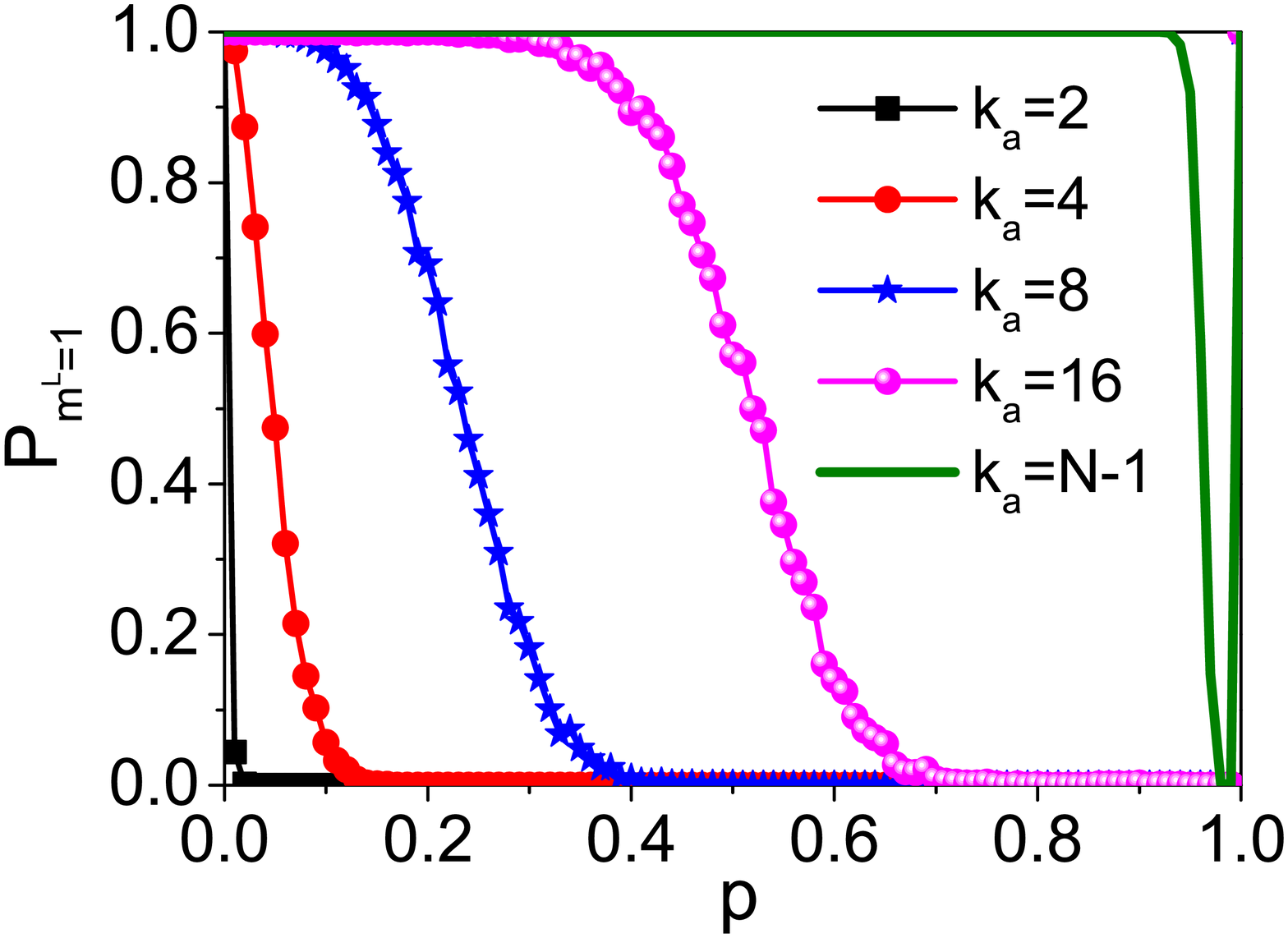}
    \includegraphics[width=0.4\textwidth]{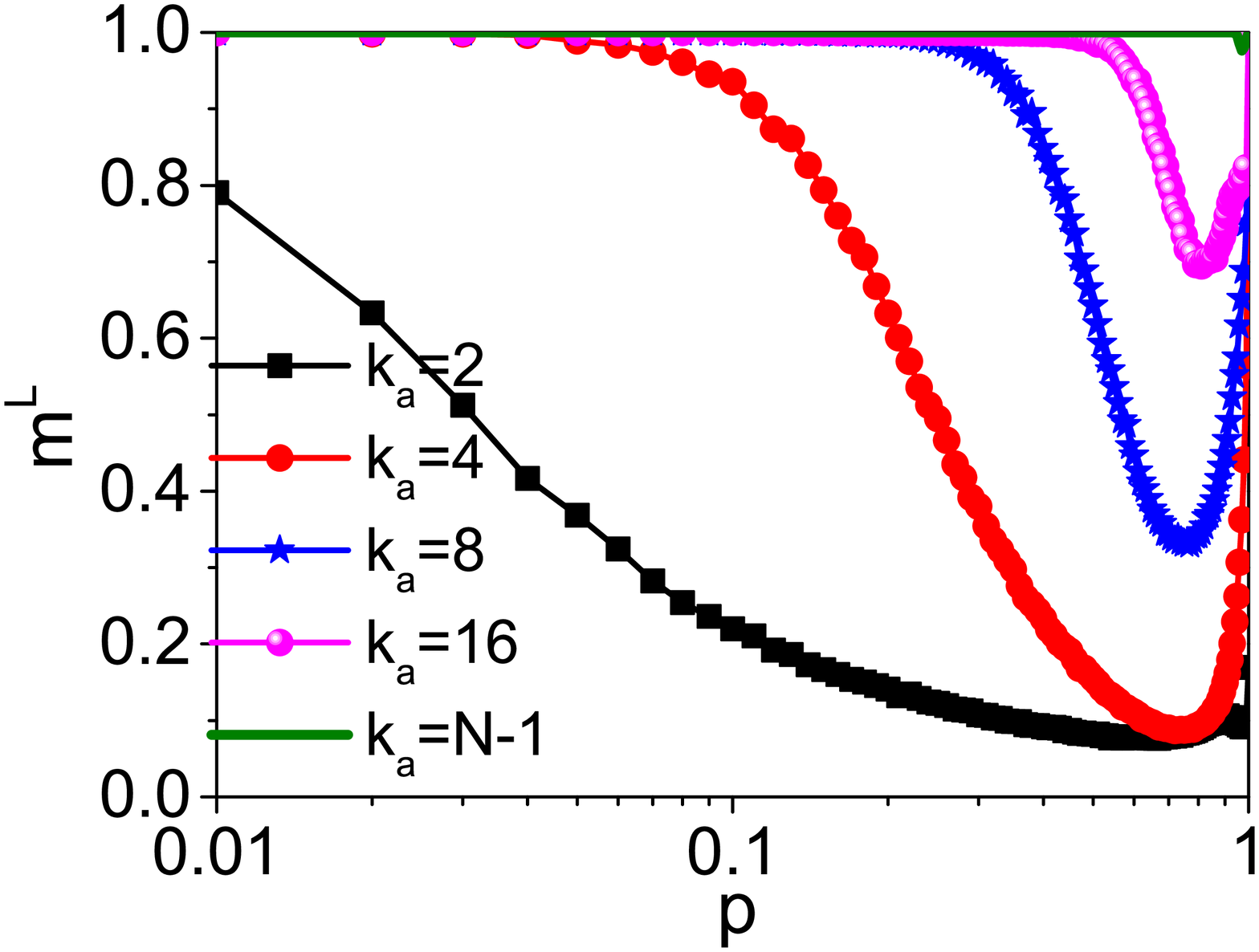}\\
    {\footnotesize $(a)$ $P_{m^L=1}$ and $m^L$ distribute with $p$ on BA scale-free networks.}\\
    \includegraphics[width=0.4\textwidth]{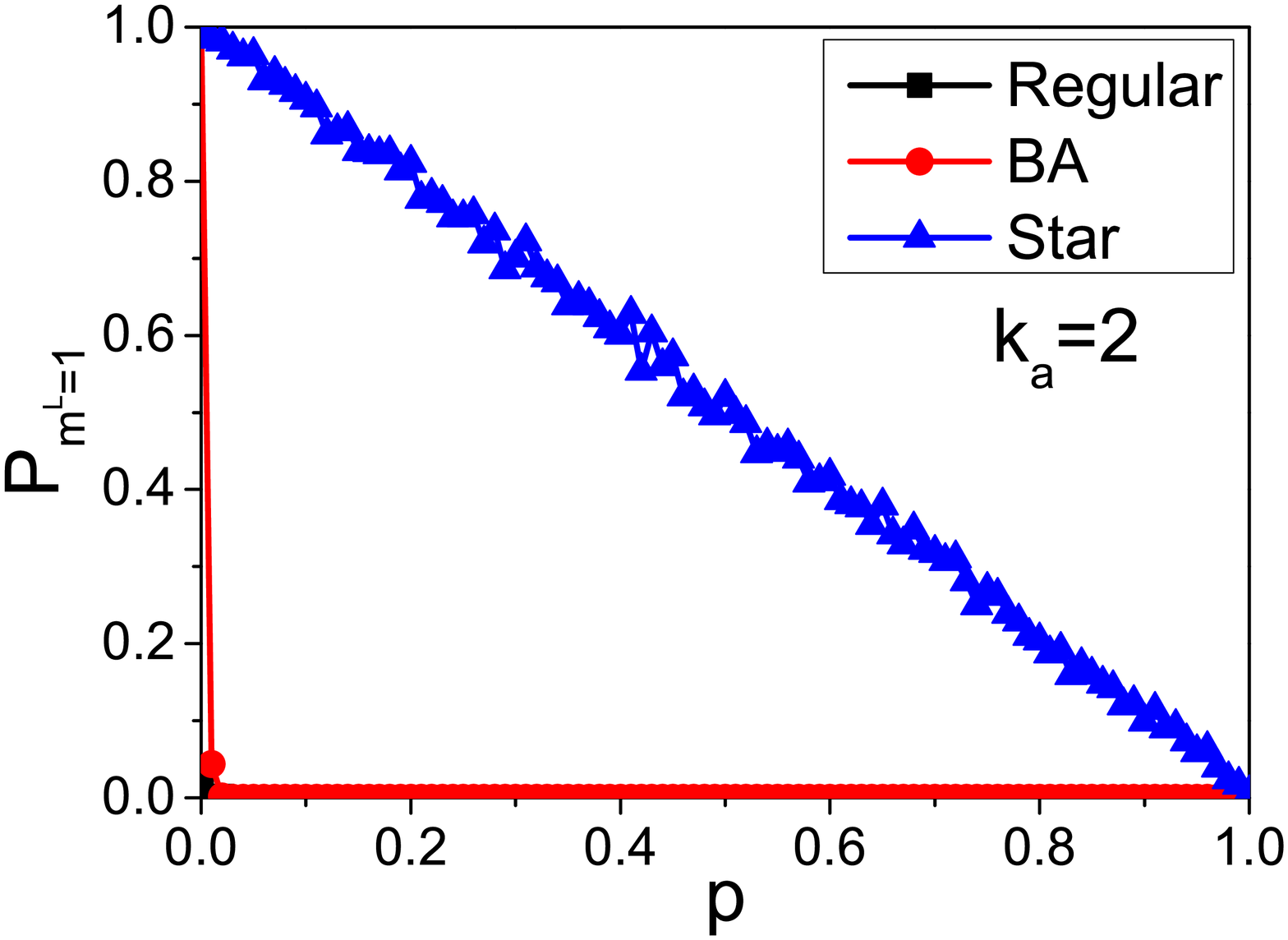}
    \includegraphics[width=0.4\textwidth]{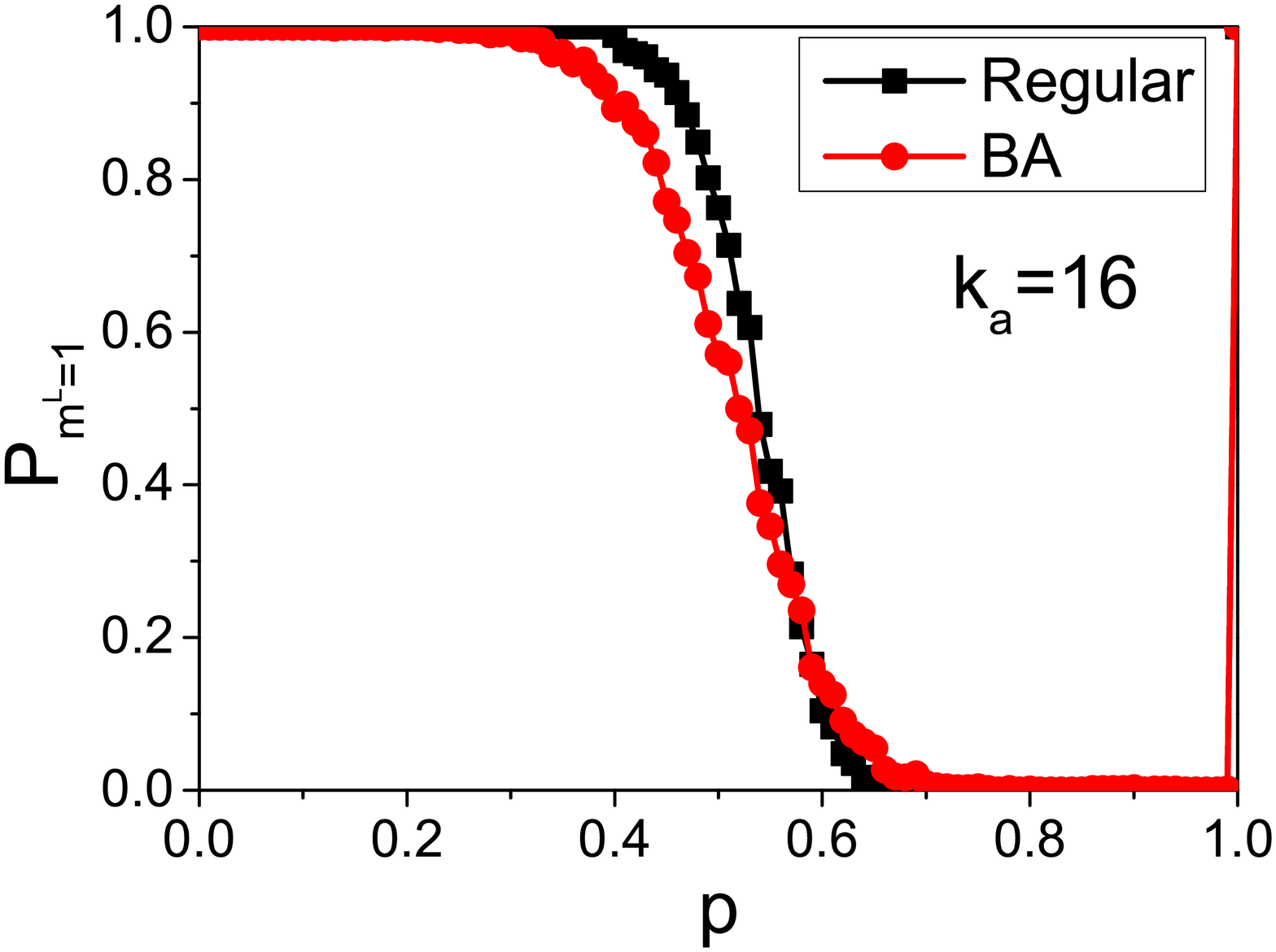}\\
    {\footnotesize $(b)$ $P_{m^L=1}$ as a function of $p$ for homogeneous and heterogeneous networks.}
	\caption{(color online) $P_{m^L=1}$ as a function of $p$ for homogeneous and heterogeneous networks of different values of $k_a$ with $\gamma=0$, $\varepsilon=2$ and $N=1000$. Each data point is averaged over 1000 realizations.}
	\label{consensus:networkdegree2}
\end{figure}

\subsubsection{The influence of parameter $p$}
As stated previously, the presence of smart agents will affect the position to reach a complete consensus. Particulary, this effect is discussed further by setting $\gamma=0$ and $\varepsilon=2$. Results in Fig. \ref{consensus:networkdegree2} $(a)$ show that, for quite sparse networks, a complete consensus happens only at two boundary conditions: $p=0$ and $p=1$. For a complete network, however, a consensus appears in almost the whole space of parameter $p$. The underlying theoretical analysis is attached in the following. Furthermore, there is a valley in the distribution of $m^L$ at $p\approx 0.7$, corresponding to the most disordered state of opinion dynamics. This phenomenon, however, will gradually disappear with the increase of $k_a$.

Analogously, the influence of network topology is also discussed on the dependence of $P_{m^L=1}$ with parameter $p$, see Fig. \ref{consensus:networkdegree2} $(b)$. The results on regular and BA scale-free networks are of negligible deviations for each specified average degree, which differ from the distribution behavior on a star network. The star network is known for its topology of special structure with a hub node connecting to the rest $N-1$ leaf nodes who do not connect to each other. If the hub node is general, smart agents will always keep their own opinions, leading finally to the coexistence of various opinions. If instead the hub node is smart, it then affects leaf nodes and further drives a complete consensus. Therefore, $P_{m^L=1}$ acts as a linear function of parameter $p$ on a star network:
\begin{equation}
P_{m^L=1}=1-p.
\end{equation}

\begin{figure}
\centering
    \includegraphics[width=0.4\textwidth]{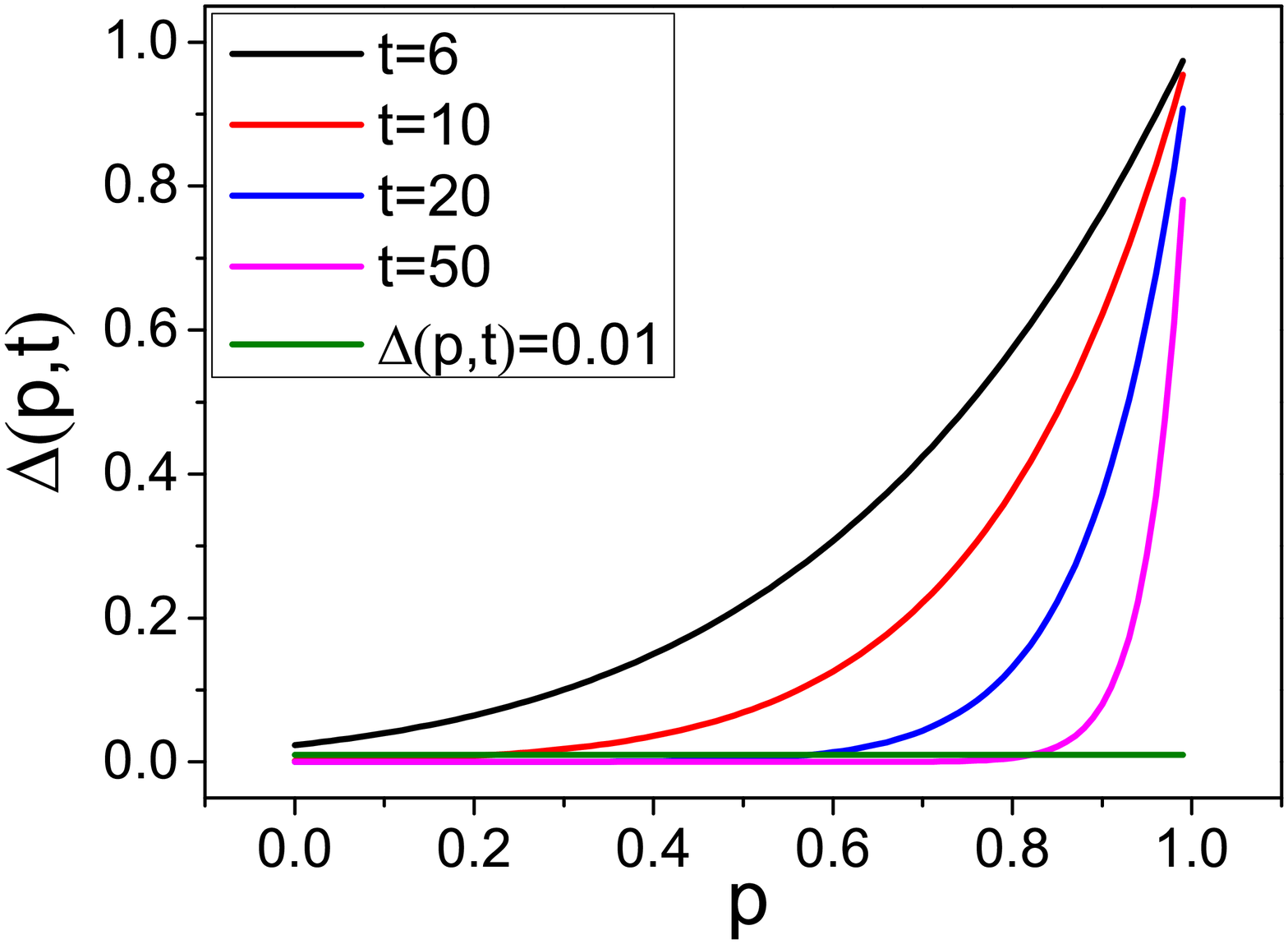}
    \includegraphics[width=0.4\textwidth]{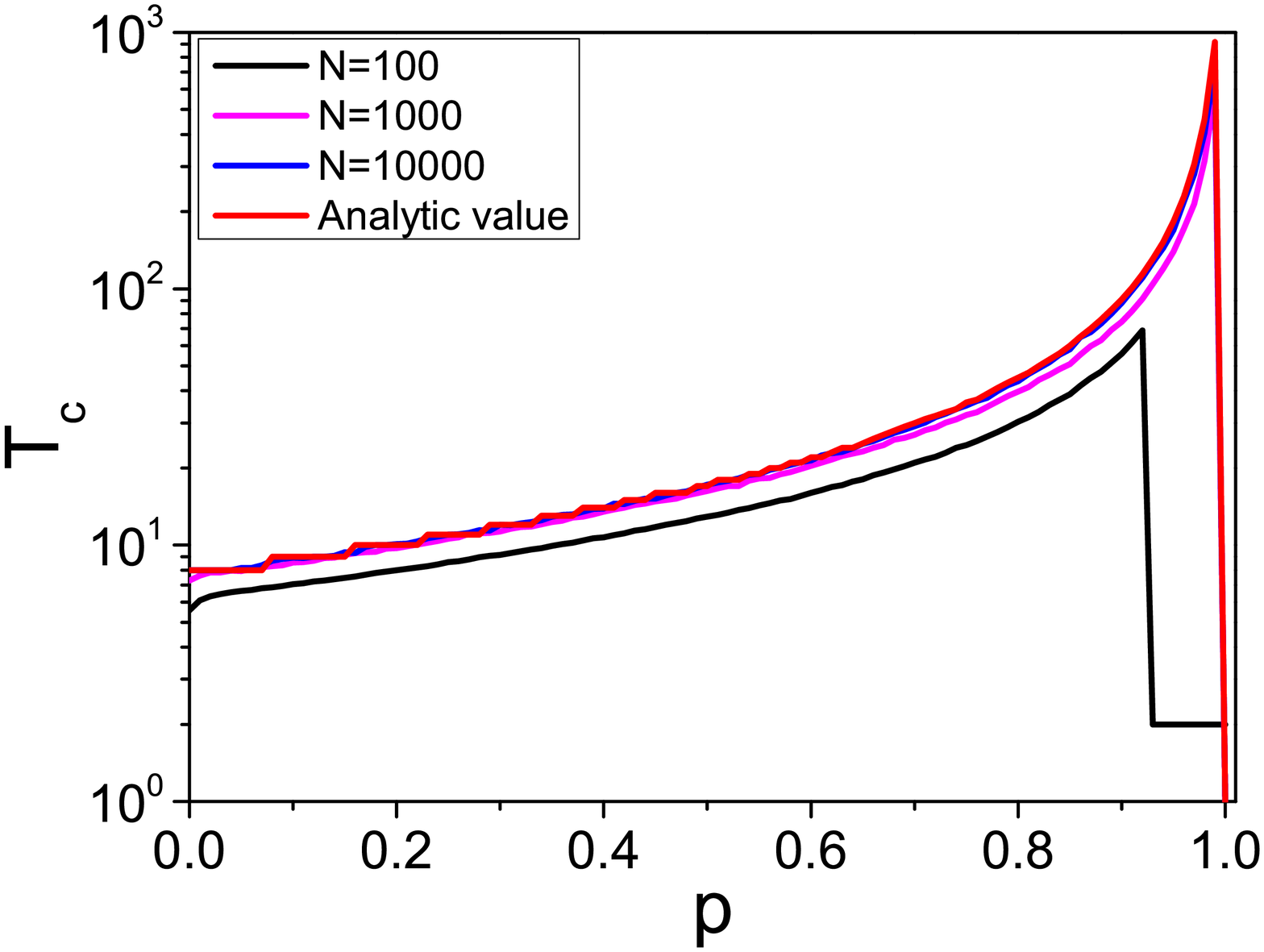}
	\caption{(color online) Analytic values for the influence of parameter $p$ on the formation of a complete consensus in the mean field limit ($k_a=N-1$) with $\gamma=0$ and $\varepsilon=2$. Left panel: $\Delta(t, p)$ acts as a function of $p$ for different values of $t$. Right panel: $T_c$ (time to reach a complete consensus) acts as a function of $p$, in good agreement with numerical results especially when the system size is large enough.}
\label{consensustheory:p}
\end{figure}

On a complete graph with $\gamma=0$ and $\varepsilon=2$, general agents update gradually their opinions by global average values in the opinion space, and smart ones adapt to the average opinions of their compatible neighbors who gained scores at $t=0$. Specifically, an absorbing state of all agents' opinions being 0 is certain to be reached at $t=1$ when all agents are general ($p=1$). If smart and general agents are coexisted ($0<p<1$), the opinion space at an arbitrary time step $t$ ($\geq 4$) can be denoted by $[-1, \varphi(t)]$ with $\varphi(t)$ following the formula (as shown in Appendix \ref{appendix:a}):
\begin{equation}
\varphi(t)=\frac{-(1-p)}{2^{t+1}}\left[\sum_{m=0}^{t-4}2^{t-m}(1+p)^m+(13+5p)(1+p)^{t-3} \right],
\end{equation}
which can be simplified as
\begin{equation}
\varphi(t)=\frac{2^{t+1}-3-F(p)}{-2^{t+1}}.
\end{equation}
$F(p)$ is a polynomial function of $p$:
\begin{equation}
F(p)=\sum_{m=1}^{t-1}c_mp^m,
\end{equation}
with coefficients $\{c_m\}$ following
\begin{equation}
 c_{t-1}=5, \text{ and } \sum_{m=1}^{t-1}c_m=2^{t+1}-3.
\end{equation}
Recalling the previous definition, a complete consensus is reached at time $t$ if and only if
\begin{equation}
\varphi(t)+1<0.01,
\end{equation}
which yields
\begin{equation}
\Delta(t, p)=\frac{3+F(p)}{2^{t+1}}<0.01.
\label{inequality:p}
\end{equation}
In Fig. \ref{consensustheory:p} it is presented that the real range of $p$ to satisfy Eq.~(\ref{inequality:p}) will be enlarged with the increase of $t$. This suggests that a complete consensus is finally to be reached for all possible values of parameter $p$, when the social graph is completely connected and $\varepsilon=2, \gamma=0$. The time $T_c$ to reach a complete consensus is well confirmed by simulation results, especially when the system size is large enough.

\subsubsection{Influence of parameter $\gamma$}

Specifically, resource allocation parameter $\gamma$ is set to be 0 in above discussion. However, this is hardly to be reached in a real-world system as it lacks some kind of fairness. To support for the fairness in resource allocation in practice, we now consider the fact that the resource held by the closed system is shared initially by both two cliques. Parameter $\gamma$ controls the biased resource allocation between two cliques in the system, as defined in section \ref{sec2}.

\begin{figure}
\centering
    \includegraphics[width=0.4\textwidth]{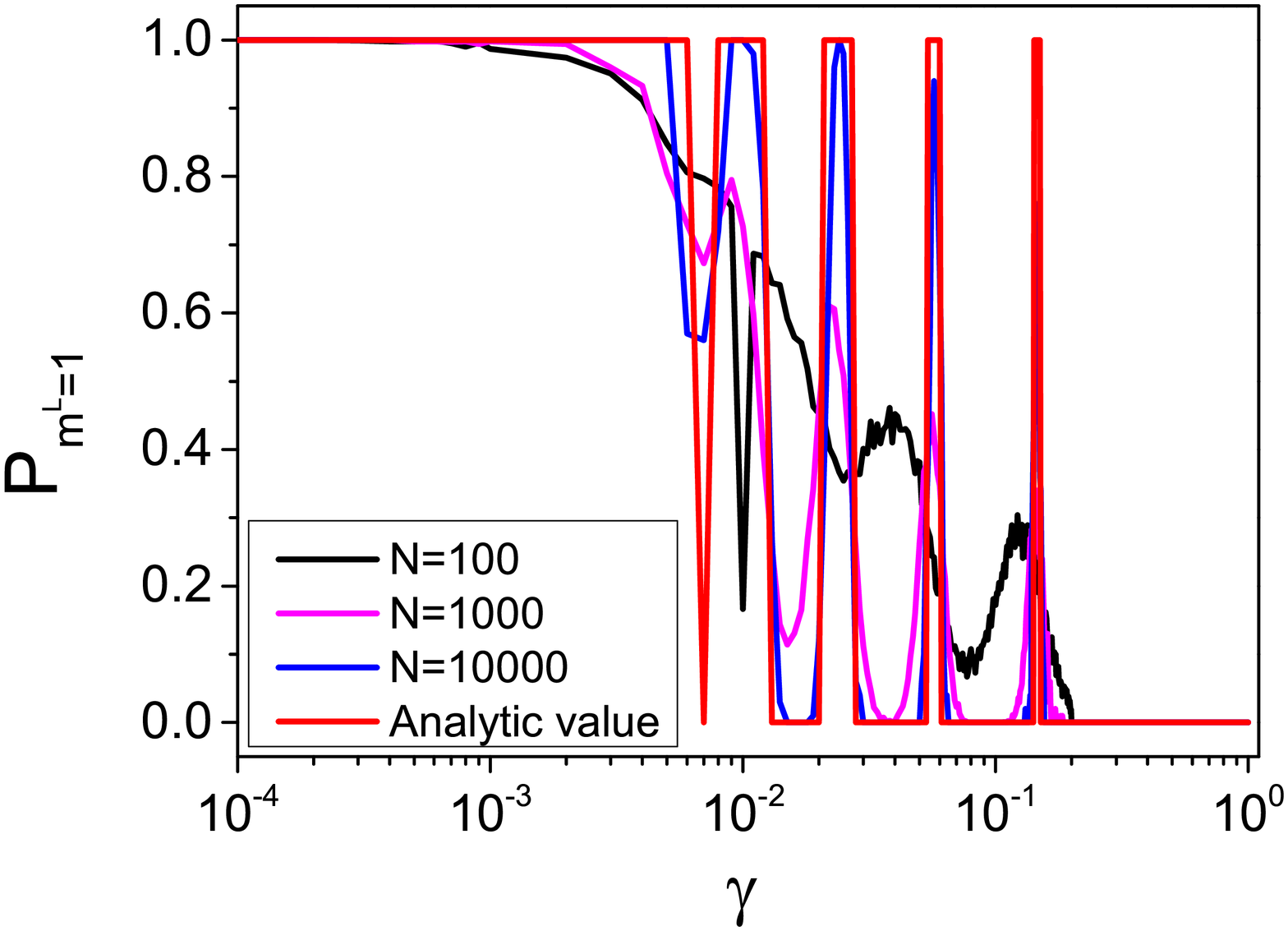}
    \includegraphics[width=0.4\textwidth]{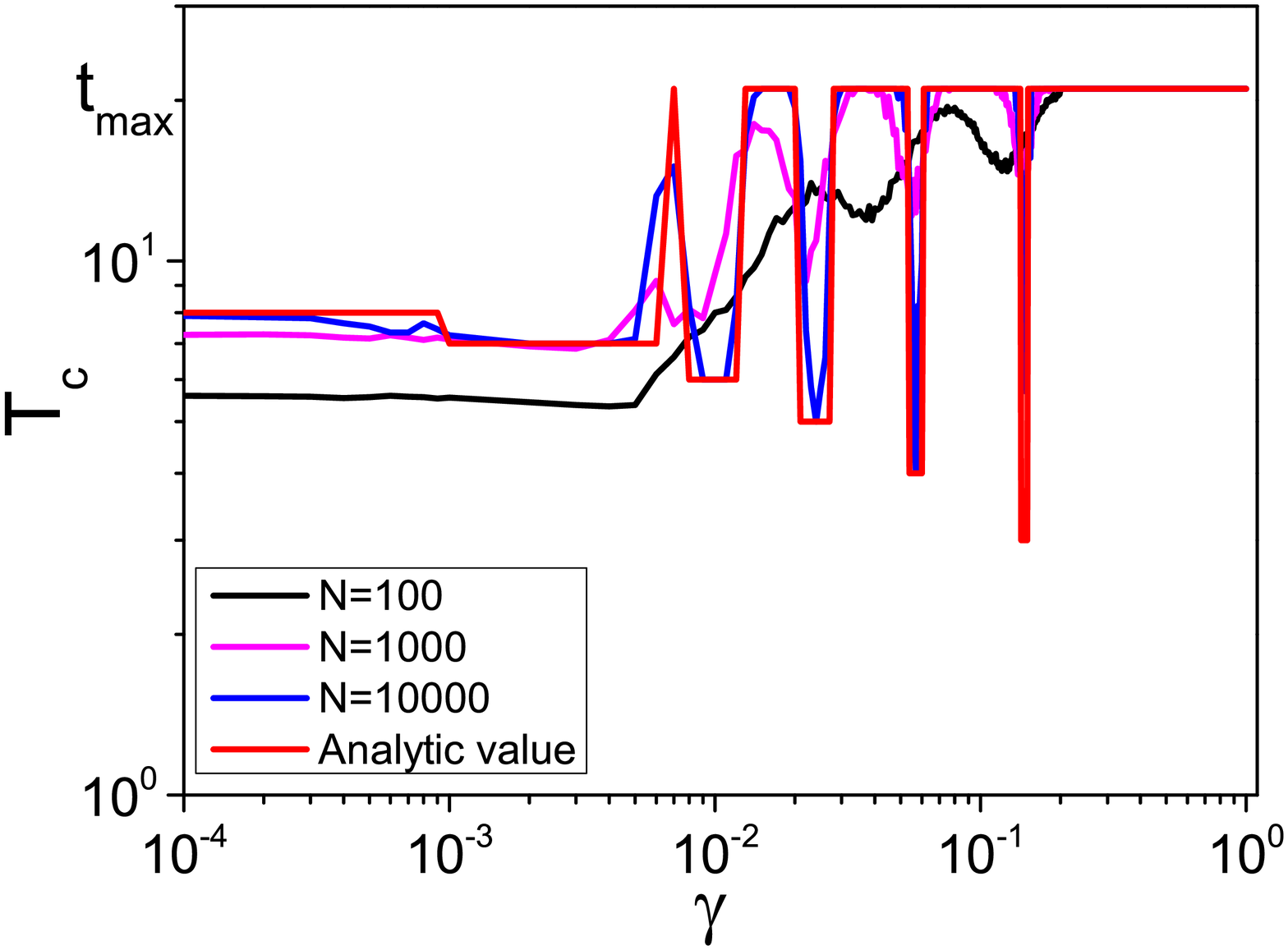}
	\caption{(color online) $P_{m^L=1}$ and $T_c$ (time to reach a complete consensus) as functions of the resource allocation parameter $\gamma$ for a complete network ($k_a=N-1$) with $p=0$ and $\varepsilon=2$. Red lines are obtained analytically by the mean filed theory. $T_c$ gets $t_\text{max}$ (the maximal time set to propose Monte Carlo simulations), representing the absence of a complete consensus. Numerical results are averaged over 1000 realizations.}
\label{timeopinion:gamma}
\end{figure}

Next, we discuss the role of parameter $\gamma$ acted in the generation of a complete consensus. For simplicity, we set $k_a=N-1$, $\varepsilon=2$, and suppose that all agents are smart. Following the gambling game, agents will update their opinions by learning from their compatible neighbors who have gained the most scores. Recalling the definition of $r_i$ (Eq. (\ref{resource})) and the condition for agent $i$ to win: $r_i\geq R/N$, we can specify the opinion of winners in both cliques at $t=0$,
\begin{equation}
o_i(0)\left\{
\begin{array}{ll}
\geq \frac{1+\gamma}{4\gamma}, & i\in G_+\\
\\
\leq -\frac{1+\gamma}{4}, & i\in G_-
\end{array}
\right.
.
\end{equation}
If winners exist in both cliques at $t=0$, the consensus state will be absent. To ensure its appearance, we should have
\begin{equation}
\frac{1+\gamma}{4\gamma}>1,
\end{equation}
yielding $\gamma<1/3$ that guarantees that winners hold opinions at $t=0$ in the real range $[-1, -\frac{1+\gamma}{4}]$. At next time step, agents with initial opinions in $[-\frac{1+\gamma}{4}, 1]$ will then update their opinions by the average opinion in $[-1, -\frac{1+\gamma}{4}]$ as opinions are uniformly distributed at $t=0$. The range of opinion for wining the gambling will be changed as more agents attracted to share the resource of clique $G_-$. With time evolution, the maximal opinion, denoted by $\phi$, will be gradually close to -1. $\phi$ is a function of both $\gamma$ and time $t$, expressed mathematically by
\begin{equation}
\left\{
\begin{array}{ll}
 \phi(t)=\frac{-\left[\sum_{m=3}^{t}2^m(1+\gamma)^{t-m+1}+(5+\gamma)(1+\gamma)^{t-1}\right]}{2^{t+1}},& t\geq 3;\\
 \\
 \phi(t)=-\frac{(1+\gamma)(5+\gamma)}{8}, & t=2;\\
 \\
 \phi(t)=-\frac{1+\gamma}{4}, & t=1.
\end{array}
\right.
\end{equation}
At any special time $t$, agents' opinions are located in the real range $[-1, \phi (t)]$. A complete consensus will be reached only when the distance of opinions satisfies the inequality:
\begin{equation}
1+\phi (t)< 0.01.
\end{equation}
Fixing $t$ and recalling two general constraints: $\gamma<\frac{1}{3}$ and $\phi (t)>-1$, we then get the range of $\gamma$ to have a complete consensus:
\begin{equation}
P_{m^L=1}=1 \Leftarrow \left \{
\begin{array}{ll}
0.142 \leq \gamma \leq 0.15, & T_c=3; \\
\\
0.054 \leq \gamma \leq 0.06, & T_c=4; \\
\\
0.021 \leq \gamma \leq 0.027, & T_c=5; \\
\\
0.008 \leq \gamma \leq 0.012, & T_c=6; \\
\\
0.0009 \leq \gamma \leq 0.006, & T_c=7; \\
\\
0 \leq \gamma < 0.0009, & T_c=8;
\end{array}
\right.
\end{equation}
For the rest values of $\gamma$, a complete consensus is absent. This suggests that the order parameter $P_{m^L=1}$ does not change continuously in the axis of $\gamma$ but acts as a rectangular pulse. The theoretical analysis is valid only when the system size is large enough, as illustrated in Fig. \ref{timeopinion:gamma}.

\subsection{Finite size effect}

In the study of the critical behavior of a physical system, it often needs to predict the thermodynamical properties near a critical point. The well developed techniques of finite size scaling allow one to extrapolate the results to $N\to \infty$ \cite{toral2006finite}. In practice, however, agents considered can never be that large. And the results in thermodynamic limit may vary with respect to those of finite-size systems. In the following, we will consider the finite size effect of the distribution behavior of order parameter $P_{m^L=1}$.

\begin{figure}
\centering
    \includegraphics[width=0.4\textwidth]{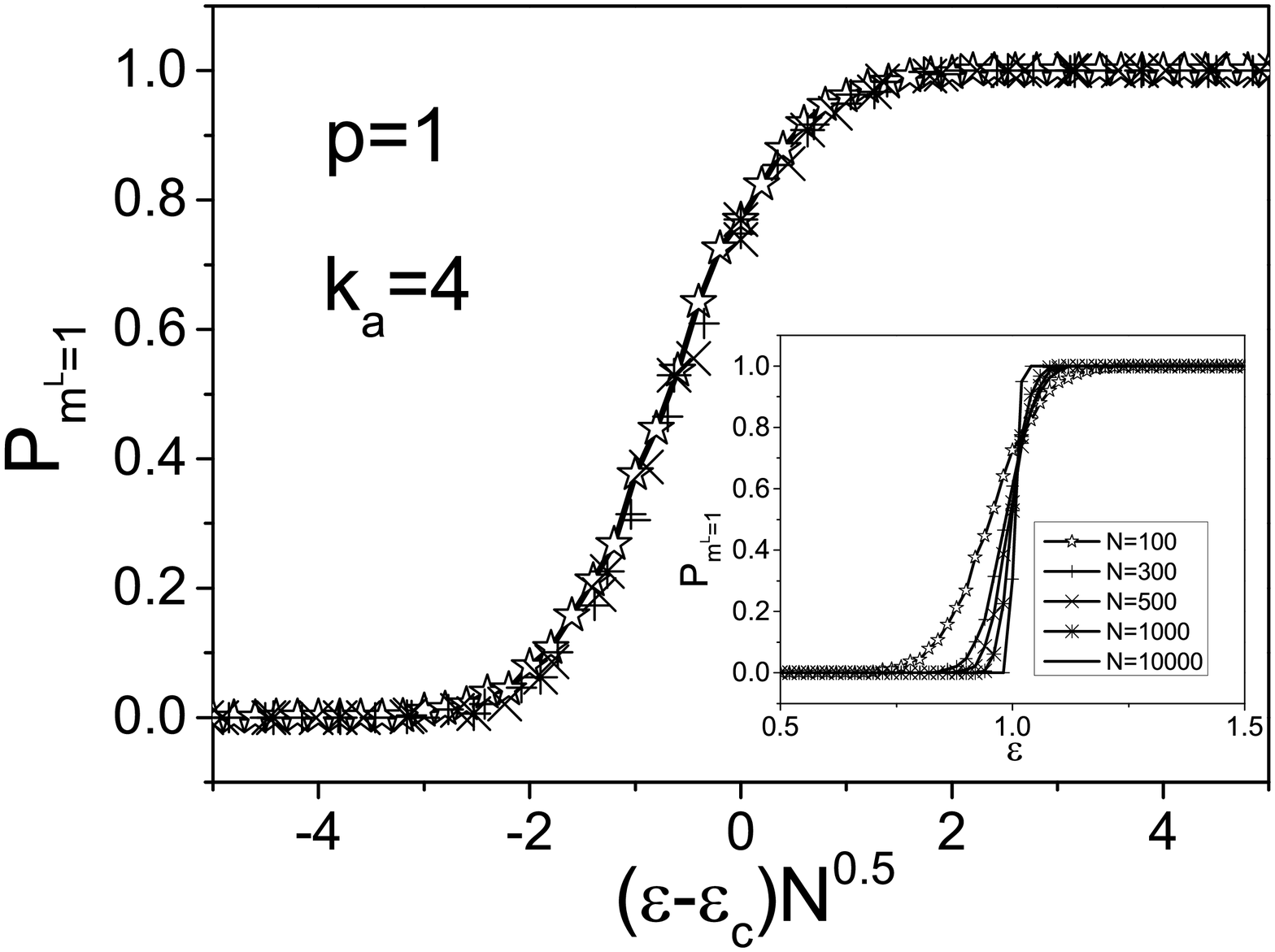}
    \includegraphics[width=0.4\textwidth]{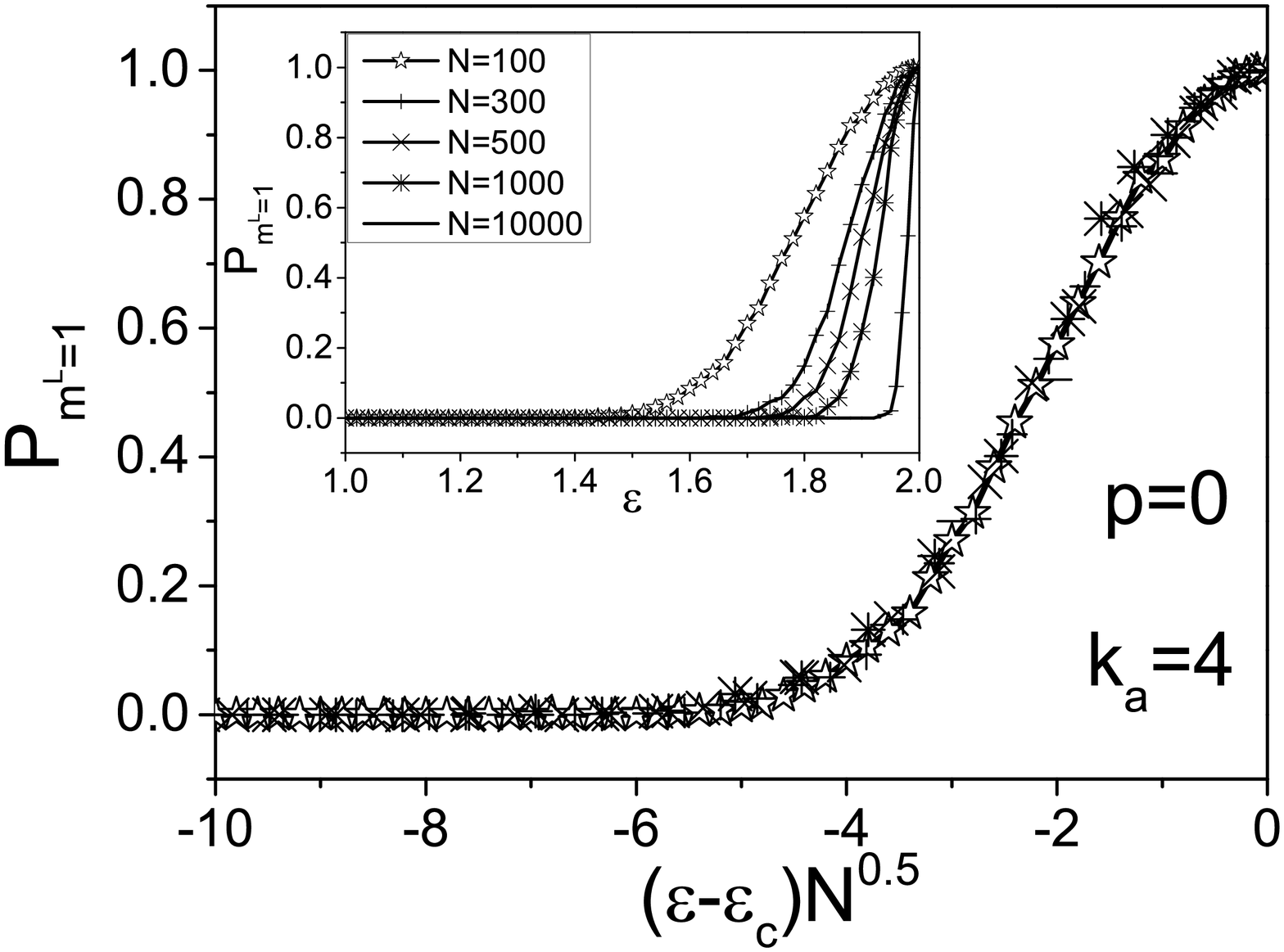}\\
   {\footnotesize $(a)$ Simulated results on a BA scare-free network with $k_a=4$.}\\
    \includegraphics[width=0.4\textwidth]{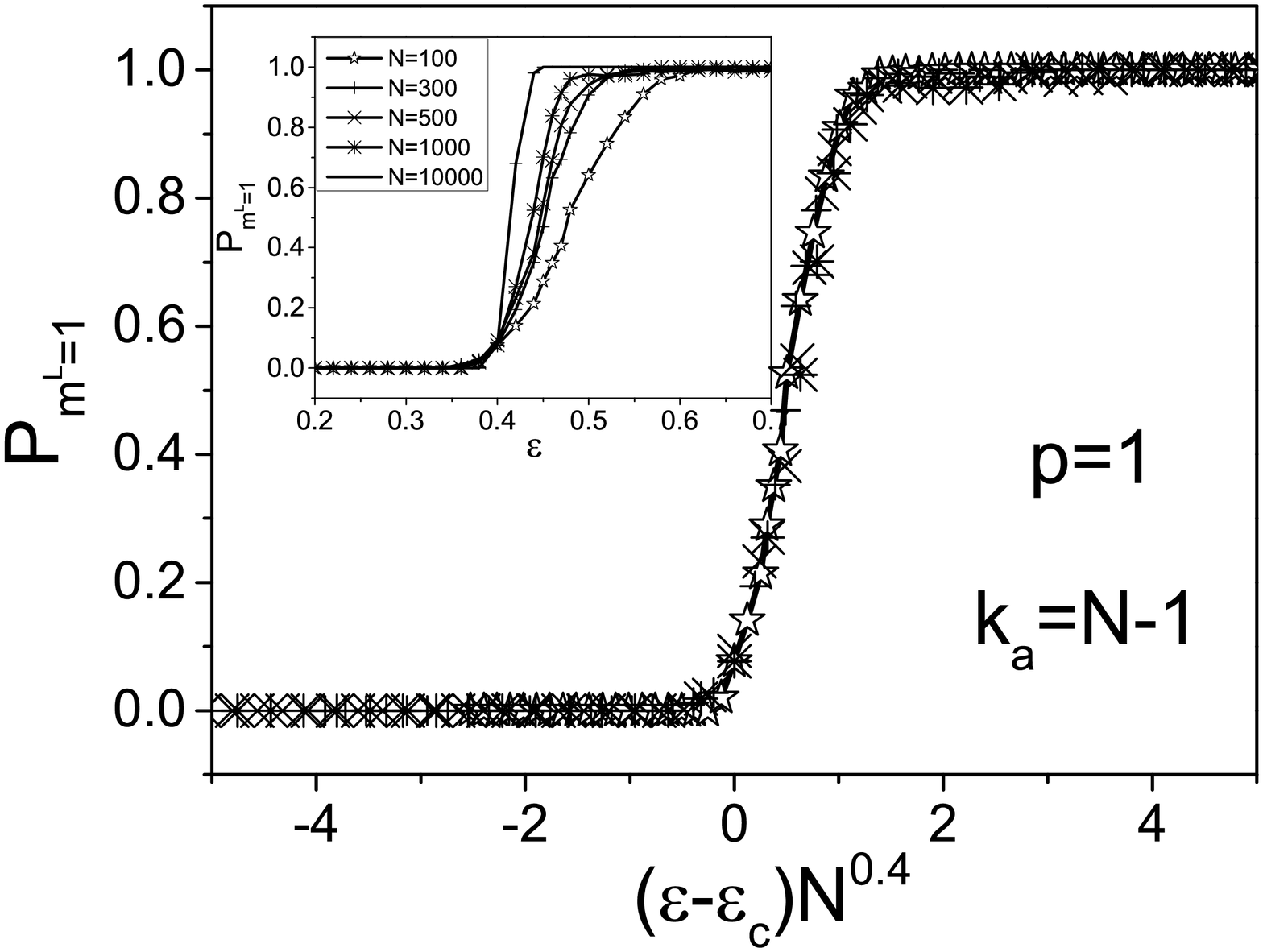}
    \includegraphics[width=0.4\textwidth]{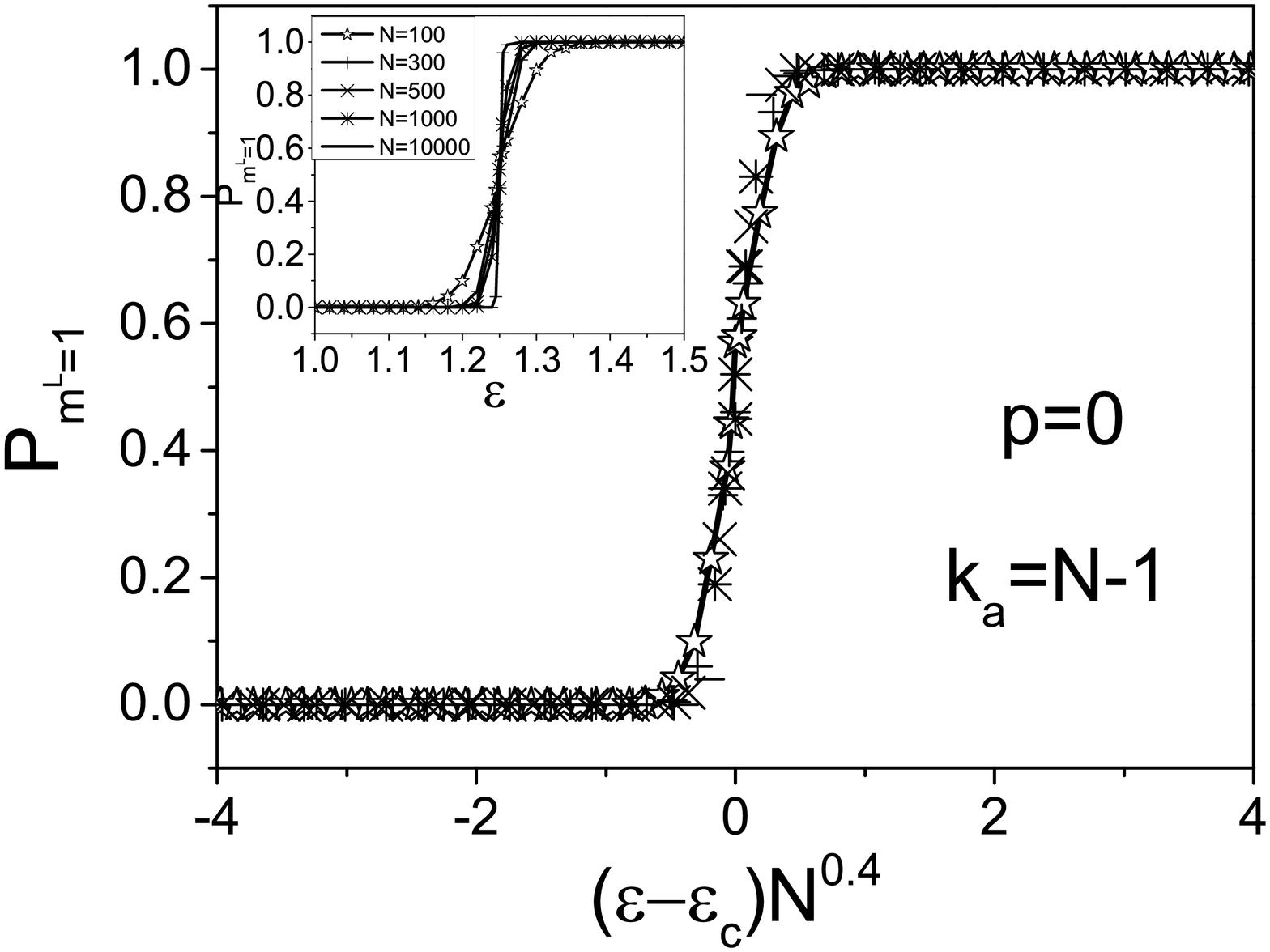}\\
   {\footnotesize $(b)$ Simulated results on a complete network ($k_a=N-1$).}
	\caption{Inset panels show $P_{m^L=1}$ as a function of $\varepsilon$ and $N$ for HK model ($p=1$) and the case of all agents being smart ($p=0$) on a BA scale-free network with $k_a=4$ and a complete network ($k_a=N-1$). Different symbols correspond to different system sizes between $N=100$ and $N=10^4$ (the transition sharpens with increasing system size). The main plots show the scaling law $P_{m^L=1}(\varepsilon, N)=P_{m^L=1}((\varepsilon-\varepsilon_c)N^{\alpha})$ with $\alpha=0.5$ for $k_a=4$, in which $\varepsilon_c=1.02$ for $p=1$ and $\varepsilon_c=2$ for $p=0$, and $\alpha=0.4$ for $k_a=N-1$, in which $\varepsilon_c=0.4$ for $p=1$ and $\varepsilon_c=1.25$ for $p=0$. Each data point is obtained by averaging over 1000 realizations with $\gamma=0$.}
	\label{consensus:networksize}
\end{figure}

We first set $\gamma$ to 0. Results of HK model are compared to that of the case with all agents being smart. For both cases, it is natural to propose the prediction that a complete consensus is certain to be reached only if the confidence threshold is above a critical value $\varepsilon_c$:
\begin{equation}
P_{m^L=1}=\left\{
\begin{array}{ll}
1 & \text{if $\varepsilon > \varepsilon_c$}\\
\\
0 & \text{if $\varepsilon < \varepsilon_c$}
\end{array}
\right.
.
\end{equation}
This analysis, however, is valid only in the thermodynamic limit. In fact, for a population of small size, the condition $\varepsilon>\varepsilon_c$ does not automatically ensure a complete consensus. There are finite size fluctuations such that the actual result of the repeated discussion process is not well established if the difference between the fixed value of confidence parameter $\varepsilon$ and the consensus threshold $\varepsilon_c$ is of order $|\varepsilon-\varepsilon_c|\sim N^{-\alpha}$. This rounding off of the sharp transition can be summarized in the following finite-size scaling law:
\begin{equation}
P_{m^L=1}(\varepsilon, N)=P_{m^L=1}\left((\varepsilon-\varepsilon_c)N^{\alpha}\right),
\end{equation}
a result well confirmed by numerical simulations with $\alpha=0.5$ for $k_a=4$ and $\alpha=0.4$ for $k_a=N-1$, as presented in Fig. \ref{consensus:networksize}. The statement is satisfied for both of the two cases: $p=1$ and $p=0$. This implies that the introduction of smart agents does not change the scaling law of HK model, but has statistically significant impact on the consensus threshold $\varepsilon_c$. Numerical results also indicate that $\varepsilon_c=1$ for $p=1$ and $\varepsilon_c=2$ for $p=0$ if the average degree $k_a$ is finite in the limit of large $N$. If instead $k_a \to \infty$ when $N \to \infty$, $\varepsilon_c=0.4$ for $p=1$ and $\varepsilon_c=1.25$ for $p=0$. These are in good agreement with analytic results illustrated in section \ref{sec3}.

\begin{figure}
\centering
	\includegraphics[width=0.3\textwidth]{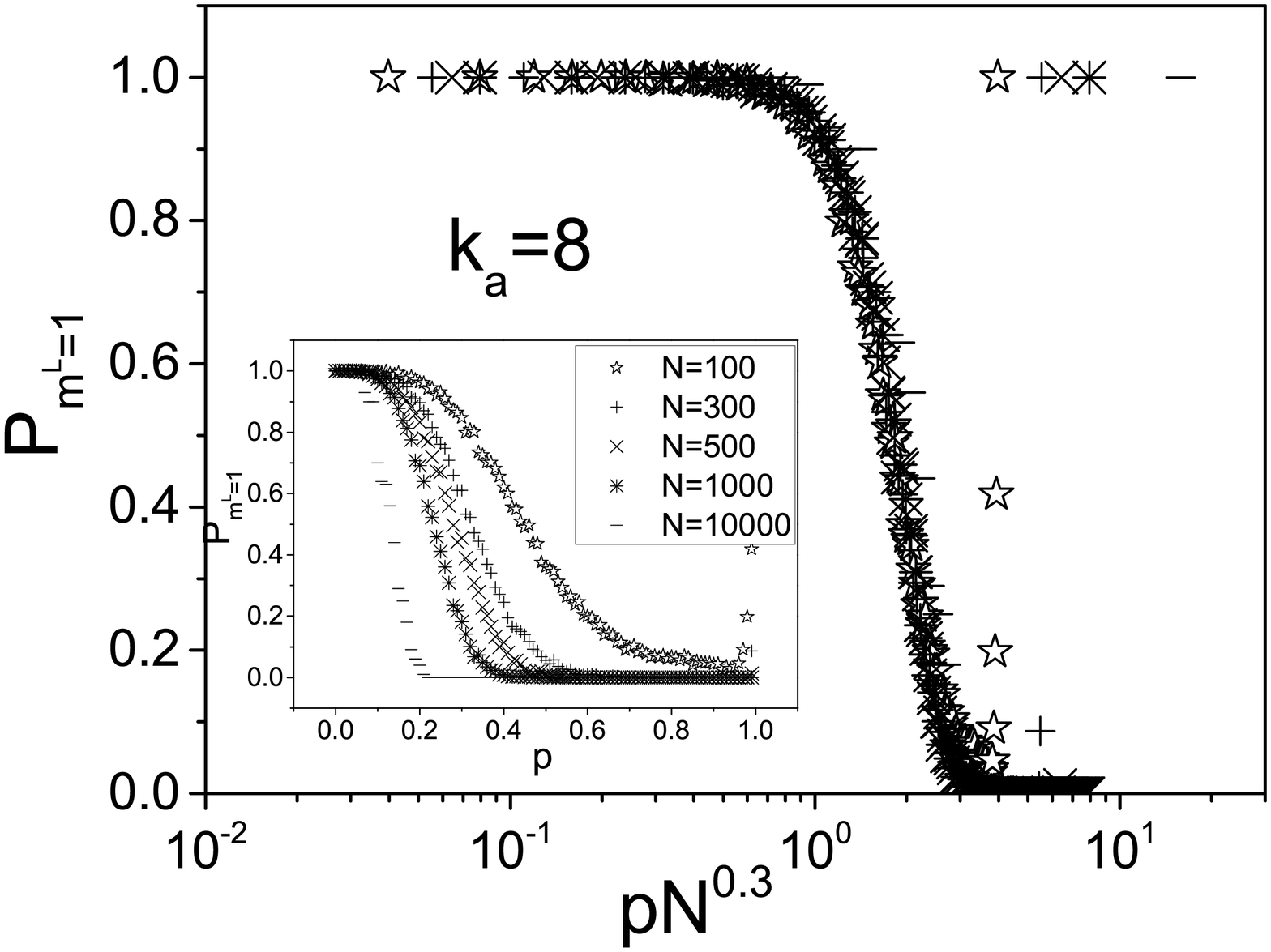}
\includegraphics[width=0.3\textwidth]{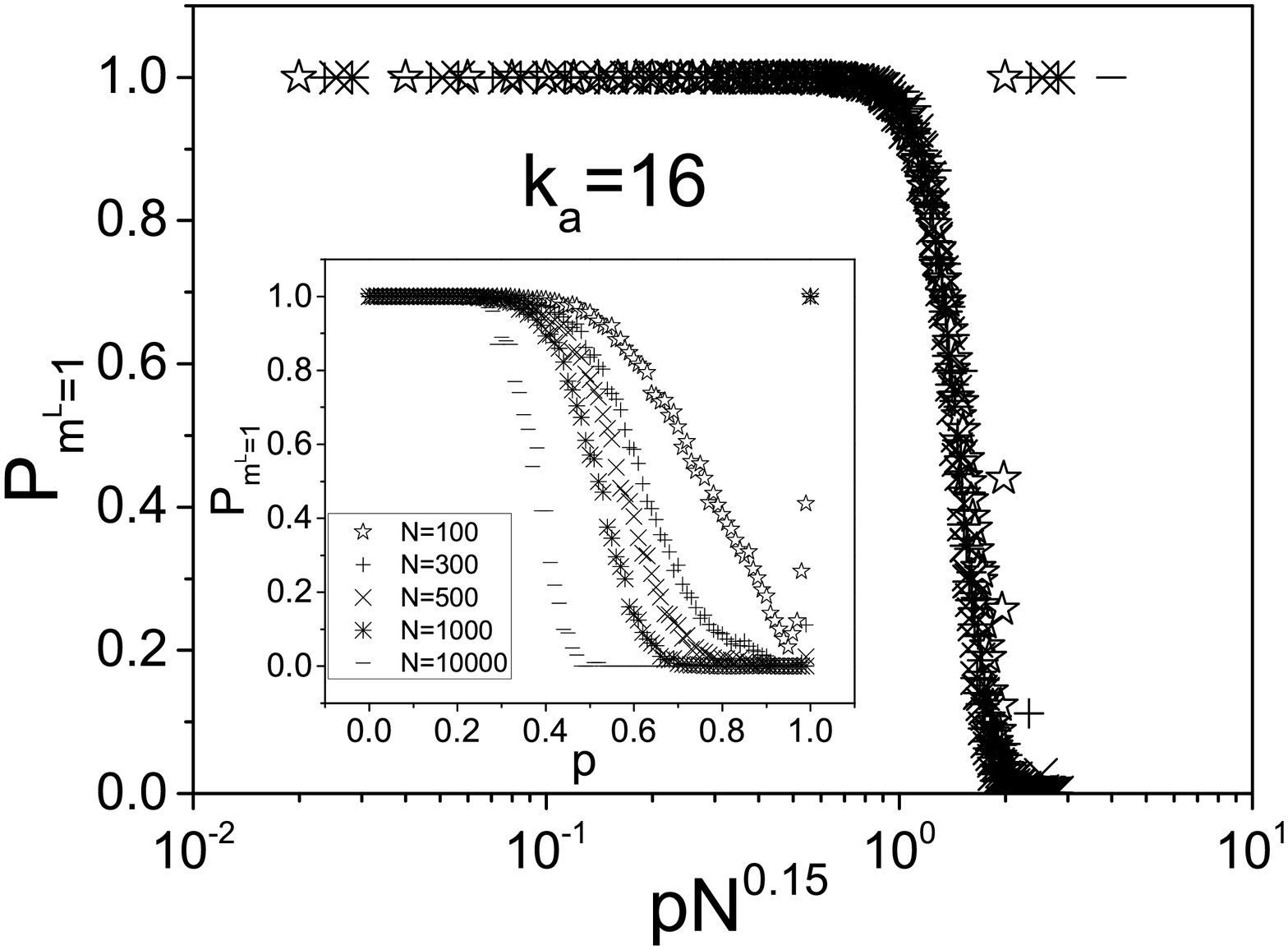}
    \includegraphics[width=0.3\textwidth]{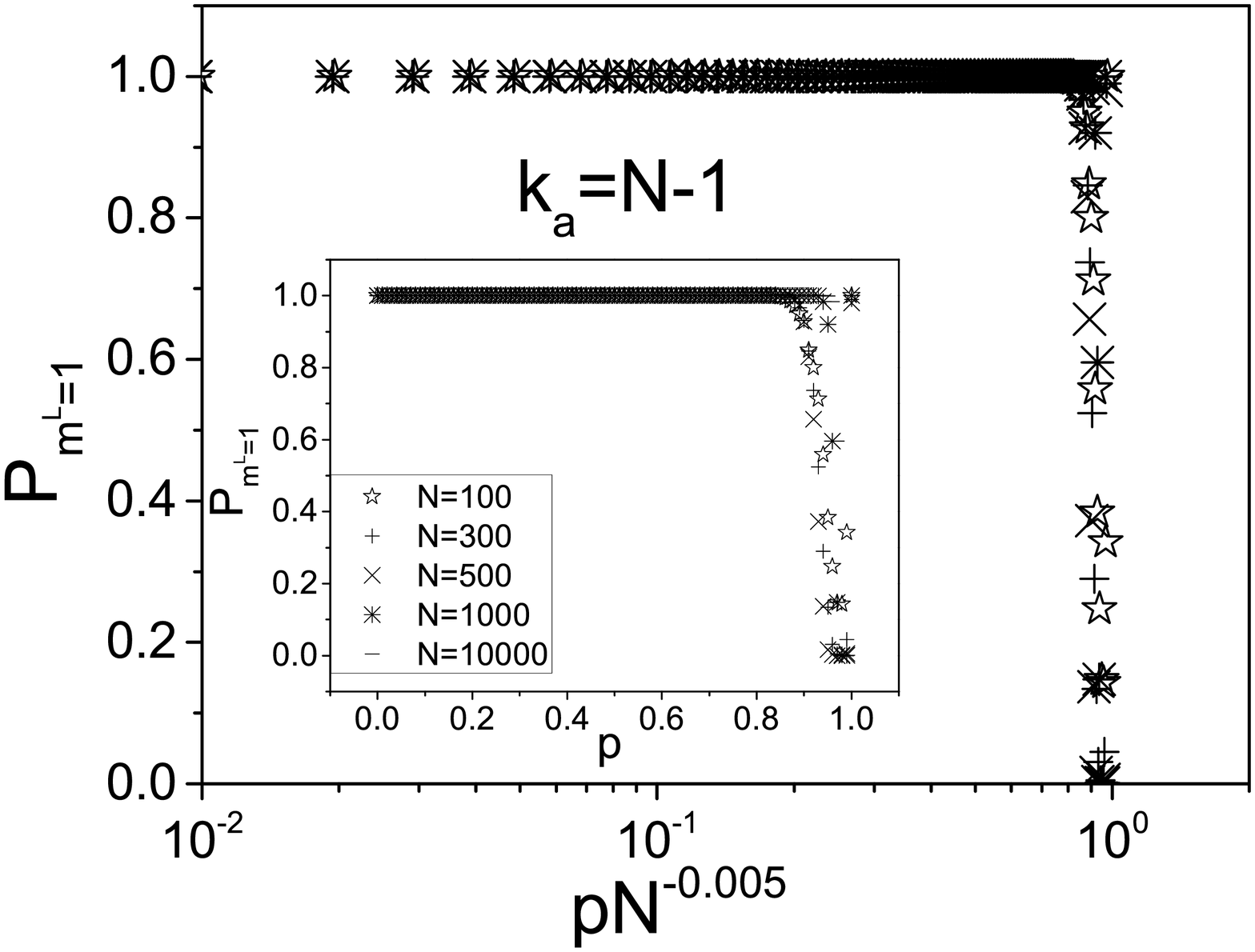}
\caption{$P_{m^L=1}$ as a function of $p$ for different values of $N$ on BA scale-free networks with $k_a=8$ and $k_a=16$, and on a complete network ($k_a=N-1$). The system size varies from $N=100$ to $N=10^4$. The main plots show the scaling law $P_{m^L=1}(p, N)=P_{m^L=1}(pN^{\beta})$ with $\beta=0.3$ for $k_a=8$, $\beta=0.15$ for $k_a=16$, and $\beta=-0.005$ for $k_a=N-1$. The insets show the unscaled results, indicating the transition point shifting toward $p=1$ as $N$ increases.}
\label{consensus:pn}
\end{figure}

More dramatic finite size effects appear if we go beyond the boundary cases ($p=0$ and $p=1$) and consider the coexistence of smart and general agents ($0<p<1$). By setting $\varepsilon=2$ and $\gamma=0$, the first feature that appears in the numerical simulations is the smooth transition between two different extreme states as a function of the proportion $p$ of general agents. This is evident in the inset of the left panel in Fig. \ref{consensus:pn}, where we plot the order parameter $P_{m^L=1}$ as a function of $p$. The transition point clearly decreases with population size $N$, when the average degree $k_a$ is small. A finite size scaling analysis shows that the data can be well fitted by the formula:
\begin{equation}
P_{m^L=1}(p, N)=P_{m^L=1}\left(pN^{\beta}\right),
\end{equation}
with $\beta=0.3$ for $k_a=8$, $\beta=0.15$ for $k_a=16$, and $\beta\approx 0$ for $k_a=N-1$. An strict statistical mechanics analysis based on the thermodynamic limit $N \to \infty$ would conclude that the actual critical point is $p\approx 1$ (but $\neq 1$) for finite $k_a$. However, for divergent $k_a$, the system size has no statistically significant impact on the phase transition behavior in the axis of parameter $p$, as shown in the right panel of Fig. \ref{consensus:pn}. It is implied that a complete consensus is certain to be reached at a finite time for almost all possible values of parameter $p$ when the social graph is completely connected. This is in good agreement with our analytic discussion in section \ref{subsec1}.

\subsection{System convergence time}

\begin{figure}
\centering
	\includegraphics[width=0.4\textwidth]{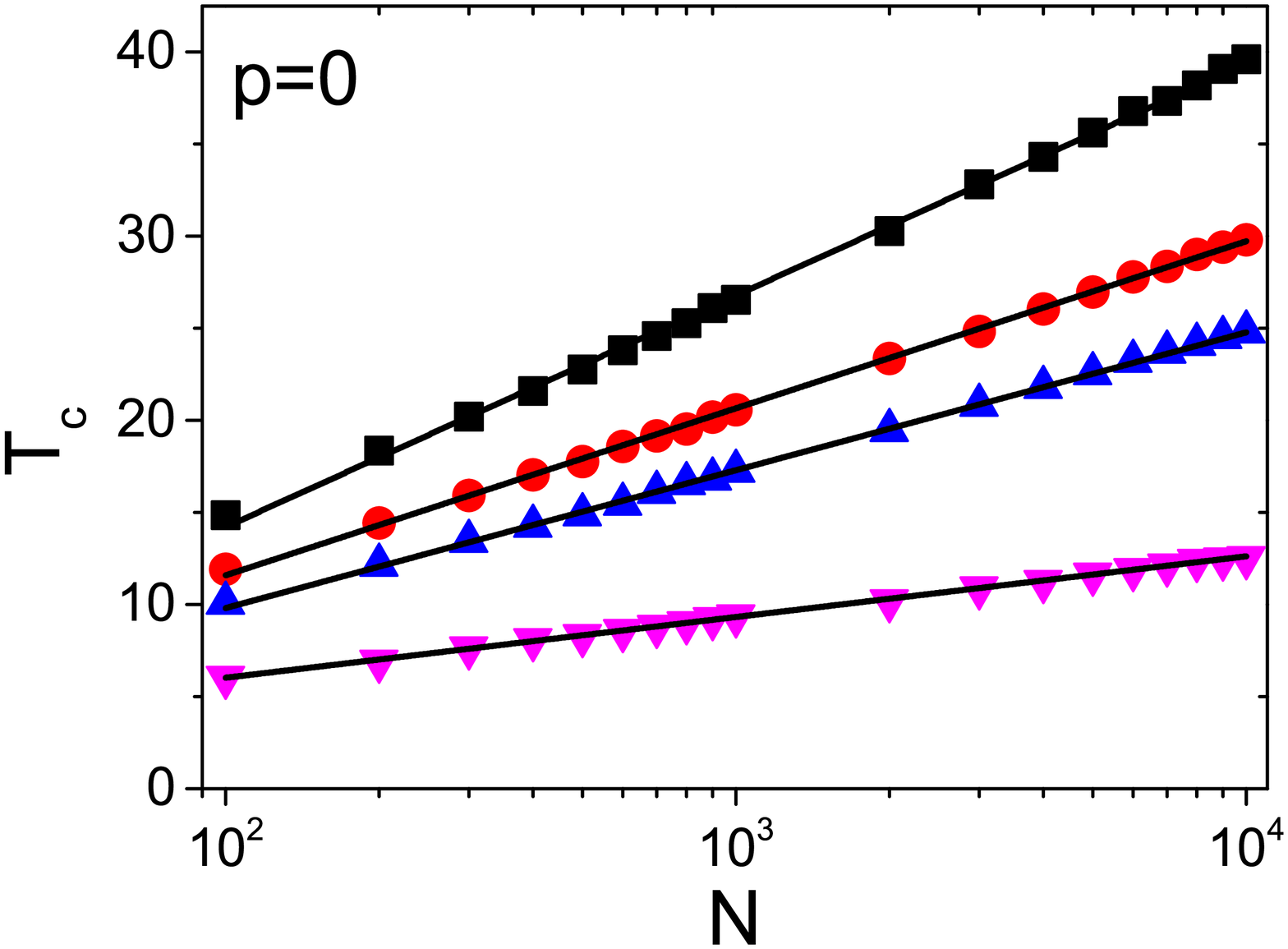}
    \includegraphics[width=0.4\textwidth]{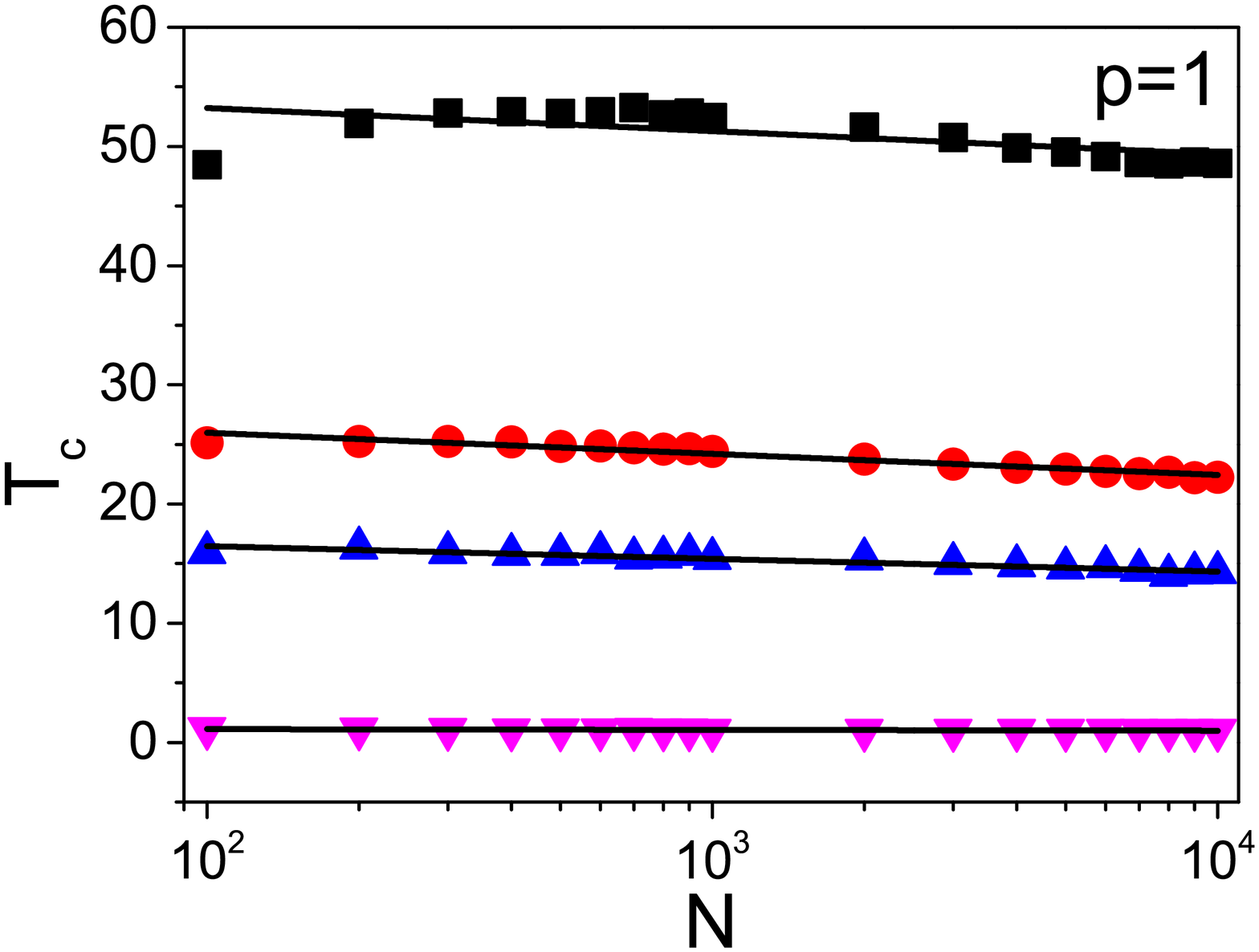}\\
    {\footnotesize $(a)$ $T_c$ as a function of $N$ for BA scale-free networks of $k_a=4$ (squares), 8 (circles) and 16 (up triangles) with $\varepsilon=2$ and $\gamma=0$. Down triangles refer to $k_a=N-1$ (the mean field limit).}\\
    \includegraphics[width=0.4\textwidth]{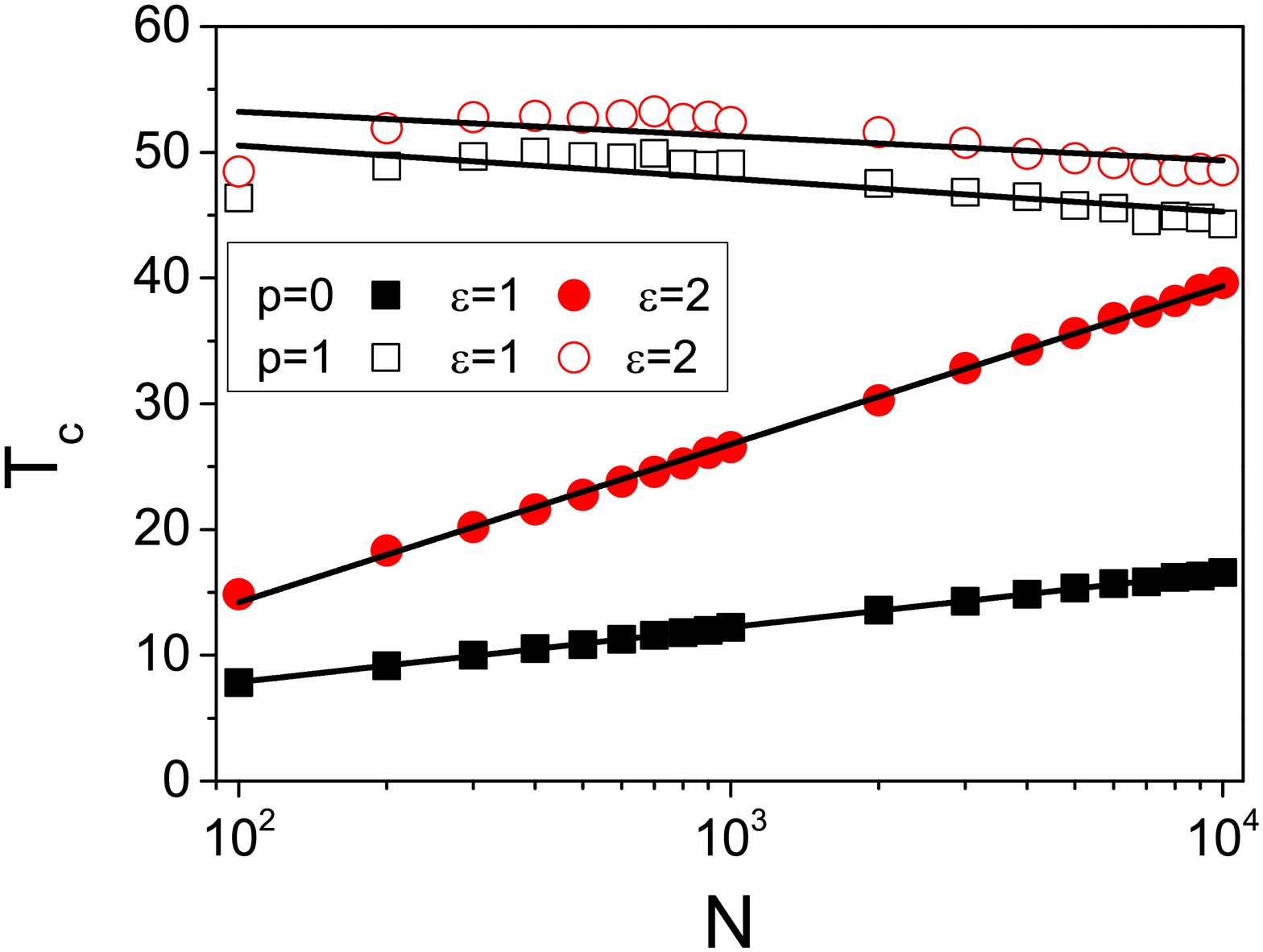}
    \includegraphics[width=0.4\textwidth]{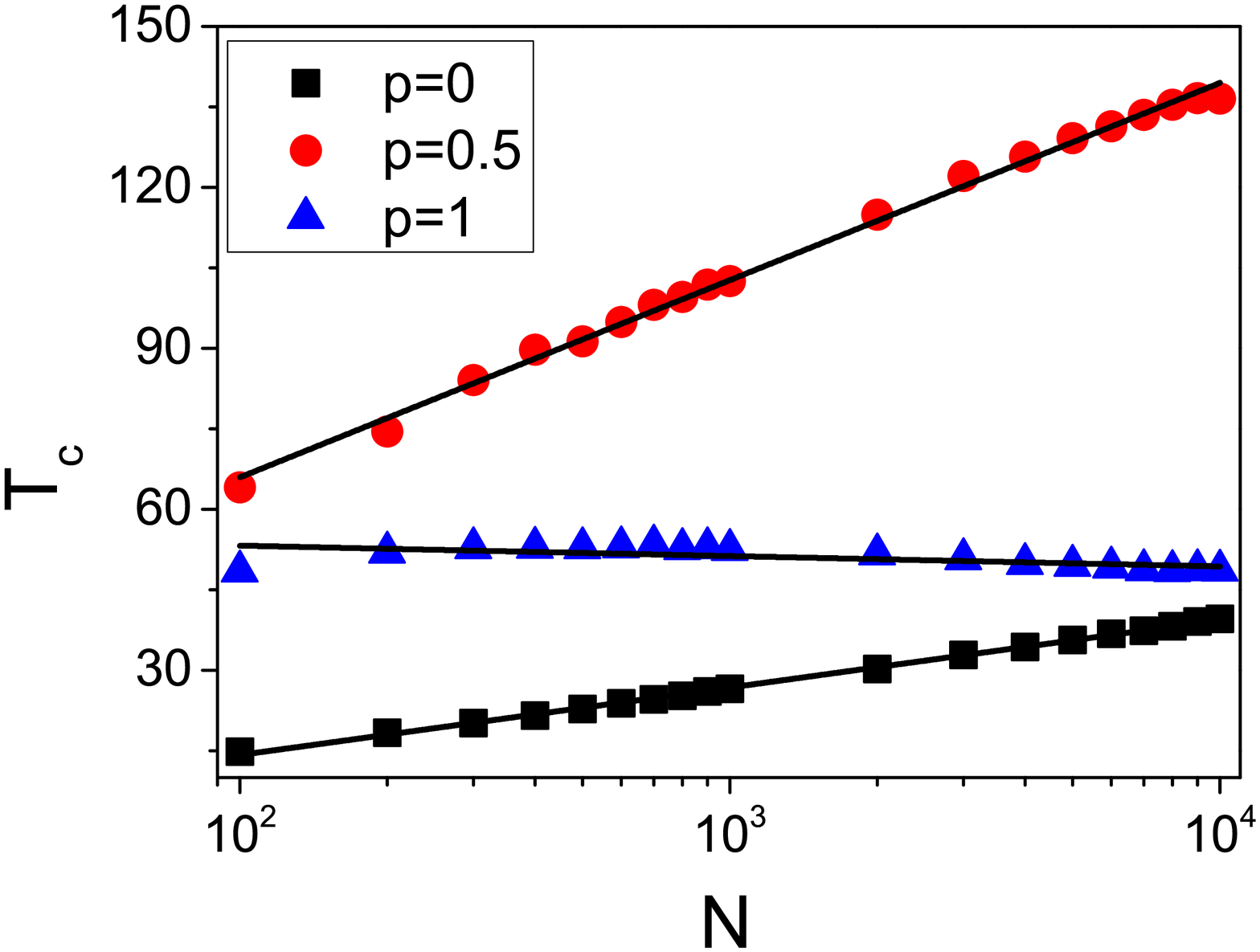}\\
    {\footnotesize $(b)$ $T_c$ as a function of $N$ for a BA scale-free network with $k_a=4$ and $\gamma=0$. $\varepsilon=2$ for the right panel.}
	\caption{(color online) Plot of the system convergence time $T_c$ as a function of network size $N$ in considering different values of $k_a$, $p$ and $\varepsilon$. Each data point is obtained by simulations averaged over 1000 realizations and the straight lines are obtained by logarithmic fittings.}
	\label{time:size}
\end{figure}

It is possible to compute the time it takes the population to reach the complete consensus. Fixing the network structure to be scale-free, we first focus on the distribution of system convergence time $T_c$ in the axis of population size $N$. In numerical simulations, the program stops if no agent changes the opinion after an iteration. Our criterion is to check whether any opinion varies by less than $10^{-8}$ after a sweep. It is shown in Fig. \ref{time:size} that $T_c$ experiences a logarithmic growth in the presence of smart agents but keeps almost unchanged in HK model, with the increase of population size:
\begin{equation}
\left\{
\begin{array}{ll}
T_c \propto \log{N} & \text{ for $p=0$}\\
\\
T_c \approx C & \text{ for $p=1$}
\end{array}
\right.
,
\end{equation}
where $C$ is a constant. In the case of $p=0$, the dependence of $T_c$ with system size $N$ experiments a decrease with the increase of average degree $k_a$. Furthermore, smart agents can drive system converge faster than general ones in HK model when the population is finite and sparsely connected. If smart and general agents are coexisted in the system, however, it is of more difficulty to get the system to a stationary state.

\begin{figure}
\centering
    \includegraphics[width=0.4\textwidth]{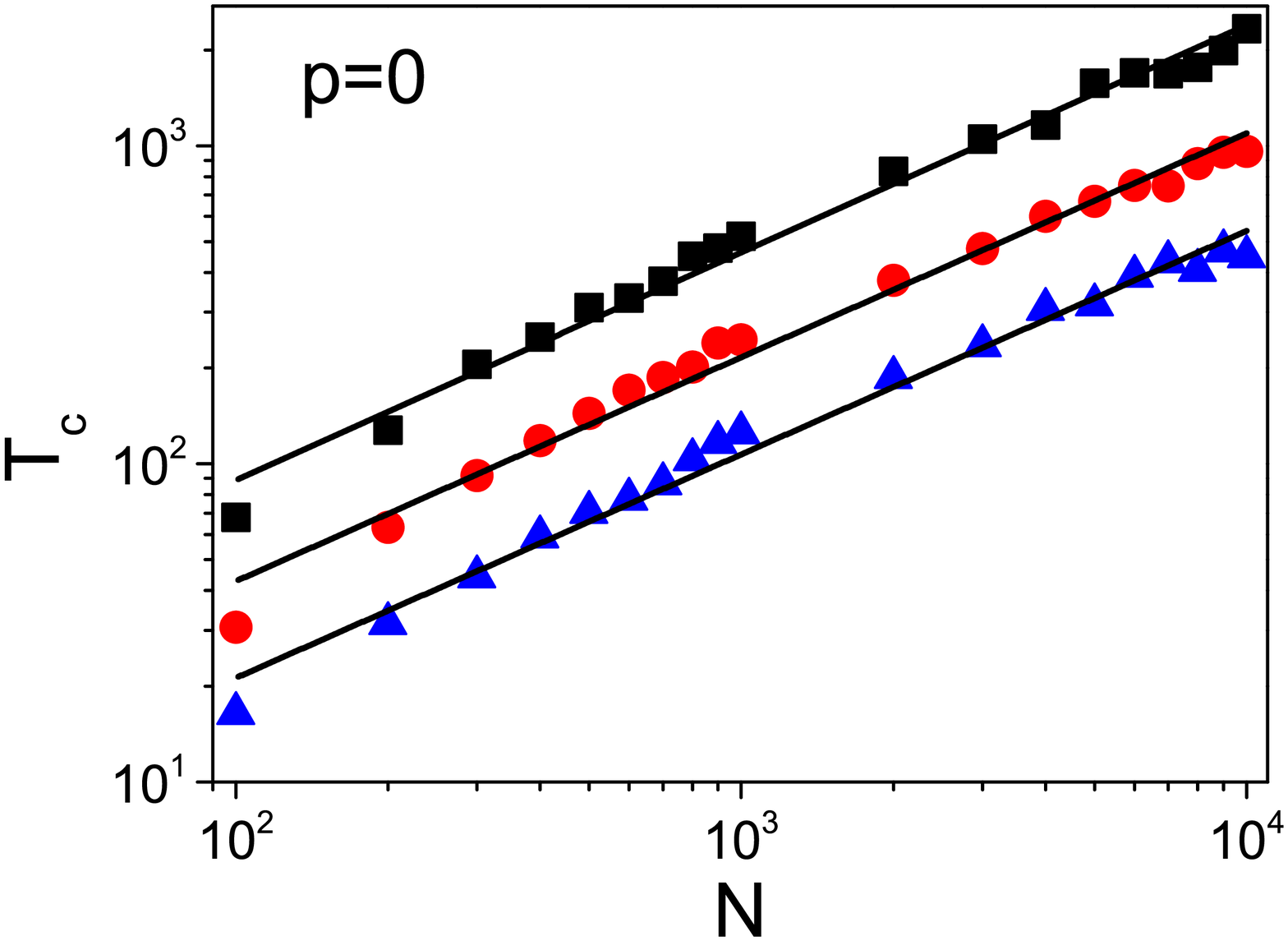}
    \includegraphics[width=0.4\textwidth]{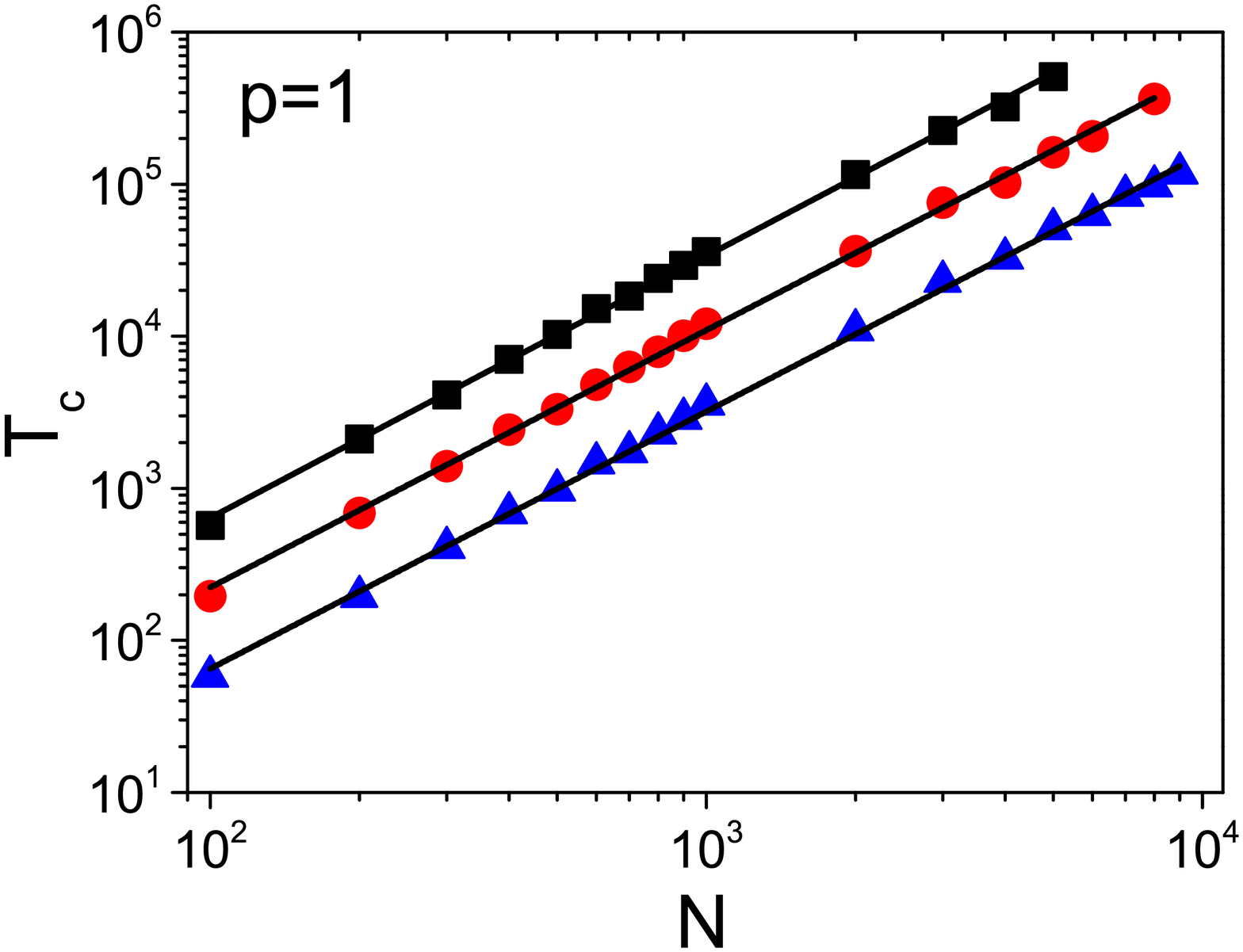}
	\caption{(color online) The time $T_c$ to reach a complete consensus as a power law function of system size $N$ on regular networks of different degrees with $\varepsilon=2$ and $\gamma=0$. Square scatters correspond to $k_a=4$, circle ones to $k_a=8$, and up triangles to $k_a=16$, which have been averaged over 1000 realizations. Straight lines are power law fittings with slope being about 0.7 for p=0 and 1.7 for p=1.}
	\label{time:size2}
\end{figure}

According to the previous analysis of HK model, a generally accepted phenomenon is that the time to reach a complete consensus is bounded by $N^{O(N)}$ and conjectured to be polynomial \cite{wedin2015quadratic}. This general phenomenon, however, is demonstrated only for regular lattices of different dimensions. To certificate again the universality of this scaling behavior, numerical simulations are proposed on general regular graphs of different degrees. It is exhibited in Fig. \ref{time:size2} that $T_c$ scales as a power law function of $N$, which is valid not only for HK model ($p=1$), but also for the case of all agents being smart ($p=0$):
\begin{equation}
T_c\propto N^{\lambda},
\end{equation}
with $\lambda\sim 1.7$ for $p=1$ and $\lambda\sim 0.7$ for $p=0$. This says that a regularly connected population of smart agents will take less time to get a complete consensus, compared to that of general agents in HK model.

\begin{figure}
\centering
	\includegraphics[width=0.4\textwidth]{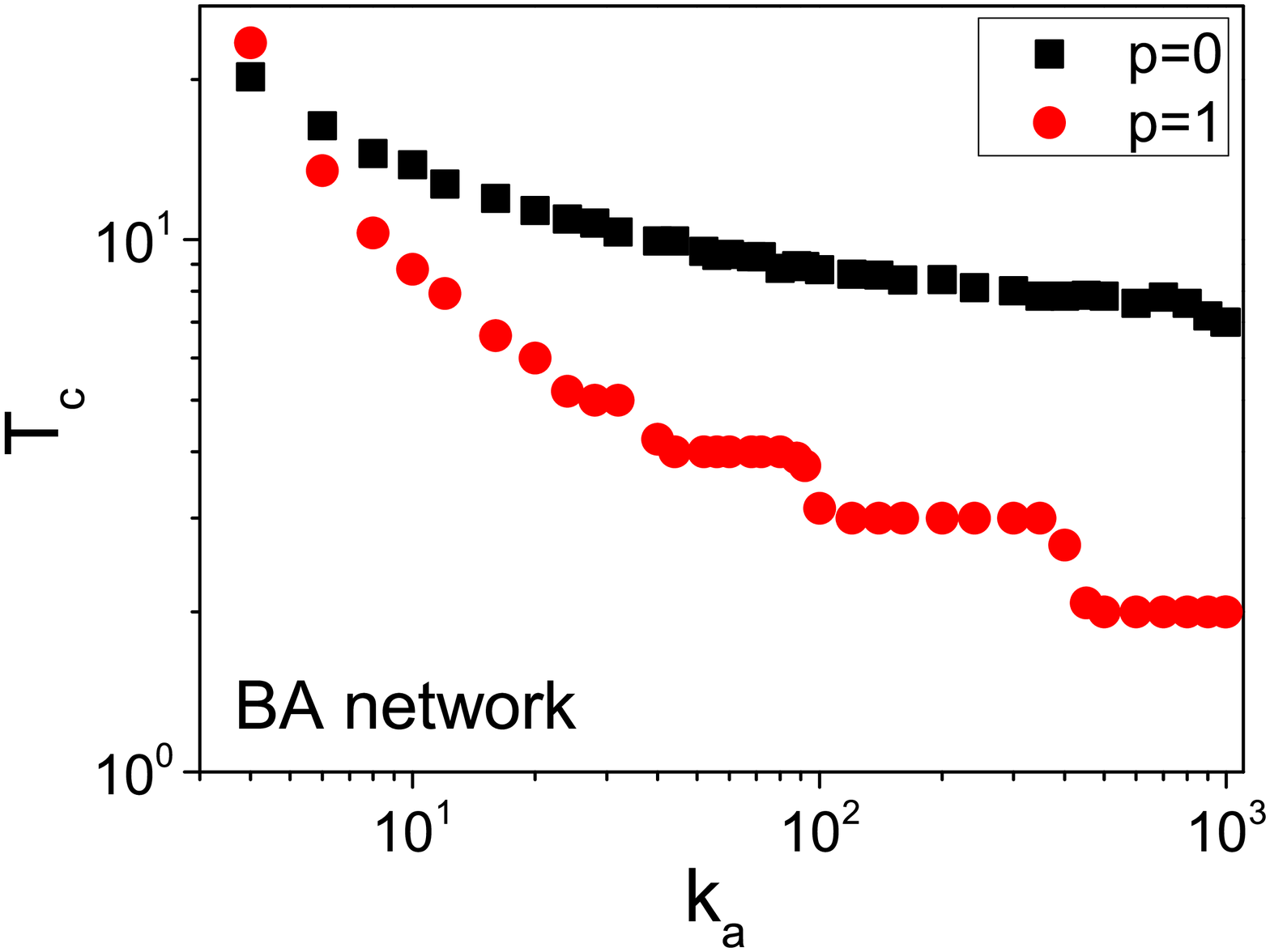}
    \includegraphics[width=0.4\textwidth]{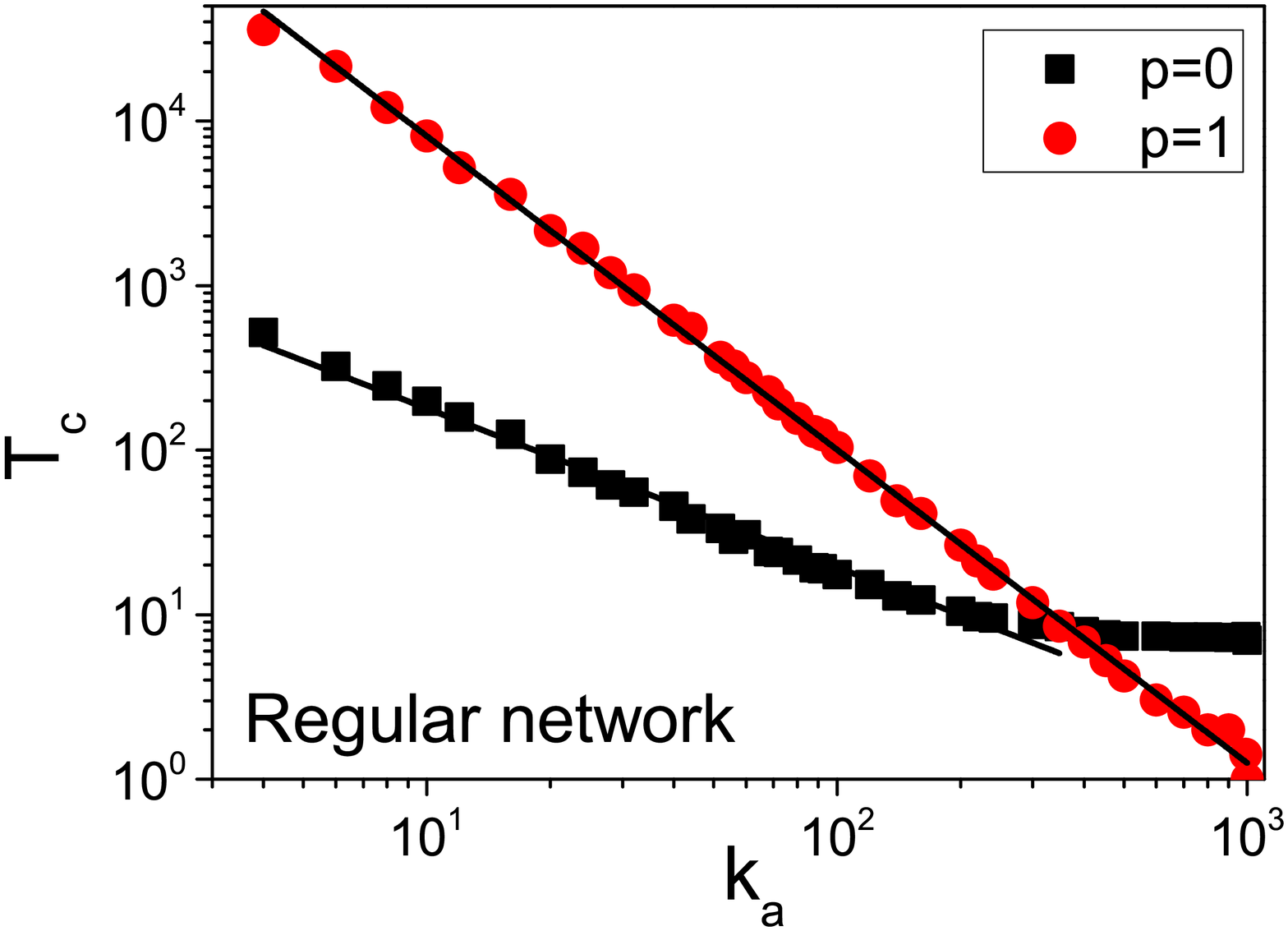}
	\caption{(color online) The time $T_c$ to reach a complete consensus as a function of the average degree $k_a$ for BA scale-free and regular networks with $\varepsilon=2$, $\gamma=0$ and $N=1000$. Notice that each data point is averaged over 1000 realizations and straight lines in the right panel are power law fittings with slope being about -0.96 for p=0 and -1.9 for p=1.}
	\label{time:degree}
\end{figure}

Different from regular networks, BA scale-free graphs will take longer time for the population to reach a complete consensus for $p=0$ than that for $p=1$, as shown in Fig. \ref{time:degree}. Actually, for BA scale-free networks, $T_c$ weakly depends on the average degree $k_a$, especially for the case of $p=0$. For regular networks, however, $T_c$ acts as a power law function of $k_a$ with exponent being about -1.9 for $p=1$ and -0.96 for $p=0$, respectively. We may draw the conclusion that the dependence of $T_c$ on the average degree $k_a$ experiences a strong change for different network structures.

\section{Conclusions}

In summary, we propose a continuous opinion model by introducing smart agents to HK model and considering also the coupling effect between human behaviors and environmental changes. It allows one to understand the propagation of human ideas or attitudes in real-world systems. Social environment is represented by a biased resource allocation among people. We identify the influence of smart agents, resource allocation parameter and network structure on the critical behavior of the underlying system.

(1) The consensus threshold $\varepsilon_c$ is first discussed for a boundary case of all agents being smart. Both theoretical and numerical results claim that $\varepsilon_c$ can only take two possible values, depending on the behavior of the average degree $k_a$ of the social graph, when the system size $N$ approaches infinity. $\varepsilon_c=2$ if $k_a$ is finite in the limit of large $N$, and $\varepsilon_c=1.25$ if instead $k_a \to \infty$ when $N \to \infty$.

(2) An order parameter $P_{m^L=1}$ is introduced, which is defined as the average, over realizations of the discussed process, of the complete consensus. There is a clear change of behavior in the sense that it is possible to identify a transition point separating a complete consensus from the coexistence of different opinions. The transition point occurs at a pseudo-critical value $\varepsilon_c$ for the bounded confidence threshold. We certify from an analytic viewpoint that the generation of a complete consensus strongly depends on the fraction of smart agents and the resource allocation parameter, but it is weakly connected with the topological structure of the network.

(3) Introducing smart agents does not change the finite size scaling law of HK model with the order parameter scaling as: $P_{m^L=1} \sim (\varepsilon-\varepsilon_c)N^{\alpha}$, but has statistically significant impact on the consensus threshold $\varepsilon_c$. The critical behavior depends on the system size in such a way that the phase transition tends to be no-continuous when the system size $N$ increases to infinity. While in the axis of parameter $p$ (the fraction of general agents), a finite size scaling analysis shows that the phase transition can be summarized by the scaling law: $P_{m^L=1} \sim pN^{\beta}$. The exponent $\beta$ decreases with the average degree and gets close to 0 when the network is completely connected. This indicates the appearance of a complete consensus for almost all possible values of $p$. Meanwhile, smart agents do not change the power law functional relationship between system convergence time and the system size on regular networks. They can drive, however, the complete consensus to be reached faster than general agents in HK model when the network structure is homogeneous and far from the mean field limit. In contrast, a heterogeneously structured network will make smart agents take more time to reach a complete consensus.

\subsection*{Acknowledgments}

We acknowledge financial support from the National Natural Science Foundation of China through the project 11947063 and from the Key Laboratory of Quark and Lepton Physics (MOE), Central China Normal University, Wuhan 430079, China.

\appendix

\numberwithin{equation}{section}
\numberwithin{figure}{section}

\section{The consensus threshold $\varepsilon_c$ of HK model}\label{appendix:1}

Denote the gap between initial maximal opinion $O_{2}$ and minimal one $O_{1}$ by $g_o$: $g_o=O_{2}-O_{1}$. With the mean field limit: $k_a=N-1$, a group opinion is formed at time $t$ only when the difference between the maximal and minimal opinions at time $(t-1)$ is no larger than the bounded confidence threshold $\varepsilon$.

\begin{figure}
\centering
	\includegraphics[width=0.3\textwidth]{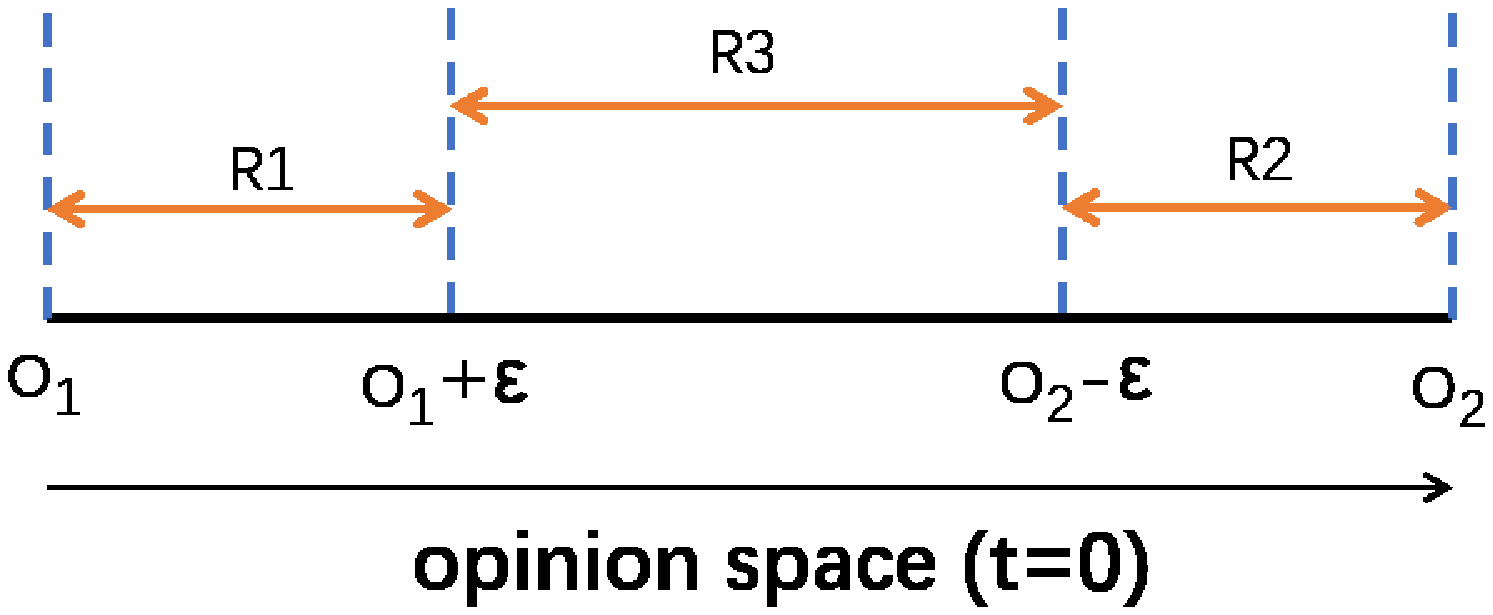}
    \includegraphics[width=0.32\textwidth]{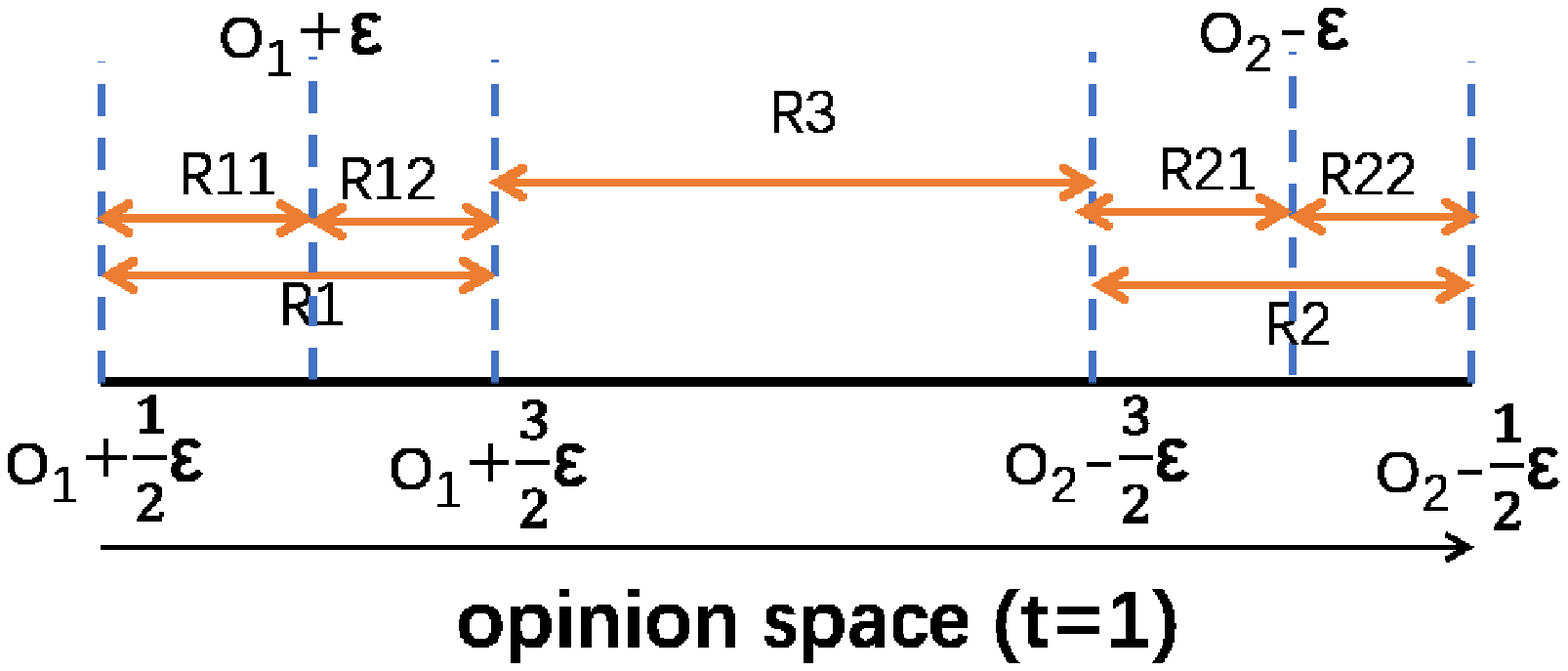}
    \includegraphics[width=0.32\textwidth]{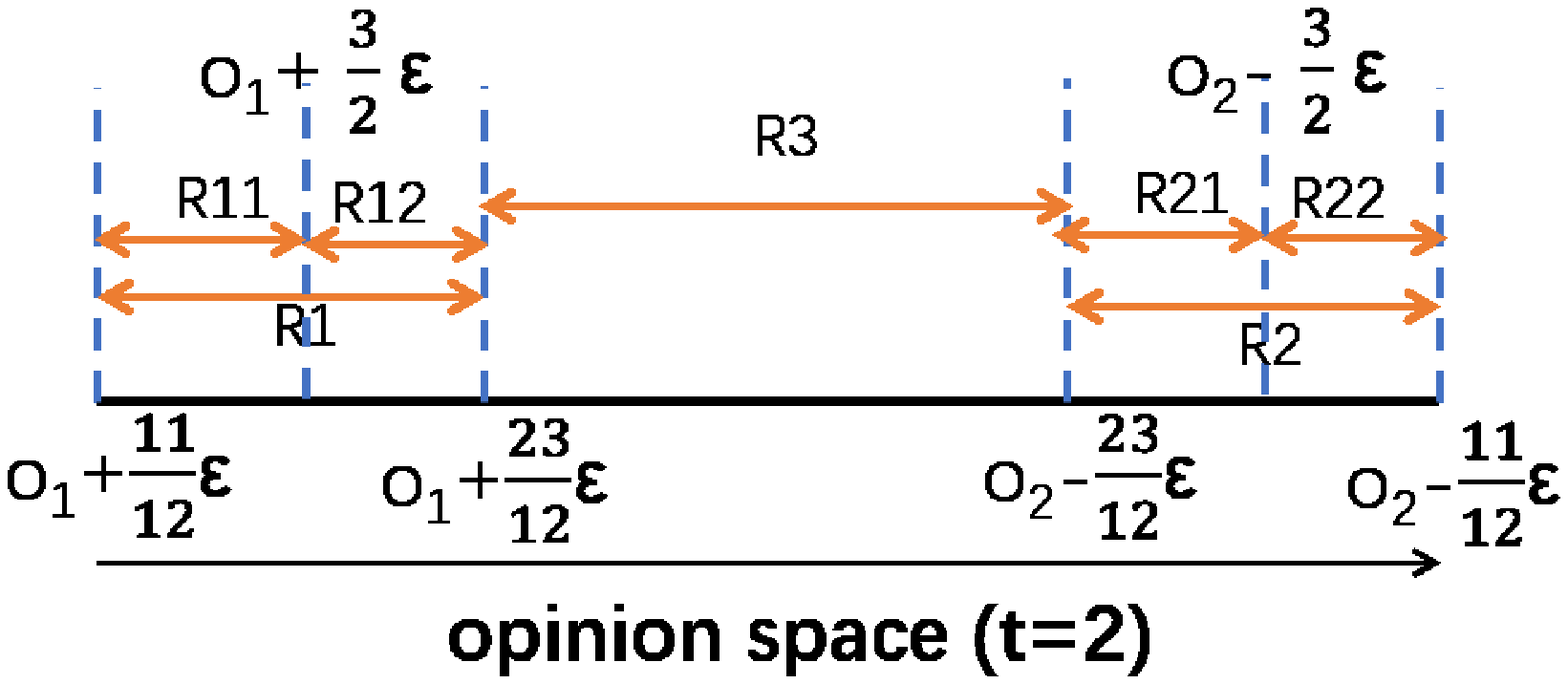}
	\caption{(color online) Division of the opinion space at different time steps.}
	\label{opinion:space}
\end{figure}

For the simplicity in theoretical analysis, we first divide the initial opinion space into three adjacent intervals, as presented in Fig. \ref{opinion:space}. With the limitation of bounded confidence, agents taking the minimal opinion $O_1$ will be affected by those ones taking opinions in interval $R1(t=0)$. Analogously, agents taking the maximal opinion $O_2$ will be affected by those ones holding opinions in interval $R2(t=0)$. Stochastic distribution of initial opinions determines the opinion space of the next step:
\begin{equation}
O_{min}(t=1)=\frac{1}{2}(O_1+O_1+\varepsilon)=O_1+\frac{\varepsilon}{2}.
\end{equation}
and
\begin{equation}
O_{max}(t=1)=\frac{1}{2}(O_2+O_2-\varepsilon)=O_2-\frac{\varepsilon}{2}.
\end{equation}
Agent $i$ of initial opinion in $R3(t=0)$ will keep its opinion as it will be affected by those ones of opinion in $[o_i-\varepsilon, o_i+\varepsilon]$, yielding
\begin{equation}
o_i(t=1)=\frac{1}{2}\left(o_i-\varepsilon+o_i+\varepsilon\right)=o_i.
\end{equation}
The average opinions in both intervals $R12(t=1)$ and $R21(t=1)$ are calculated as
\begin{equation}
\Omega_{R12}(t=1)=\frac{1}{2}({O _1+\varepsilon+O_1+\frac{3}{2}\varepsilon})=O_1+\frac{5}{4}\varepsilon,
\end{equation}
and
\begin{equation}
\Omega_{R21}(t=1)=\frac{1}{2}({O_2-\varepsilon+O_2-\frac{3}{2}\varepsilon})=O_2-\frac{5}{4}\varepsilon.
\end{equation}
An agent of opinion in $R1(t=0)$ ($R2(t=0)$), however, will be persuaded by agents holding opinions in $[O_1, o_i+\varepsilon]$ ($[o_i-\varepsilon, O_2]$). This guides, respectively, the average opinions in intervals $R11(t=1)$ and $R22(t=1)$ to be
\begin{equation}
\Omega_{R11}(t=1)=\frac{1}{\varepsilon}\int_{O_1}^{O_1+\varepsilon}\frac{1}{2}(O_1+o_i+\varepsilon){\rm d}o_i=O_1+\frac{3}{4}\varepsilon,
\end{equation}
and
\begin{equation}
\Omega_{R22}(t=1)=\frac{1}{\varepsilon}\int_{O_2-\varepsilon}^{O_2}\frac{1}{2}(O_2+o_i-\varepsilon){\rm d}o_i=O_2-\frac{3}{4}\varepsilon.
\end{equation}
The opinion space at $t=1$ is then updated by the average opinions in intervals $R1(t=1)$ and $R2(t=1)$:
\begin{equation}
O_{min}(t=2)=\frac{\left[(O_1+\frac{3}{4}\varepsilon)\varepsilon+(O_1+\frac{5}{4}\varepsilon)\frac{1}{2}\varepsilon\right]}{\frac{3}{2}\varepsilon}\approx O_1+0.91\varepsilon,
\end{equation}
\begin{equation}
O_{max}(t=2)=\frac{\left[(O_2-\frac{3}{4}\varepsilon)\varepsilon+(O_2-\frac{5}{4}\varepsilon)\frac{1}{2}\varepsilon\right]}{\frac{3}{2}\varepsilon}\approx O_2-0.91\varepsilon.
\end{equation}
They will be further influenced by opinions in intervals $R1(t=2)$ and $R2(t=2)$.

\begin{figure}
\centering
\includegraphics[width=0.4\textwidth]{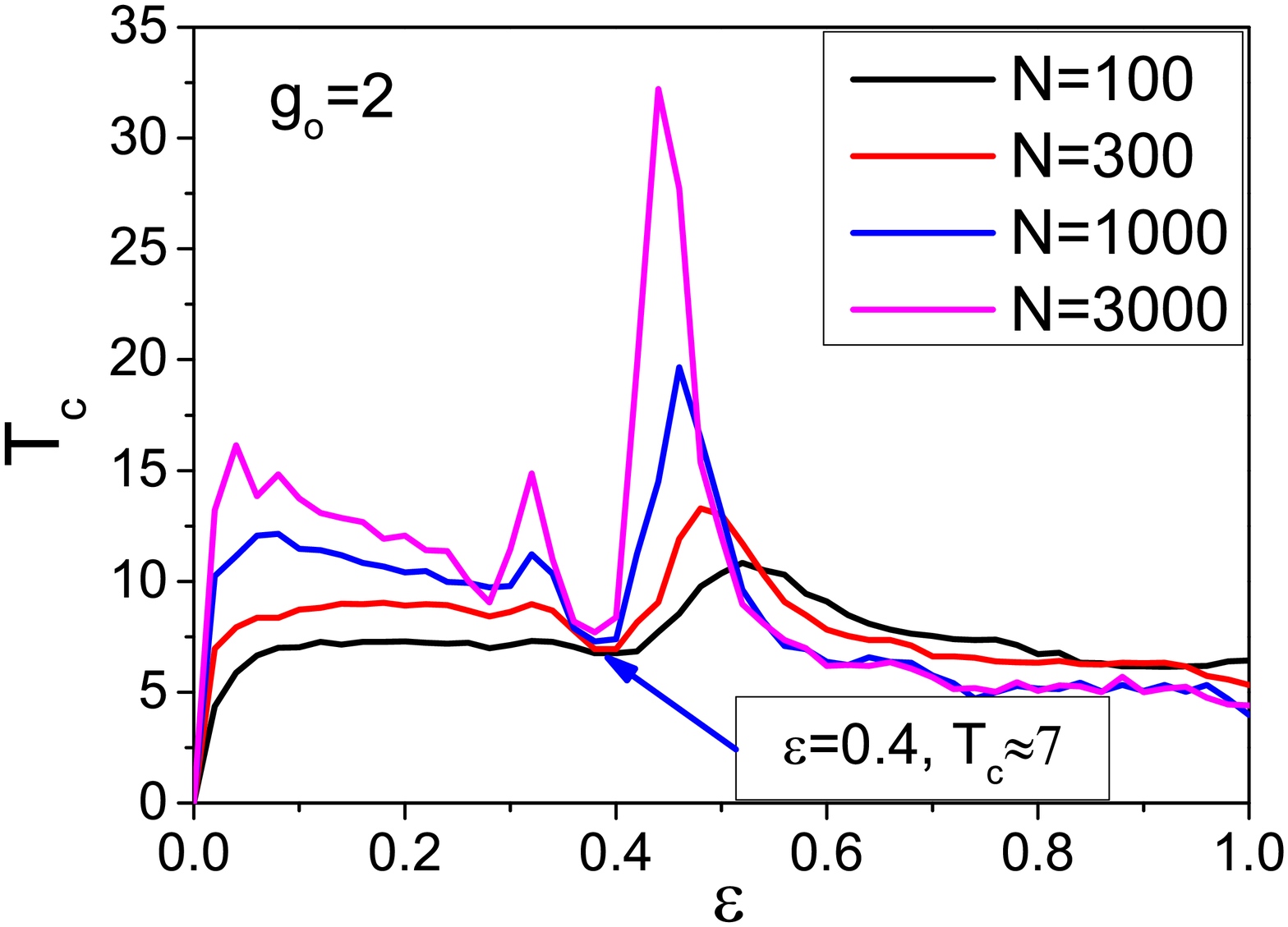}
\includegraphics[width=0.4\textwidth]{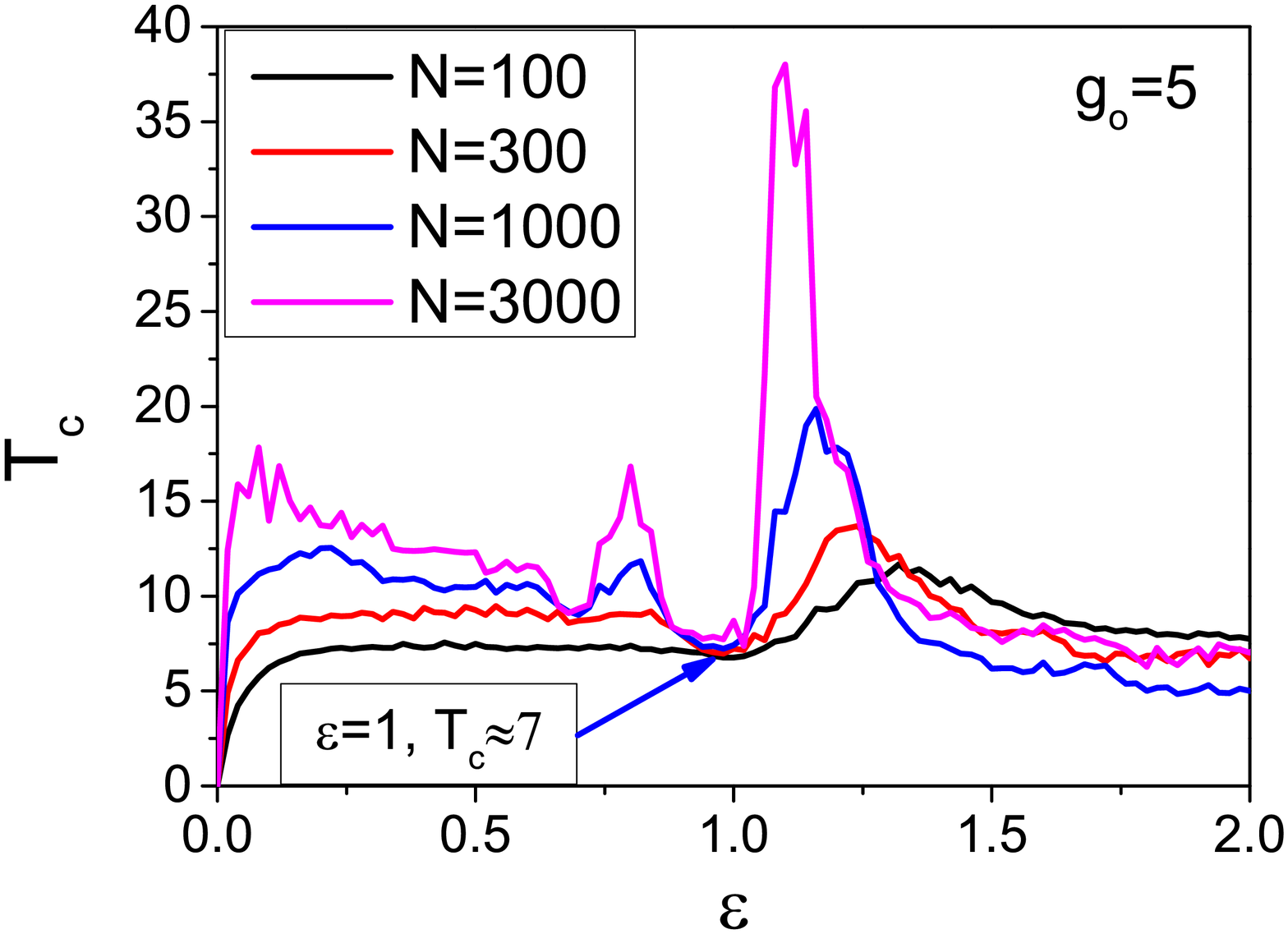}
\caption{(color online) Simulation results for the distribution of the time $T_c$ to reach a complete consensus with $k_a=N-1$ and $p=1$. $g_o$ denotes the opinion distance at $t=0$.}
\label{time:gap}
\end{figure}

The average opinions in both $R11(t=2)$ and $R22(t=2)$ can be obtained as
\begin{eqnarray}
\Omega_{R11}(t=2)=\frac{1}{\varepsilon}\int_{O_1+\frac{1}{2}\varepsilon}^{O_1+\frac{3}{2}\varepsilon} \frac{(O_1+\frac{3}{4}\varepsilon)\varepsilon+\frac{1}{2}(O_1+o_i+2\varepsilon)(o_i-O_1)} {\varepsilon+o_i-O_1}{\rm d}o_i=O_1+(\frac{1}{4}\ln{\frac{5}{3}}+1)\varepsilon,\\
\Omega_{R22}(t=2)=\frac{1}{\varepsilon}\int_{O_2-\frac{3}{2}\varepsilon}^{O_2-\frac{1}{2}\varepsilon} \frac{(O_2-\frac{3}{4}\varepsilon)\varepsilon+\frac{1}{2}(O_2+o_i-2\varepsilon)(O_2-o_i)} {O_2+\varepsilon-o_i}{\rm d}o_i=O_2-(\frac{1}{4}\ln{\frac{5}{3}}+1)\varepsilon.
\end{eqnarray}
We then get respectively the average opinions in both $R12(t=2)$ and $R21(t=2)$:
\begin{eqnarray}
\Omega_{R12}(t=2)= \frac{1}{\frac{5}{12}\varepsilon}\int_{O_1+\frac{3}{2}\varepsilon}^{O_1+\frac{23}{12}\varepsilon} \frac{(O_1+\frac{3}{4}\varepsilon)\cdot \frac{O_1+2\varepsilon-o_i}{\frac{1}{2}\varepsilon}\varepsilon+\frac{1}{2}(O_1+2\varepsilon+o_i)(o_i-O_1)} {\frac{O_1+2\varepsilon-o_i}{\frac{1}{2}\varepsilon}\varepsilon+o_i-O_1}{\rm d}o_i = O_1-(\frac{113}{48}+\frac{108}{5}\ln{\frac{5}{6}})\varepsilon,\\
\Omega_{R21}(t=2)= \frac{1}{\frac{5}{12}\varepsilon}\int_{O_2-\frac{23}{12}\varepsilon}^{O_2-\frac{3}{2}\varepsilon} \frac{(O_2-\frac{3}{4}\varepsilon)\cdot \frac{o_i+2\varepsilon-O_2}{\frac{1}{2}\varepsilon}\varepsilon+\frac{1}{2}(O_2-2\varepsilon+o_i)(O_2-o_i)} {\frac{o_i+2\varepsilon-O_2}{\frac{1}{2}\varepsilon}\varepsilon+O_2-o_i}{\rm d}o_i = O_2+(\frac{113}{48}+\frac{108}{5}\ln{\frac{5}{6}})\varepsilon,.
\end{eqnarray}
Computing further the average opinions in intervals $R1(t=2)$ and $R2(t=2)$, we have the opinion space at $t=3$:
\begin{eqnarray}
O_{min}(t=3) &=& \frac{\left[O_1+(\frac{1}{4}\ln{\frac{5}{3}}+1)\varepsilon\right]\cdot \frac{3}{2}\varepsilon+\left[O_1-(\frac{113}{48}+\frac{108}{5}\ln{\frac{5}{6}})\varepsilon\right]\cdot \frac{5}{12}\varepsilon}{\frac{23}{12}\varepsilon}  \approx O_1+1.23\varepsilon,\\
O_{max}(t=3) &=& \frac{\left[O_2-(\frac{1}{4}\ln{\frac{5}{3}}+1)\varepsilon\right]\cdot \frac{3}{2}\varepsilon+\left[O_2+(\frac{113}{48}+\frac{108}{5}\ln{\frac{5}{6}})\varepsilon\right]\cdot \frac{5}{12}\varepsilon}{\frac{23}{12}\varepsilon} \approx O_2-1.23\varepsilon.
\end{eqnarray}
Following the above time sequence of the opinion space, it is natural to derive the general approximations:
\begin{eqnarray}
O_{min}(t+1)-O_{min}(t)=0.5-0.1\cdot t,\\
O_{max}(t+1)-O_{max}(t)=-0.5+0.1\cdot t.
\end{eqnarray}
To have a complete consensus, we should first make $O_{min}(t+1)-O_{min}(t)=0$ and $O_{max}(t+1)-O_{max}(t)=0$. This yields $t=5$ and
\begin{equation}
O_{min}(t=5)=O_1+1.5\varepsilon, \quad  O_{max}(t=5)=O_2-1.5\varepsilon.
\end{equation}
These two boundary opinions are then updated for the coming time step as
\begin{eqnarray}
O_{min}(t=6)=\frac{1}{2}(O_1+1.5\varepsilon+\frac{O_1+O_2}{2}),\\
O_{max}(t=6)=\frac{1}{2}(O_2-1.5\varepsilon+\frac{O_1+O_2}{2}).
\end{eqnarray}
A complete consensus requires
\begin{equation}
O_{max}(t=6)-O_{min}(t=6)\leq \varepsilon,
\end{equation}
which yields
\begin{equation}
\varepsilon_c=\frac{1}{5}g_0, \qquad T_c=7.
\end{equation}
a result well confirmed by the numerical simulations, as shown in Fig. \ref{time:gap}.

\section{Distribution of the number of opinion clusters} \label{opinion:clustering}

\begin{figure}
\centering
    \includegraphics[width=0.4\textwidth]{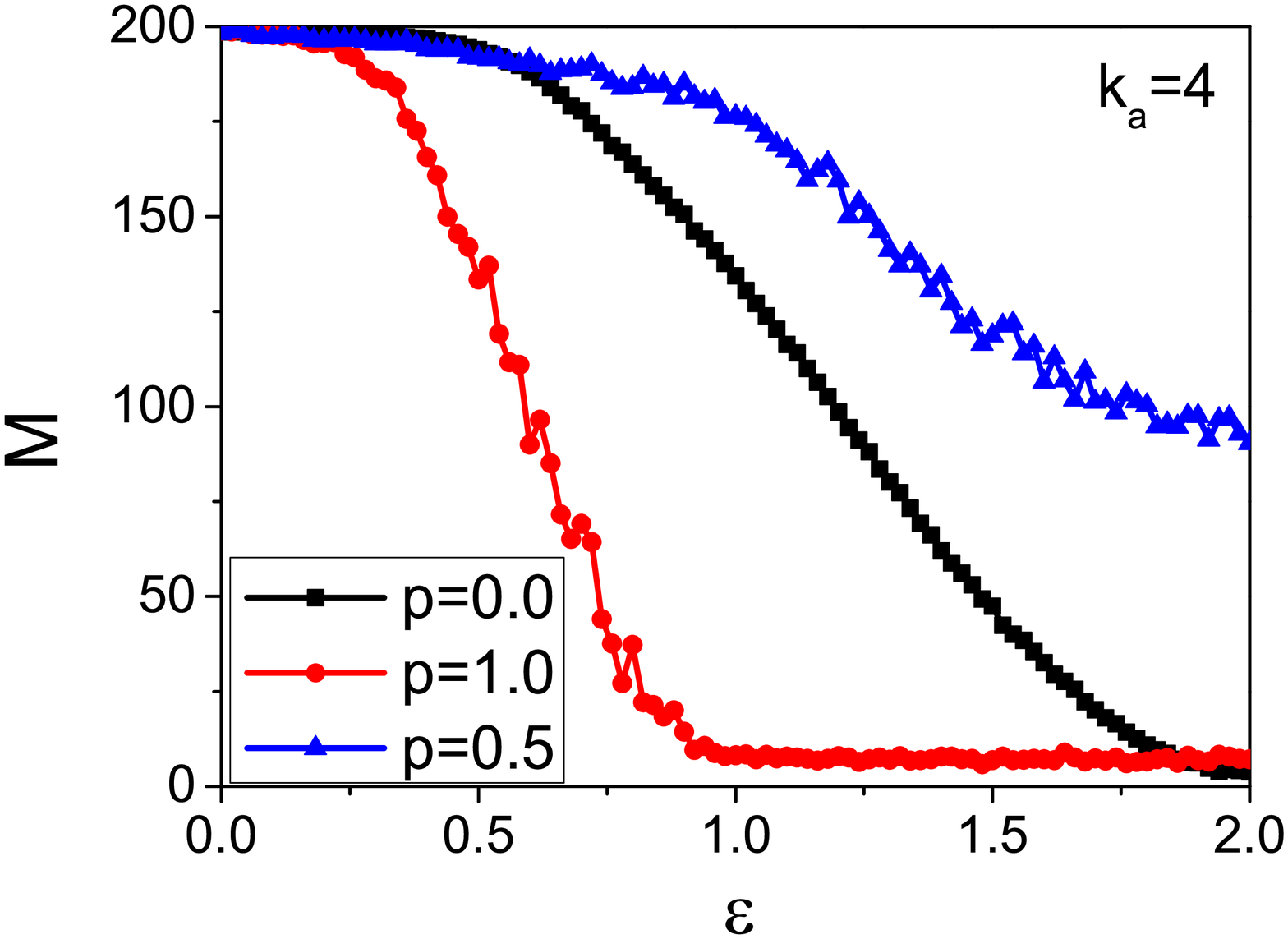}
    \includegraphics[width=0.4\textwidth]{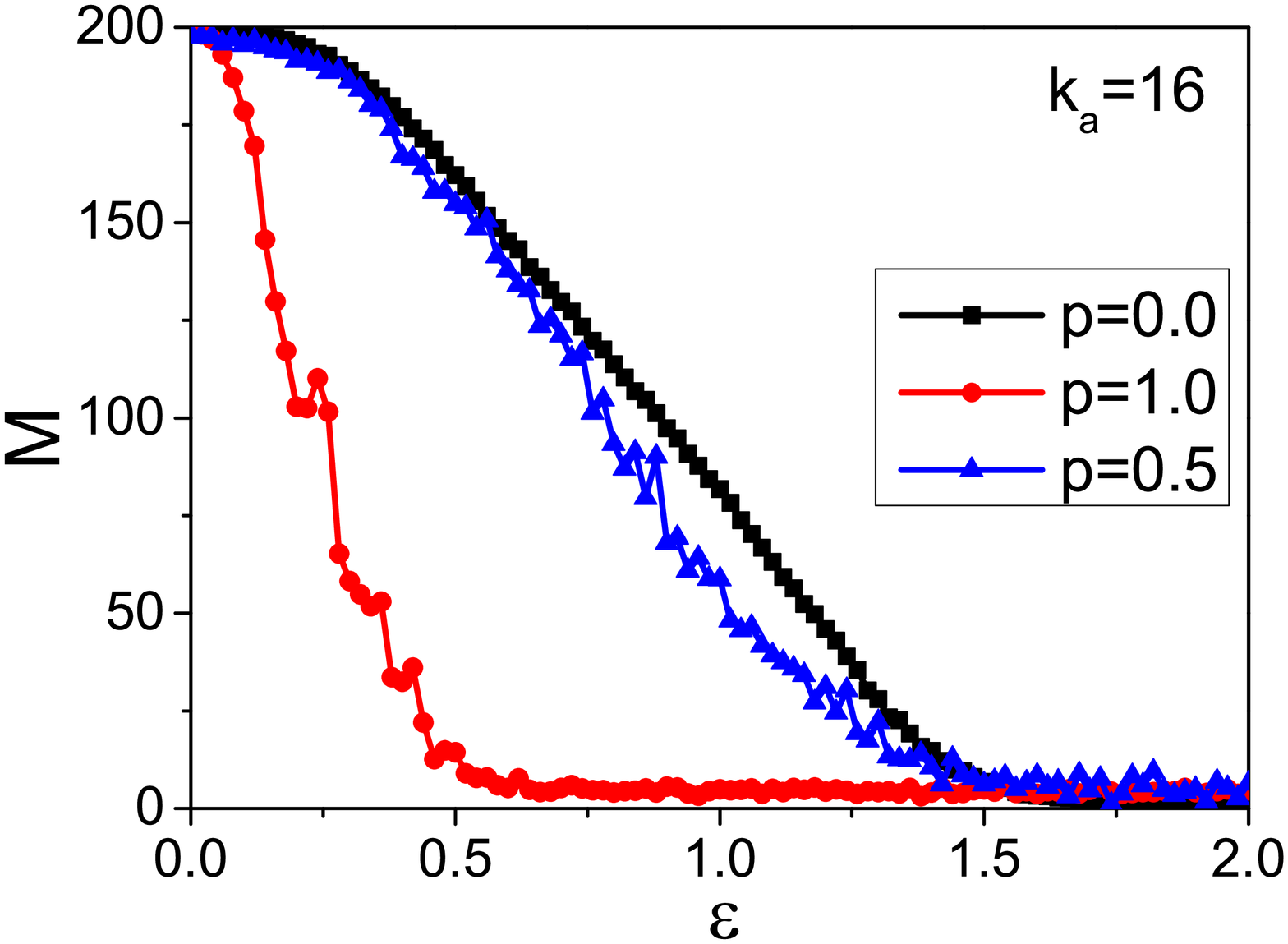}\\
    {\footnotesize $(a)$ Results for regular networks.}\\
    \includegraphics[width=0.4\textwidth]{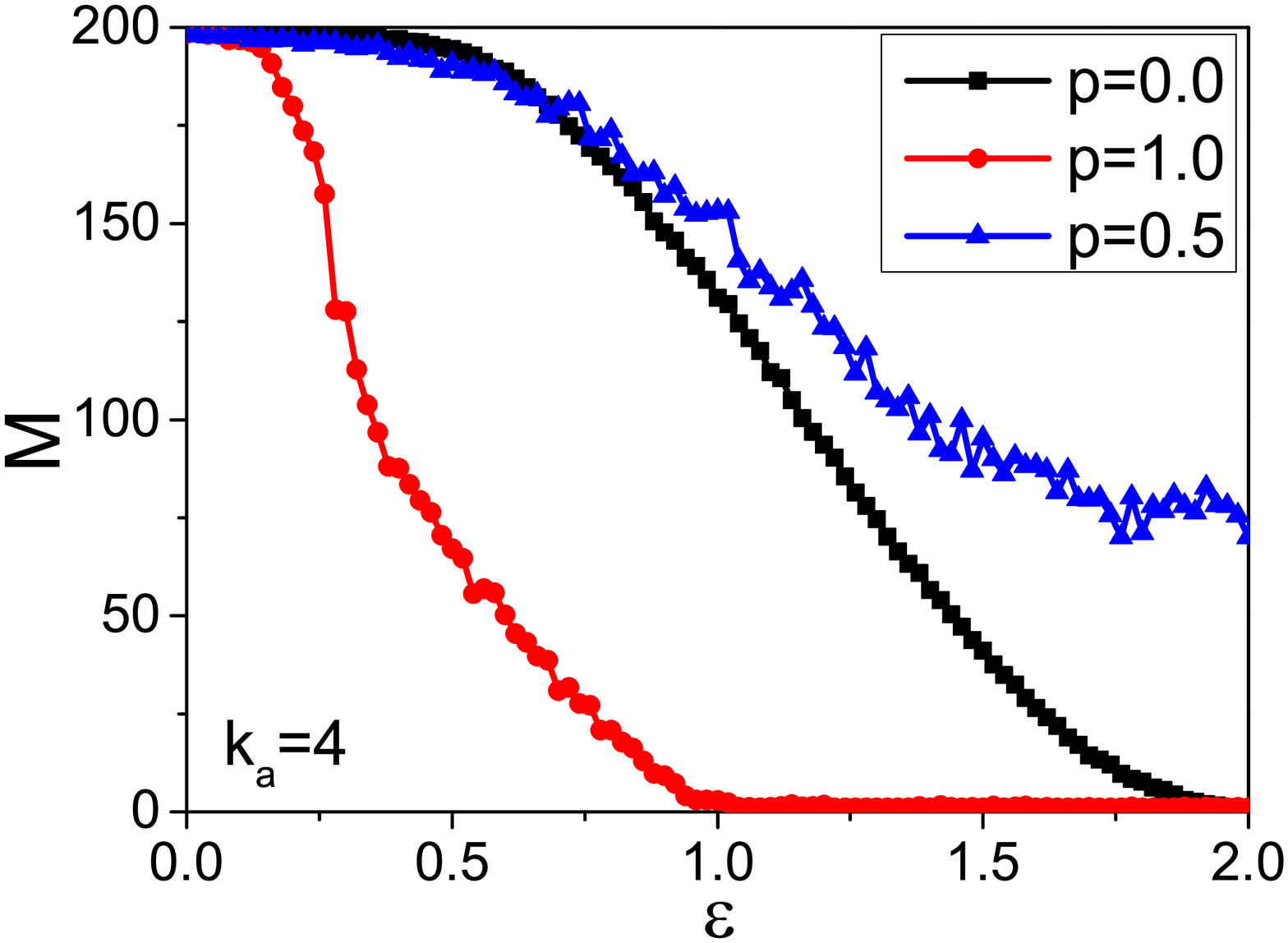}
    \includegraphics[width=0.4\textwidth]{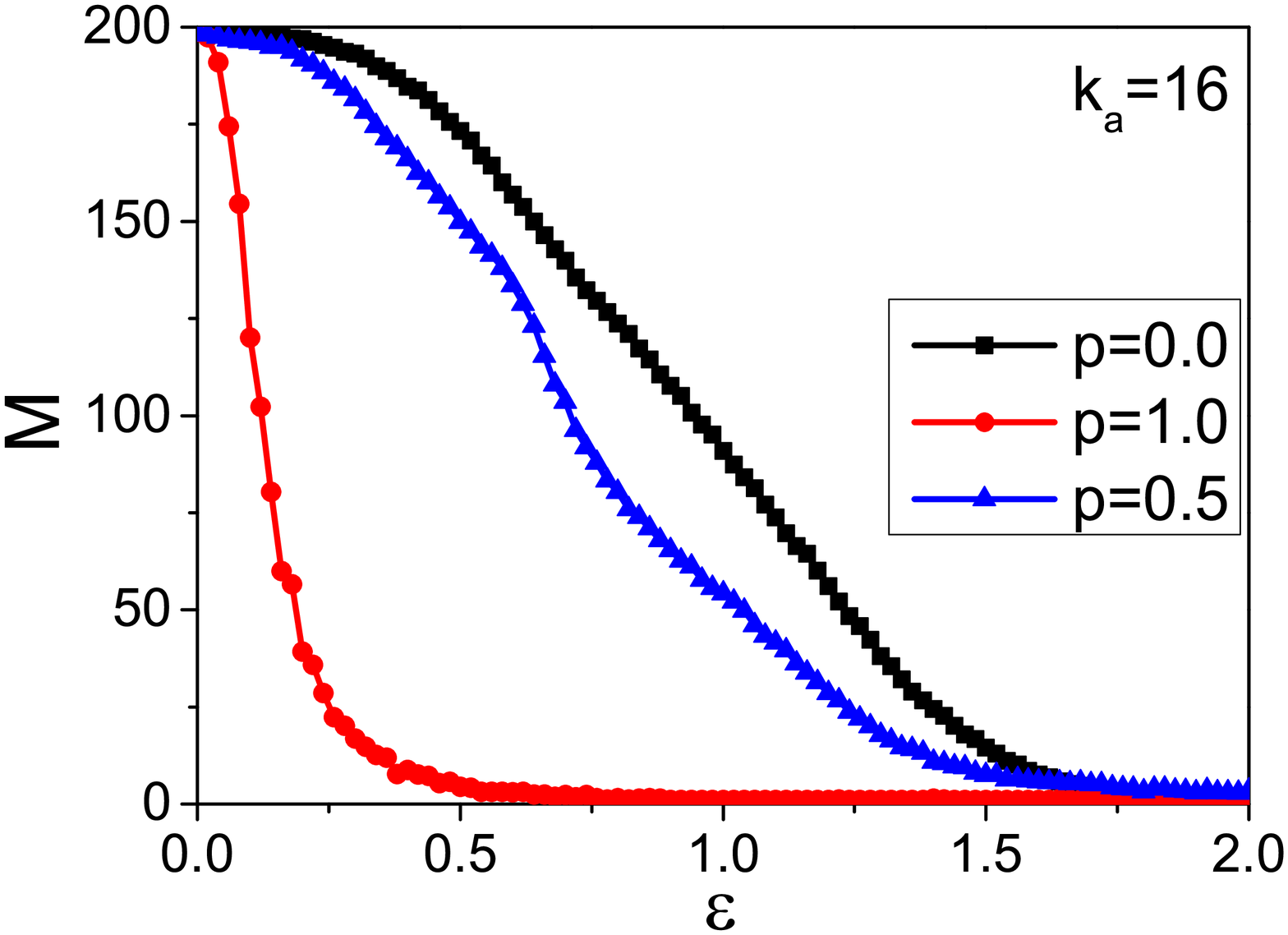}\\
    {\footnotesize $(b)$ Results for BA scale-free networks.}
	\caption{(color online) The number $M$ of opinion clusters in the stable state as a function of the confidence threshold $\varepsilon$ for regular and BA scale-free networks with $\gamma=0$ and $N=1000$. Results are averaged over 1000 realizations.}
	\label{cluster:varepsilon}
\end{figure}

It is of great interest to reveal how smart agents affect the opinion clustering in the dimension of confidence threshold $\varepsilon$ when the resource is concentrated to one clique: $\gamma=0$. Three situations are discussed here: $p=1$ (HK model), $p=0$ (all agents being smart), and $p=0.5$ (coexistence of both general and smart agents). Apparent opinion clustering requires higher confidence threshold in the presence of smart agents, compared to HK model. Furthermore, when the average degree $k_a$ is quite small, there will be no apparent opinion clustering if both general and smart agents are coexisted in the system. This is different from the case where only one kind of agents presents. However, with the increase of $k_a$, coexistence of both kinds of agents can also gradually drive apparent opinion clustering. And the fraction of smart agents has ignorable influence on the system's clustering behavior in opinion for large $k_a$ only if there is smart agents. Above conclusions are applicable to the underlying opinion dynamics on both homogeneous and heterogeneous network structures. This suggests again network structure has no statistically significant impact on the opinion clustering whether there is smart agent or not. More detailed information is presented in Fig. \ref{cluster:varepsilon}.

\section{The influence of parameter $p$ on the generation of a complete consensus} \label{appendix:a}

Particularly, we set $\gamma=0$ and $\varepsilon=2$ in the following analysis. We first divide the opinion space into several adjacent regions. For the simplicity in analysis, the underlying system is assumed to satisfy the mean field limit ($k_a=N-1$). Initially, agents can gain a constant number of scores from gambling if their opinions locate in the real range $[-1, -\frac{1}{4}]$. $Np$ general agents are then update their opinions for the next time by the global average opinion 0 as opinions are uniformly distributed at $t=0$. Smart agents holding opinions outside the real range $[-1, -\frac{1}{4}]$ will adapt to the average opinion of this range at next time. For those agents holding opinions inside this range, they will keep their own opinions.

The update rule for the opinion of agent $i$ at time $t$ can be expressed as
\begin{equation}
o_i(t+1)=\left \{
\begin{array}{ll}
\frac{1}{N}\sum_{j=1}^No_j(t); & i\in N_g,\\
\\
\frac{1}{|N_w(t)|}\sum_{j\in N_w(t)}o_j(t); & i\in N_s,
\end{array}
\right.
\end{equation}
where $N_g$, $N_s$, and $N_w(t)$ denote in sequence the set of general agents, the set of smart ones, and the set of winners:
\begin{equation}
\left \{
\begin{array}{ll}
N_g=\{j, j\in N | j \text{ is a general agent }\},\\
N_s=\{j, j\in N | j \text{ is a smart agent }\},\\
N_w(t)=\{j, j\in N | r_j(t)\geq R/N\}.
\end{array}
\right.
\end{equation}
$|\star|$ returns the number of elements in set $\star$. Uniform distribution of initial opinions yields
\begin{equation}
N_w(0)=\{j, j\in N | -1\leq o_j(0)\leq -\frac{1}{4}\},
\end{equation}
and
\begin{equation}
o_i(1)=\left \{
\begin{array}{ll}
0; & i\in N_g,\\
\\
-\frac{5}{8}; & i\in N_s \text{ and } o_i(0)\in (-\frac{1}{4}, 1],\\
\\
o_i(0); & i\in N_s \text{ and } o_i(0) \in [-1, -\frac{1}{4}].
\end{array}
\right.
\end{equation}
This specifies the opinion space: $[-1, 0]$ at $t=1$. A winner's opinion is limited to the real range $[-1, -\frac{5}{8}(1-p)]$ according to the condition
\begin{equation}
\frac{o_i(1)}{(1-p)N\cdot \frac{-5}{8}}\cdot \frac{R}{N}\geq \frac{R}{N}
\end{equation}
The opinion of an arbitrary agent $i$ is further updated by
\begin{equation}
o_i(2)=\left \{
\begin{array}{ll}
-\frac{5}{8}(1-p), & i\in N_g;\\
\\
-\frac{13-5p}{16}, & i\in N_s \text{ and } o_i(0)\in (-\frac{5}{8}(1-p), -\frac{1}{4}];\\
\\
o_i(1), & i\in N_s \text{ and } o_i(0) \in [-1, -\frac{5}{8}(1-p)],
\end{array}
\right.
\end{equation}
driving opinion space to be $[-1, -\frac{5}{8}(1-p)$ at $t=2$, and winner's opinion space to be $[-1, -\frac{(13+5p)(1-p)}{16}]$. For the next time step, we have
\begin{equation}
o_i(3)=\left \{
\begin{array}{ll}
-\frac{13+5p}{16}(1-p); & i\in N_g,\\
\\
-\frac{16+(13+5p)(1-p)}{32}; & i\in N_s^1,\\
\\
o_i(1); & i\in N_s^2 ,
\end{array}
\right.
\end{equation}
in which
{\small
\begin{eqnarray}
N_s^1=\{j, j\in N_s| \frac{-(13+5p)(1-p)}{16}<o_j(0)\leq \frac{-5(1-p)}{8}\},\\
N_s^2=\{j,j\in N_s|-1\leq o_j(0)\leq -\frac{(13+5p)(1-p)}{16}\}.
\end{eqnarray}
}
This determines the opinion space to be $[-1, -\frac{(13+5p)(1-p)}{16}]$ at $t=3$, and for winners: $[-1, -\frac{16(1-p)+(13+5p)(1+p)(1-p)}{32}]$. Denote the opinion space at an arbitrary time $t$ by $[-1, \varphi(t)]$. We then have the general expression in statistics:
\begin{equation}
\varphi(t)=\left\{
\begin{array}{ll}
1; & t=0,\\
\\
0; & t=1, \\
\\
-\frac{5(1-p)}{8}, & t=2, \\
\\
-\frac{(13+5p)(1-p)}{16}, & t=3,\\
\\
\frac{-(1-p)\left[\sum_{m=0}^{t-4}2^{t-m}(1+p)^m+(13+5p)(1+p)^{t-3} \right]}{2^{t+1}}, & t\geq 4.
\end{array}
\right.
\end{equation}

\nocite{*}
\bibliography{manuscript_revised}

\end{document}